\pdfoutput=1
\documentclass[letterpaper,compsoc]{IEEEtran}

\usepackage[utf8]{inputenc}
\usepackage[T1]{fontenc}
\usepackage[english]{babel}
\usepackage{graphicx}
\usepackage{times}
\usepackage{mathptmx}

\usepackage{etoolbox}
\AtBeginEnvironment{thebibliography}{\linespread{1}\selectfont}

\usepackage[protrusion=true,expansion,tracking=false,kerning,spacing]{microtype}
\microtypecontext{spacing=nonfrench}

\PassOptionsToPackage{warn}{textcomp}
\usepackage{textcomp}

\usepackage[authoryear,round,longnamesfirst]{natbib}
\renewcommand{\cite}[1]{\citep{#1}}

\usepackage[hang,footnotesize]{subfigure}

\usepackage[cmex10]{amsmath}
\usepackage{amssymb}
\usepackage[table]{xcolor}
\usepackage{tabularx,booktabs}
\usepackage{enumitem}
\usepackage{footnote}

\usepackage{caption}
\usepackage{algorithm}
\usepackage{algpseudocode}

\newcommand{\mytitle}{Analysis of Locally Coupled 3D~Manipulation Mappings Based on Mobile Device Motion}
\newcommand{\mysubject}{3D~Manipulation Based on Mobile Device Motion}
\newcommand{\myauthor}{Paul~Issartel, Florimond~Gu\'{e}niat, Tobias~Isenberg, and~Mehdi~Ammi}
\newcommand{\mykeywords}{3D~Interaction, Mobile~Devices, Control-Display~Mappings.}

\title{\mytitle}

\usepackage[final]{hyperref}
\hypersetup{
  pdftitle={\mytitle},
  pdfsubject={\mysubject},
  pdfauthor={\myauthor},
  pdfcreator={LaTeX2e},
  pdfkeywords={\mykeywords},
  plainpages=false,
  hypertexnames=false,
  bookmarksnumbered=false,
  bookmarksopen=true,
  bookmarksopenlevel=3,
  pdfpagemode=UseNone,
  pdfstartview=Fit,
  pdfborder={0 0 0},
  breaklinks=true,
  colorlinks=true,
  linkcolor=black,
  citecolor=black,
  filecolor=black,
  urlcolor=black,
  nesting=true}

\ifx\pdfoutput\undefined%
  \renewcommand{\pdfbookmark}[3][]{}
\fi

\usepackage{authblk}
\author[$*$]{\normalsize Paul Issartel}
\author[$\dagger$]{\normalsize Florimond Gu\'{e}niat}
\author[$\ddagger$]{\normalsize Tobias Isenberg}
\author[$\mathsection$]{\normalsize Mehdi Ammi}
\affil[$*$]{LIMSI-CNRS, Univ.~Paris-Sud, \href{mailto:paul.issartel@limsi.fr}{paul.issartel@limsi.fr}}
\affil[$\dagger$]{Dept.~of~Mathematics, Florida~State~University, \href{mailto:contact@gueniat.fr}{contact@gueniat.fr}}
\affil[$\ddagger$]{INRIA Saclay, \href{mailto:tobias.isenberg@inria.fr}{tobias.isenberg@inria.fr}}
\affil[$\mathsection$]{LIMSI-CNRS, Univ.~Paris-Sud, \href{mailto:mehdi.ammi@limsi.fr}{mehdi.ammi@limsi.fr}%
\vspace{-5mm}}

\newcommand{\eg}{e.\,g.}
\newcommand{\ie}{i.\,e.}
\newcommand\nolabel[1]{\nonumber}
\newcommand*\sceneref[1]{Scene~\ref{#1}}
\renewcommand*\vec[1]{#1}
\newcommand*\quat[1]{#1}
\DeclareMathOperator{\slerp}{slerp}
\DeclareMathOperator{\gain}{gain}
\DeclareMathOperator{\dist}{dist}

\renewcommand\footnotemark[1][]{$^\mathit{#1}$}
\renewcommand\footnotetext[2][]{\footnotesize\textit{#1}.\enskip#2}

\begin{document}

\newcommand{\subfigureautorefname}{Figure}
\renewcommand{\figureautorefname}{Figure}
\renewcommand{\tableautorefname}{Ta\-ble}
\renewcommand{\chapterautorefname}{Chapter}
\renewcommand{\sectionautorefname}{Section}
\renewcommand{\subsectionautorefname}{Section}
\renewcommand{\subsubsectionautorefname}{Section}
\renewcommand{\equationautorefname}{Eq.}

\maketitle

{\small
  \textbf{Abstract}---We examine a class of techniques for 3D~object manipulation on mobile devices, in which the device's physical motion is applied to 3D~objects displayed on the device itself. This ``local coupling'' between input and display creates specific challenges compared to manipulation techniques designed for monitor-based or immersive virtual environments. Our work focuses specifically on the \emph{mapping} between device motion and object motion. We review existing manipulation techniques and introduce a formal description of the main mappings under a common notation. Based on this notation, we analyze these mappings and their properties in order to answer crucial usability questions. We first investigate how the 3D~objects should move on the screen, since the screen also moves with the mobile device during manipulation. We then investigate the effects of a limited range of manipulation and present a number of solutions to overcome this constraint. This work provides a theoretical framework to better understand the properties of locally-coupled 3D~manipulation mappings based on mobile device~motion.%
\par}

\section{Introduction}
\label{sec:introduction}

Mobile devices differ from traditional computers in that they combine input, display, and processing capabilities into a single handheld object. Recent technological advances have made it possible to run 3D~applications directly on mobile devices. One of the fundamental tasks~\citep{bowman04} in such applications is object manipulation, \ie~the translation and rotation of objects in 3D~space. A major challenge for 3D manipulation tool design is thus to create efficient 3D~manipulation techniques, tailored to the unique characteristics of this portable and integrated environment.

Currently, the most common way to interact with mobile devices is by means of an integrated touch screen. Each contact on a touch screen provides two degrees of freedom~(DOF). While this type of input is well suited to 2D~interaction, 3D~manipulation requires three degrees of freedom for translations and three for rotations. The constraint of 2-DOF input often leads to complex and unnatural 3D~manipulation techniques. An alternative type of input exists in the form of tangible interaction: manipulating physical objects around the mobile device~\citep{issartel14}. The motion of these physical objects is then mapped to the 3D~objects displayed on the device's screen. Tangible input integrates all six degrees of freedom required for 3D~interaction into a simple and natural way that takes advantage of real-world manipulation skills~\citep{ishii08}. One important drawback, though, is that the user must carry and handle several objects in addition to the mobile device.

\begin{figure}[t]
  \centering
  \vspace{-1mm}
  \includegraphics[width=\linewidth]{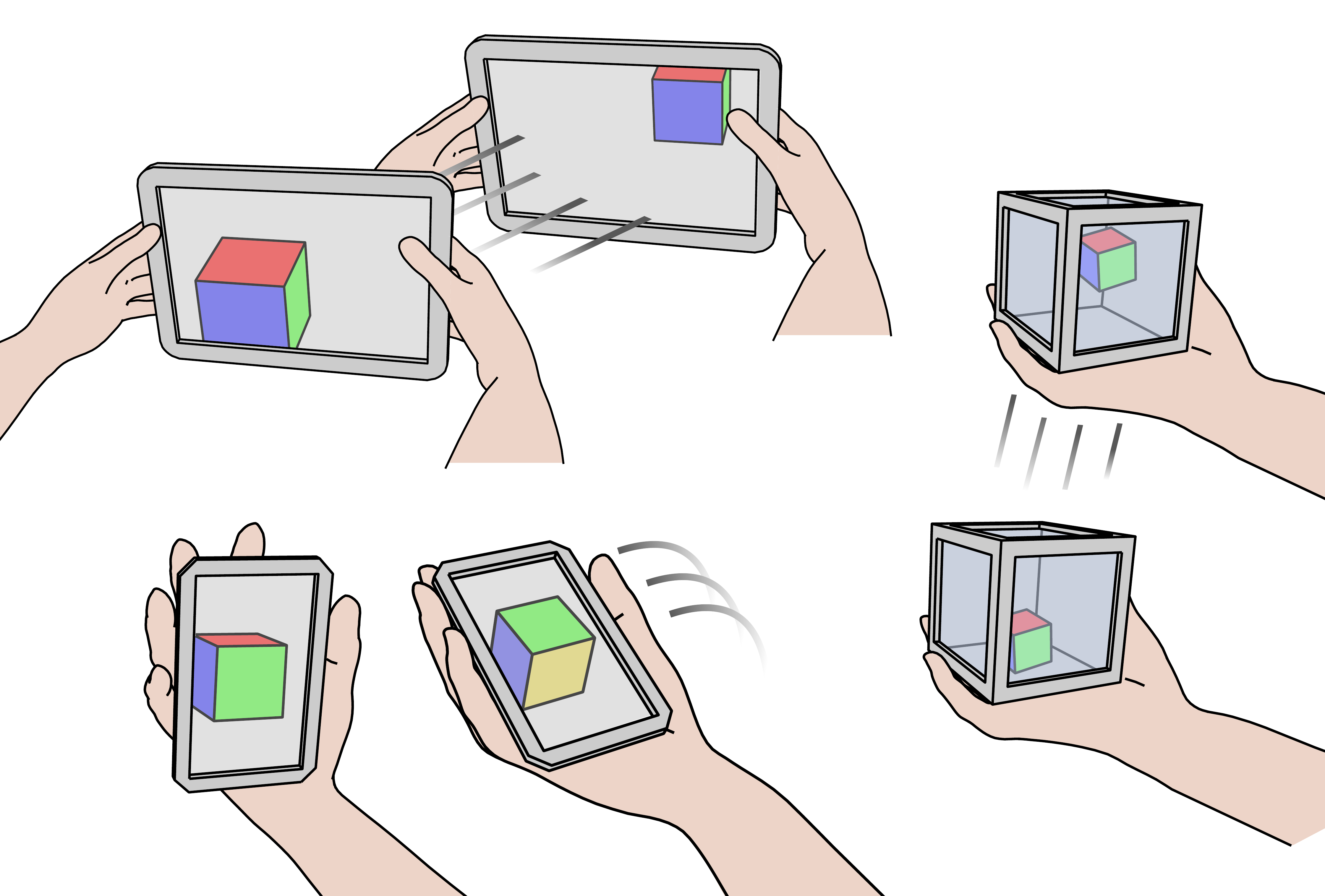}
  \caption[Illustration of locally-coupled manipulation]{\hyphenpenalty=10000
    Using the motion of a mobile device to translate and rotate a 3D~object displayed on the device itself (locally coupled manipulation). This is illustrated here on different types of mobile devices. In this paper, we specifically focus on the \emph{mapping} between device motion and object motion.
  }
  \label{fig:intro}
  \vspace{-2mm}
\end{figure}

In this paper, we investigate a different class of techniques which retain the advantages of tangible interaction but do not require any external objects. They consist in using the \emph{\mbox{mobile device} itself} as a tangible input device, by measuring its own motion relative to the environment. In other words, these techniques use the physical motion of the mobile device in the real world to control a 3D~object on the device's screen\footnote{Note that this refers to object manipulation in \emph{screen space}, rather than manipulation of objects located in an external reference frame.} (\autoref{fig:intro}). Compared to the previously mentioned interaction modes, this approach has clear advantages. Unlike touch input, it provides sufficient degrees of freedom for 3D~interaction. Unlike the tangible interfaces described above, it does not require any separate objects.

However, this configuration also presents important challenges. The screen on which the manipulated object is displayed is coupled with the input device---a ``locally coupled'' configuration~\citep{rahman09}. Therefore, \emph{the screen moves and rotates along with the input device}, a fundamental difference from the typical case of a fixed monitor with a separate input peripheral. This coupling raises crucial usability questions. The first question is how to match visual feedback to device motion. Objects displayed on the screen appear to move from the user's point of view, since the screen is moving during manipulation. This raises the issue of how the manipulated object should move on the \emph{screen} itself, so that its apparent motion remains consistent with the device's motion. Another question is whether users see the device as a ``handle'' that controls the object, or a ``window'' that controls the viewpoint. A third issue is the limited range of manipulation. As with any handheld input device, this range is limited by the space reachable by the user. But the screen is attached to the device itself. Since the screen provides visual feedback, it must remain legible during manipulation, which further reduces the usable range of motion.

In order to address the previous questions, it is essential to understand well the \emph{mapping} between device motion and object motion. In this analysis, we thus specifically focus on the mappings themselves. As we will see, several researchers have proposed manipulation techniques that were based on mobile device motion. Many of them, however, have emphasized the application rather than the mapping. We thus aim to provide an explicit discussion and detailed description of the possible mappings, facilitating a comprehensive understanding of their properties.

In this work, we contribute a \emph{theoretical framework} for locally-coupled 3D~manipulation mappings based on mobile device motion. We begin with a review of existing manipulation techniques, followed by a discussion of their common aspects. We then introduce a formalization of the main mappings and unify them under a common notation. Using this formalism, we proceed with an analysis of these mappings in order to demonstrate their properties. Our analysis addresses two main questions: how the object should move on the screen to match device motion, and how to address the constraints of a limited motion space. For each property of the mappings, we first examine existing arguments from previous work. However, where previous evidence is lacking or inconclusive, we contribute new theoretical and experimental results to answer the above questions. Based on this analysis, we finally discuss possible adaptations and improvements for each mapping. By providing a comprehensive, formalized, and substantiated overview of these mappings, our framework assists designers in making more informed choices when implementing such techniques.

\section{Existing manipulation techniques}

As a first step to establish our theoretical framework, we review ex\-is\-ting 3D~manipulation techniques based on mobile device motion.

\subsection{3D~manipulation through physical objects}

The idea of using a handheld physical object---in this case, a mobile device---to manipulate virtual 3D~objects can be related to graspable user interfaces~\citep{fitzmaurice96} and, more generally, to tangible interaction~\citep{ishii08}. One of the earliest examples was the PassProps prototype by \citet{hinckley94} in which tangible objects are tracked in real space and their position and orientation are mapped to 3D~objects shown on an external display. Similar examples are the Cubic~Mouse~\citep{frohlich00} and the CID~device~\citep{vanrhijn06}. The use of tangible objects for manipulation is a rather natural mode of interaction since it exploits the user's real-world manipulation skills~\citep{ishii08}. The projects mentioned above, however, require custom-made objects and specific sensors for input and tracking.

With the increasing availability of mobile devices, many projects have proposed to use handhelds as readily-available tangible objects with built-in sensors~\citep{katzakis09,ha13,benzina12,song11,ha11,liang13,du11}. These interfaces allow users to manipulate virtual 3D~objects through the motion of a mobile device. Although device motion provides interactive control, the manipulated objects are still displayed on an external screen. Thus, the manipulation does not actually occur on the mobile device itself.

Alternatively, tangible objects can be used in combination with a mobile device \citep{issartel14,liang13}: tangible objects serve as input, while the mobile device processes and renders the manipulated 3D~objects on its integrated screen. With this approach, the manipulation takes place on the mobile device as the entire interface is portable and self-contained. However, the user also has to handle several ob\-jects during the manipulation which can be ergonomically chal\-len\-ging. Moreover, external tangible objects need to be inconveniently carried with the mobile device to wherever the interface is used.

The next logical step is to use a mobile device as tangible input to manipulate objects displayed \emph{on} the device. We survey these types of approaches and discuss them within our framework.

\subsection{On-device interaction based on device motion}

A number of existing mobile interaction techniques exploit the motion of a mobile device to translate and rotate objects on its own screen. Many such techniques are tailored for~1D or 2D~interaction, but some of them are actually designed for 3D~manipulation.

\subsubsection{Tilt-based interaction}

In one of the first works on the subject, \citet{rekimoto96} proposed to use device inclination (``tilt-based interaction'') to navigate menus on a palmtop computer. According to the given description, the current position within the menu directly depends on the device angle. \citet{weberg01} also described an interface that uses the device's tilt to navigate menus and select menu items on a PDA~device. In this case, however, the device inclination controls the \emph{rate} of motion within the menu. \citet{oakley05} evaluated both approaches for menu selection. We can thus identify two ways of mapping mobile device motion to a manipulated object: one that directly controls the position of the object (position control), and another that controls its rate of motion (rate control).

Many other works have investigated tilt-based interaction. Scrolling in lists, documents, and images by tilting a mobile device seems to be a frequently studied task. Early works~\citep{small97,harrison98,bartlett00} appear to use rate control, but the exact mapping is only informally described. Unfortunately, the lack of formalization makes these mappings ambiguous and difficult to compare to each other. Subsequent works on tilt-to-scroll~\citep{hinckley00,eslambolchilar08,cho07} then introduced more formally described rate control mappings. \citet{rahman09} present a thorough study of tilt-based position control mappings for 1-DOF discrete input. Tilt-based interaction has also been used for 2D~panning and zooming. The RotoView technique~\citep{feinstein02}, for example, facilitates map navigation with a rate control mapping. \citet{joshi12} present a hybrid position-rate control mapping to visualize 360\textdegree{} panoramas. Finally, tilt-based interaction has been studied for pointing. \citet{tsandilas13} compared rate control, position control and hybrid control for 1D~pointing, with formal descriptions of each mapping. \citet{teather14} compared position control and rate control mappings for a 2D~pointing task. This task is closer to a 3D~manipulation than previous examples, since it involves accurate manipulation of an object (pointer) on the screen with multiple degrees of freedom.

\subsubsection{Spatially-aware displays}

The tilt-based techniques mentioned so far only use device orientation as input. Interfaces where the position of a mobile device serves as an input modality tend to be categorized as \emph{spatially-aware displays}. For example, \citet{small97} presented a system to visualize long paintings, using a wheel-mounted monitor which scrolls its contents when rolled on the floor. Its mapping is not described in detail but appears to be position-controlled. \citet{yee03} presented the ``peephole display'' in which movements of a PDA---tracked with te\-thers---al\-low the user to pan and navigate workspaces larger than the device's screen. Again, the mapping is position-controlled but not formally described. \citet{spindler14} demonstrated a similar approach with an infrared-tracked mobile device. \citet{wang06} used a mobile device's internal camera to track its own translations and rotations, mapping them to various 2D~interaction tasks. In one of the only works to mention both position and rate control mappings in a spatially-aware display, \citet{eriksson06} also used the integrated camera to track the device position and orientation, for several possible applications.

Overall, there seems to be fewer works that exploit device position than device orientation. This fact may be due to the complexity of tracking a position compared to an orientation. The device orientation can be easily tracked with integrated and inexpensive sensors, such as gyroscopes and magnetometers. Such sensors have long been found in many mobile devices. In contrast, tracking the device position is more difficult. Some of the above projects use wheels, wires, or external infrared~(IR) sensors which are unwieldy and impractical in a mobile setting. Other projects choose to use an integrated camera. Now that cameras are becoming ubiquitous and embedded processing power becomes sufficient for real-time image analysis, inside-out optical tracking seems to be the most promising solution for small scale position tracking on a mobile device. The recently launched Tango project\footnote{\url{{http://www.google.com/atap/project-tango/}}}, a tactile tablet featuring inside-out motion tracking, may open the way for more applications of position tracking.

\subsubsection{3D~manipulation based on device motion}

Although the mapping of device motion to~1D or 2D~tasks can serve as a basis for 3D~manipulation mappings, there is no substitute for studies focusing on actual 3D~tasks. Only such studies can highlight the constraints and challenges specific to 3D~interaction.

\citet{fitzmaurice93} described the Chameleon system in which the position of a handheld monitor controls the viewpoint on a displayed 3D~scene. Subsequent works~\citep{tsang02} later improved this concept by tracking the device orientation in addition to its position, facilitating a full control of the viewpoint. These projects, however, primarily simulated a window on a virtual scene---restricting the possible mappings to an isomorphic position control and excluding other mappings that might be useful for 3D~object manipulation.

Other projects use the motion of a mobile device for actual 3D~manipulation. Some of them demonstrate 3D~object manipulation in augmented reality~(AR). \citet{henrysson05} and \citet{marzo14} described a ``grasping'' metaphor in which, during the manipulation, the object remains fixed relative to the mobile device. A drawback of this approach is that it makes it difficult to rotate the manipulated object without translating it. Since the virtual scene is fixed in an external reference frame and the manipulated object is fixed in the device reference frame, the mobile device must be moved across an arc. The HOMER-S technique~\citep{mossel13} eliminates this issue by separately applying device rotations to the manipulated object. As a consequence, however, the object is no longer fixed relative to the mobile device and can thus leave the field of view during large rotations. These approaches cannot avoid both of these problems as they are caused by the intrinsic separation between the object's reference frame and the device's reference frame in normal perspective rendering. A different approach is the concept proposed by \citet{spindler12} which uses a head-coupled perspective to let the device intersect the manipulated object, thus greatly reducing its separation from the object. Assuming head tracking is available---which can be challenging to accomplish in a truly mobile interface---this approach can solve the rotation issue. Yet, all the ``grasping'' techniques share another drawback: the object must remain fixed relative to the device, thus the translation mapping is restricted to isomorphic position control even though different mappings might be desirable in some situations (\autoref{sec:motion-space}).

The alternative is to perform 3D~manipulation entirely in the device reference frame, \ie~in screen space, avoiding the constraints caused by an external reference frame. \citet{kratz10} compared a tilt-based rotation mapping with a two-sided touch metaphor on a smartphone. Their tilt-based mapping uses rate control but only supports rotation on two axes. \citet{neale13} presented an interface to visualize museum artifacts on a tactile tablet. They compared touchscreen input to both tilt-based position control and rate control mappings. This interface, however, only supports object rotation and the mappings are not described in detail. \citet{daiber12} presented an interface to translate and rotate 3D~objects on a smartphone. The tilt-based rotation mapping appears to be position-controlled. Their translations, however, are not based on device motion but on touch gestures. The PDDM device by \citet{noma96} is a rare example of using both device translations and rotations for screen-space 3D~manipulation. The device is a palmtop monitor mounted on a mechanical arm. The authors presented four different mappings for screen-space object manipulation, all based on position control. The mappings are explained and illustrated, but not formally described. Furthermore, the study of the mappings themselves was still limited in scope. Important questions such as the frame of reference of manipulation were only mentioned as future work.

As we can see, a few existing works use the motion of a mobile device for actual screen-space 3D~manipulation. But each of them only addresses a small subset of the possible mappings. Some only consider rotations and ignore translations, others only use position control, and yet others only consider rate control. The authors generally do not provide a formal description of the proposed mappings, making it difficult to generalize the results. In particular, the lack of a formal notation makes it impossible to assess key usability properties such as the matching between visual feedback and device motion and how well the proposed mappings make use of the limited motion space. In the rest of the paper we, therefore, conduct an in-depth analysis of the mappings themselves and their properties. We start by presenting a formalization of the main mappings and then use this notation to determine how well they address the above-mentioned usability questions.

\section{About the term ``mobile device''}
\label{sec:mobile-device-definition}

Early approaches that used device motion as an input modality associated the mobile device concept with technologies such as por\-ta\-ble~TVs, PDAs, and palmtop computers. Today, the term ``mobile device'' generally refers to smartphones, tablets, phablets, or a size variation thereof. These devices all share a similar form factor: they are mostly flat, rectangular, and have a single screen on one side.

There is no reason, however, why a 3D~manipulation mapping could not work with other device shapes (\eg,~\autoref{fig:intro}). There have been proposals for adding a screen on the other side of current mobile devices \citep[\eg,][]{kratz10}, for creating highly deformable mobile devices \citep[\eg,][]{ramakers14}, and for creating tiltable devices \citep[\eg,][]{alexander12}. The recent interest for ``smart watches'' is driving the industry and academia toward the development of wristband-shaped displays \citep[\eg,][]{lyons12}. There are prototypes of small, portable cubic displays~\citep{lopez-gulliver09,stavness10} with a screen on each face, capable of displaying a 3D~scene as if it were inside the cube. Spherical screens~\citep{benko08} and volumetric globes~\citep{grossman06} are also being investigated. These remain too large to be considered ``mobile'', but could be down-sized as technology advances. Future mobile devices might thus have a radically different shape than current ones.

In this article we thus define a mobile device in a rather generic way: any interactive physical (tangible) object that can be easily carried by one person's own hands and is capable of displaying virtual objects on its surface or inside its volume. All our conclusions remain applicable to any device which corresponds to this definition, unless otherwise specified.

There is still an additional requirement for a mobile device to be compatible with the 3D~manipulation mappings discussed here. The manipulated virtual object \emph{must} appear to have a single defined position and orientation within the device reference frame. The reason for this additional constraint is that it would be impossible to know the exact location of the object, and object manipulation would no longer make sense, if multiple copies of a single virtual object were to appear at conflicting locations.

This constraint, however, normally does not pose a problem for devices that have a single flat screen since there is only one view of the virtual scene. For devices with multiple non-coplanar screens or devices covered with a curved screen, special care must be taken to ensure that a virtual object does not appear at multiple conflicting locations. This can be accomplished with perspective correction, \ie~by~ensuring that each screen (or each point of the surface) shows the virtual scene from a different perspective such that a virtual object appears at a fixed location within the device volume. This solution requires either autostereoscopic displays~\citep{lopez-gulliver09}, true volumetric displays~\citep{grossman06}, or a way to continuously track the position of the user's eyes in order to update the perspective~\citep{stavness10}.

\section{Formalization of the main mappings}
\label{sec:formalization}

A \emph{mapping}, also called transfer function, describes how device motion is mapped to object motion on the screen and we now present the main ways to perform such a mapping. We express them in a unified formal notation, allowing us to compare them and assess their properties in the next sections. Unlike many previous works, we consider both translations and rotations in our formal model. We also provide a pseudocode description in the appendix.

\subsection{Basic notation}

The values~$\vec{pc_t}$ and~$\quat{qc_t}$ represent the position and orientation, respectively, of the mobile device at time~$t$. They are the \emph{control} values, obtained from tracking and expressed in an arbitrary tracking-specific coordinate system. The position~$\vec{pc_t}$ is a 3D~vector, while the orientation~$\quat{qc_t}$ is a quaternion that represents the rotation of the device relative to some base orientation.

The values~$\vec{pd_t}$ and~$\quat{qd_t}$ represent the position and orientation of the manipulated object at time~$t$. They are the \emph{display} values, expressed in the screen coordinate system. The position~$\vec{pd_t}$ is a 3D~vector, while the orientation~$\quat{qd_t}$ is a quaternion that represents the rotation of the object relative to some base orientation on the screen. The display values~$\vec{pd_t}$ and~$\quat{qd_t}$ are computed from the control values by applying the mapping function.

Time~$t\,{=}\,0$ designates the beginning of manipulation, \ie~the time when the user starts manipulating the object.\footnote{Users should be able to explicitly engage or disengage manipulation mode with, \eg, a dedicated button to move the device without affecting the object.} The values~$\vec{pc_0}$ and~$\quat{qc_0}$ thus represent the initial position and orientation of the mobile device. Similarly, $\vec{pd_0}$ and~$\quat{qd_0}$ represent the initial position and orientation of the manipulated object. Each subsequent time~$t$ indicates the time when a new sample is obtained from the tracking system. Time increments are unitary in our notation.

\subsection{From tracker coordinates to screen coordinates}
\label{sec:tracker-to-screen-space}

\begin{figure}[t]
  \centering
  \includegraphics[width=\linewidth]{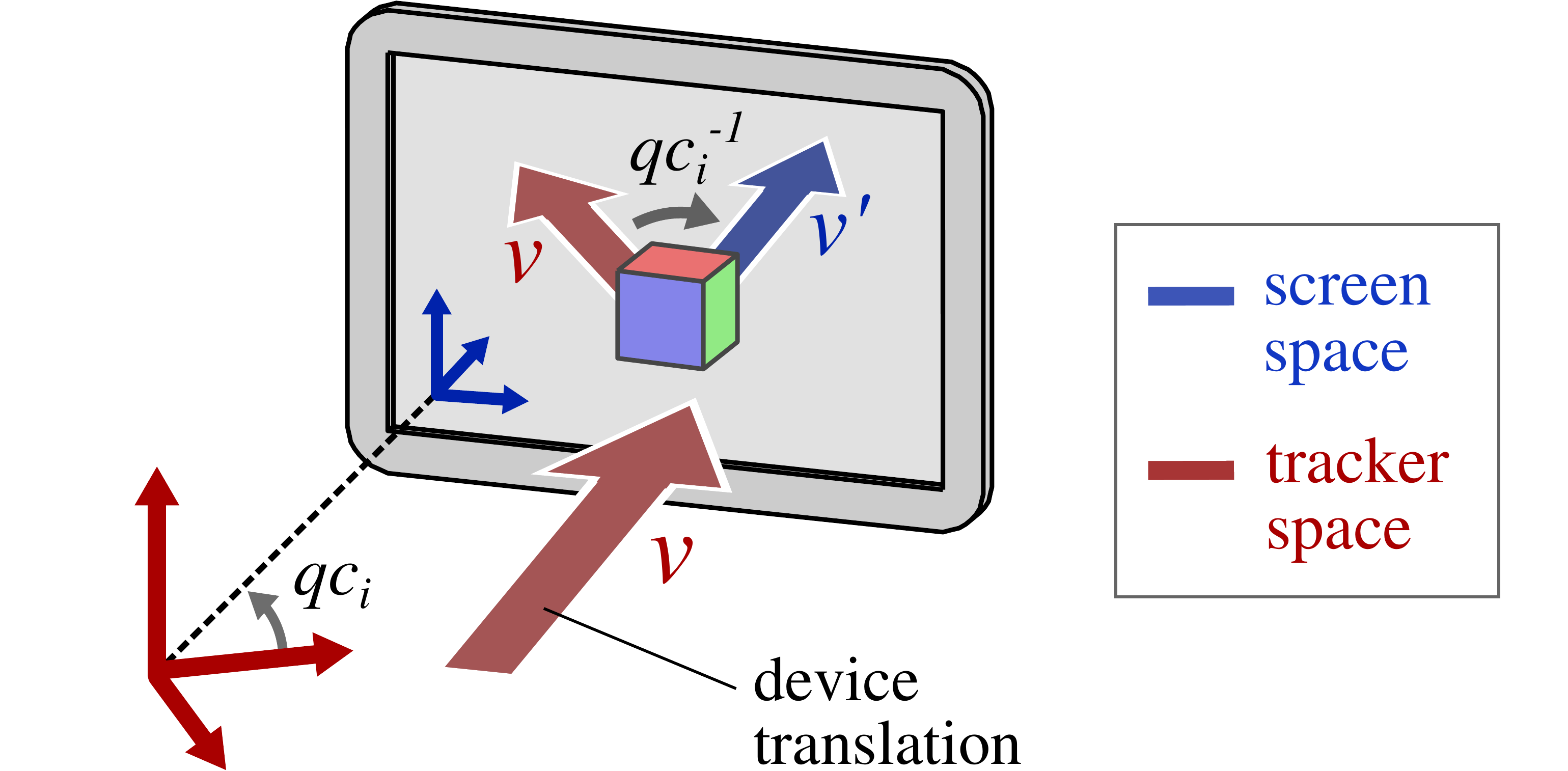}
  \caption[Accounting for device orientation during manipulation]{Directly applying the measured device translation~$\vec{v}$ to the manipulated object would move it in unexpected directions, depending on the device orientation~$\quat{qc_i}$ in tracker space. To get a more predictable behavior, the vector~$\vec{v}$ should be rotated by the inverse rotation~$\quat{qc_i}^{-1}$, producing the correct translation~$\vec{v'}$. The same process is applied to rotations.}
  \label{fig:trackertoscreen}
\end{figure}

Control values (positions~$\vec{pc_t}$ and orientations~$\quat{qc_t}$ of the mobile device) are measured by the tracking system in a tracking-specific reference frame. Consequently, the motion of the mobile device is also expressed in this reference frame. But the manipulated object belongs to the screen reference frame. Therefore, the mapping function must convert device motion into the screen reference~frame.

In the locally coupled configuration we study here, the screen is attached to the input device itself. Thus, a rotation of the device during manipulation also rotates the screen in relation to the tracking system, causing the screen and tracking reference frames to become \emph{misaligned} (\autoref{fig:trackertoscreen}). If the measured device motion was directly applied to an object on the screen, the object would move in unexpected directions. Converting device motion into the screen reference frame requires to compensate for this misalignment.

Consider a translation~$\vec{v}$ and a rotation~$\quat{r}$ of the mobile device, measured in the tracking reference frame between times~$i$ and~$j$. At the beginning of movement, the device orientation in the tracking reference frame is~$\quat{qc_i}$. Since the screen is attached to the mobile device, the screen orientation is also~$\quat{qc_i}$ (for the sake of simplicity, we assume a null offset between the device and the screen). To map $\vec{v}$ and~$\quat{r}$ to the screen reference frame, this orientation must be \emph{canceled}, hence re-aligning the reference frames (\autoref{fig:trackertoscreen}). We achieve this re-alignment by applying the inverse rotation~$\quat{qc_i}^{-1}$ to $\vec{v}$ and~$\quat{r}$. We thus apply this inverse rotation to the direction (the vector part) of the quaternion~$\quat{r}$ using the conjugation operation $\quat{qc_i}^{-1}\;\quat{r}\;(\quat{qc_i}^{-1})^{-1}$, shortened to $\quat{qc_i}^{-1}\;\quat{r}\;\quat{qc_i}$. For this purpose we consider the translation vector~$\vec{v}$ as a quaternion whose real part is zero, and apply the same operation. In summary, the new translation~$\vec{v'}$ and rotation~$\quat{r'}$ (corresponding to the translation~$\vec{v}$ and the rotation~$\quat{r}$ measured by the tracking system) are hence obtained as follows:
\begin{align}
  \begin{split}
    \vec{v'} &= \quat{qc_i}^{-1}\;\vec{v}\;\quat{qc_i}\\
    \quat{r'} &= \quat{qc_i}^{-1}\;\quat{r}\;\quat{qc_i}
    \label{eq:trackertoscreen}
  \end{split}
\end{align}

\begin{figure*}
  \includegraphics[width=\textwidth]{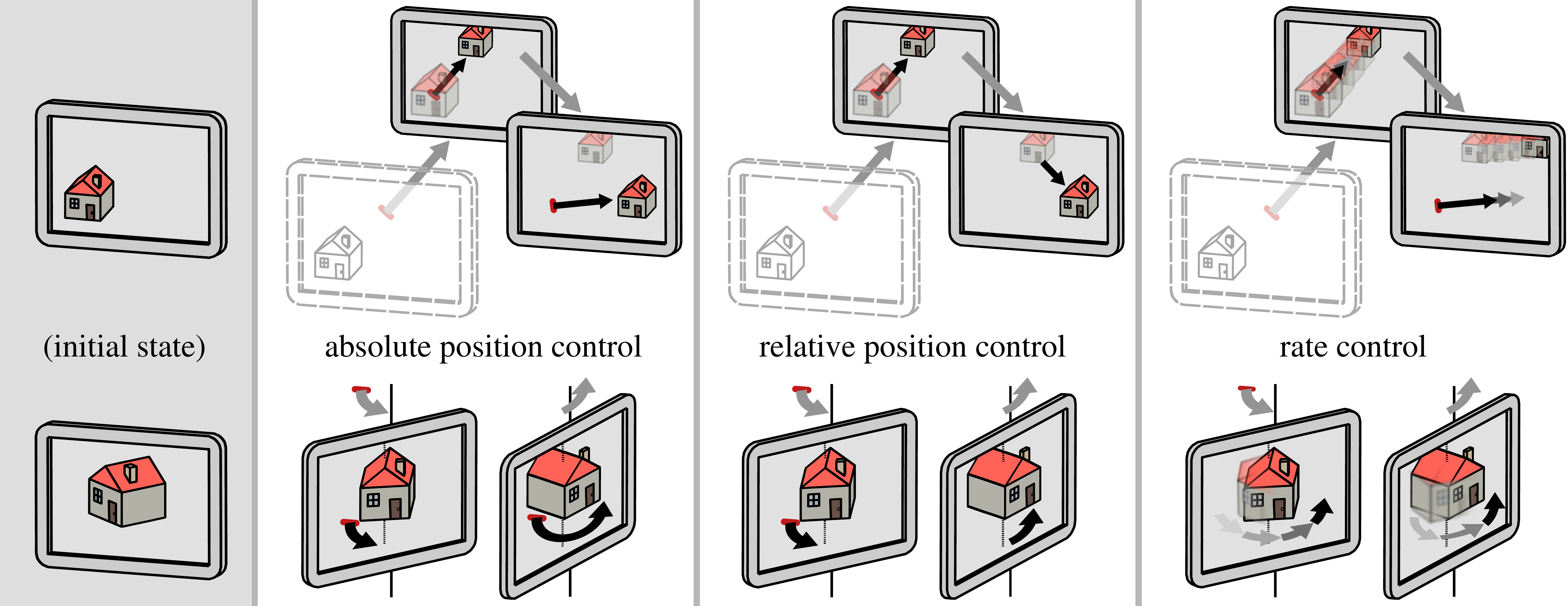}%
  \caption[Illustration of the three main mappings studied here]{%
    Three main ways to map mobile device motion to a 3D~object's motion, shown for a tablet.
    Absolute~position control: the device displacement \emph{from its initial location}
    is applied to the object.
    Relative position control: the \emph{incremental} device displacement is applied to the object.
    Rate control: the device displacement from its initial location controls the object's \emph{velocity}.\hspace{-.8ex}
  }%
  \label{fig:base-mappings}
\end{figure*}

This transformation expresses device translations and rotations in a stable reference frame. We can now apply these transformations to a 3D~object displayed on the screen. As demonstrated in previous work, however, there are different ways to apply device motion to a manipulated object (\autoref{fig:base-mappings}) as we will see next.

\subsection{Position control mappings}

In a ``position control'' (or zero-order) mapping, the motion of an input device  directly controls the position and orientation of the manipulated object~\citep{zhai95}. In our case this means that translations of the mobile device control the position of the displayed object, while rotations of the device control the orientation of the object.

Two main ways of mapping exist to control an object's position and orientation. The first one---an ``absolute'' mapping---directly assigns the position and orientation of the mobile device to the object, as measured in some fixed reference frame. The other way---a ``relative'' mapping---applies \emph{incremental} translations and rotations of the device (\ie,~its change in position and orientation between each times~$t$ and~$t{-}1$) to the object. Our notion of absolute and relative mappings reuses the terminology proposed by \citet{poupyrev00} for rotations, which we extend to also include translations.

\textbf{Absolute and relative sensors}\quad
The distinction between absolute and relative mappings has a practical significance. Some tracking sensors measure an ``absolute'' position or orientation, \ie~expressed in a static reference frame outside of the mobile device. For example, a mechanical arm can measure the location of the device relative to its base, and a magnetometer can measure the orientation of the device in the Earth's reference frame. An embedded camera can track a fixed marker in the environment~\citep{eriksson06} or sense IR~light reflected from the surrounding environment~(Tango project) to measure the device's position and orientation. Other sensors only measure relative motion such as the gyroscopes or accelerometers found on many current devices. An absolute sensor can be used with both absolute and relative mappings, whereas a relative sensor is not suitable for absolute mappings due to drift. However, relative inside-out sensors are generally fully contained in the mobile device itself and do not depend on the external environment, which is a strong benefit for portability. Although some absolute sensors can be physically embedded in the device (\eg,~cameras and magnetometers), they are easily disrupted by some environmental conditions (lack of visual markers, insufficient or excessive ambient light, presence of magnetic materials, etc.). Thus, the use of relative sensors rather than absolute ones might be dictated by technical~constraints.

Aside from these practical aspects, the distinction between absolute and relative tracking also has an impact on usability since---as we demonstrate next---absolute and relative position control mappings do not have the same properties.

\subsubsection{Absolute mapping}

An absolute mapping~\citep{poupyrev00,bowman04} directly applies the device position and orientation ($\vec{pc_t}$ and $\quat{qc_t}$) to the manipulated object position and orientation ($\vec{pd_t}$ and $\quat{qd_t}$).
To make incremental manipulation possible, the mapping must take into account the initial position and orientation of the object ($\vec{pd_0}$ and~$\quat{qd_0}$). To ensure that the object does not move unexpectedly at the beginning of manipulation, it is also necessary to subtract the initial position and orientation of the mobile device ($\vec{pc_0}$ and~$\quat{qc_0}$). This corresponds to a translation $\vec{pc_t}-\vec{pc_0}$ and a rotation $\quat{qc_t}\:\quat{qc_0}^{-1}$ which are to be applied to the object's initial position $\vec{pd_0}$ and orientation~$\quat{qd_0}$. As we explained in \autoref{sec:tracker-to-screen-space}, the device translations and rotations must be converted into screen space. Since they are measured from the initial device location at~$t=0$, a rotation of~$\quat{qc_0}^{-1}$ is applied. The absolute mapping is thus given by:
\begin{equation}
  \begin{split}
    \vec{\Delta pc_t} &= \quat{qc_0}^{-1}\; (\vec{pc_t}-\vec{pc_0})\; \quat{qc_0}\\
    \quat{\Delta qc_t} &= \quat{qc_0}^{-1}\; (\quat{qc_t}\:\quat{qc_0}^{-1})\; \quat{qc_0}\\
    \vec{pd_t} &= \vec{\Delta pc_t} + \vec{pd_0}\\
    \quat{qd_t} &= \quat{\Delta qc_t}\; \quat{qd_0}
    \label{eqn:absolute-mapping}
  \end{split}
\end{equation}

\subsubsection{Relative mapping}

Rather than directly applying the device position and orientation to the object, we can also apply \emph{incremental} translation and rotation offsets. A relative mapping~\citep{poupyrev00,bowman04} applies incremental device translations and rotations, measured between times~$t{-}1$ and~$t$ ($\vec{pc_t}-\vec{pc_{t-1}}$ and~$\quat{qc_t}\:\quat{qc_{t-1}}^{-1}$, respectively), to the current object position and orientation ($\vec{pd_{t-1}}$ and $\quat{qd_{t-1}}$, resp.). Again, device translations and rotations must be converted into screen space. Since they are measured from time~$t{-}1$, a rotation of~$\quat{qc_{t-1}}^{-1}$ is applied. The relative mapping is thus given by:
\begin{equation}
  \begin{split}
    \vec{\Delta pc_t} &= \quat{qc_{t-1}}^{-1}\; (\vec{pc_t}-\vec{pc_{t-1}})\; \quat{qc_{t-1}}\\
    \quat{\Delta qc_t} &= \quat{qc_{t-1}}^{-1}\; (\quat{qc_t}\:\quat{qc_{t-1}}^{-1})\; \quat{qc_{t-1}}\\
    \vec{pd_t} &= \vec{\Delta pc_t} + \vec{pd_{t-1}}\\
    \quat{qd_t} &= \quat{\Delta qc_t}\; \quat{qd_{t-1}}
    \label{eqn:relative-mapping}
  \end{split}
\end{equation}

In order to unify all the main mappings under a common formalism, our notation assumes the availability of \emph{absolute} tracking information. However, relative sensors embedded into the device (such as gyroscopes) provide \emph{incremental} device translations and rotations. Since those incremental translations and rotations are already expressed in the device's reference frame, they do not have to be converted into screen space. Therefore, the values returned by such sensors can be directly used in place of the~$\vec{\Delta pc_t}$ and~$\quat{\Delta qc_t}$ terms in \autoref{eqn:relative-mapping}.

As a consequence, only the relative position-control mapping should be used with relative sensors, since it does not require the absolute~$\vec{pc_t}$ and~$\quat{qc_t}$ values to be known (unless the sensor values are integrated, ultimately leading to drift).

\subsection{Rate control mapping}

In a ``rate control'' (or first-order) mapping, the motion of the input device controls the \emph{velocity} (linear or angular) of the object~\citep{zhai95}. The mobile device can be translated and rotated in 3D~space from its initial position~$\vec{pc_0}$ and orientation~$\quat{qc_0}$. In a rate control mapping, the linear velocity of the manipulated object increases when the mobile device moves away from its initial position, and decreases when returning to this point. The linear velocity thus depends on the translation vector $\vec{pc_t}-\vec{pc_0}$. Similarly, the angular velocity of the manipulated object depends on the rotation of the device from its initial orientation, i.e.,~$\quat{qc_t}\:\quat{qc_0}^{-1}$. Since those displacements are relative to the initial device location (at~$t\,{=}\:0$), a rotation of~$\quat{qc_0}^{-1}$ is applied to convert them to screen space. Applying a linear and angular velocity to an object means adding these to its current position and orientation ($\vec{pd_{t-1}}$ and~$\quat{qd_{t-1}}$) at each time~$t$. The rate control mapping is thus given by:
\begin{equation}
  \begin{split}
    \vec{\Delta pc_t} &= \quat{qc_0}^{-1}\; (\vec{pc_t} - \vec{pc_0})\; \quat{qc_0}\\
    \quat{\Delta qc_t} &= \quat{qc_0}^{-1}\; (\quat{qc_t}\:\quat{qc_0}^{-1})\; \quat{qc_0}\\
    \vec{pd_t} &= \vec{\Delta pc_t} + \vec{pd_{t-1}}\\
    \quat{qd_t} &= \quat{\Delta qc_t}\; \quat{qd_{t-1}}
    \label{eqn:rate-mapping}
  \end{split}
\end{equation}

\subsection{Higher-order control}

Position control is a zero-order mapping: it directly maps device positions and orientations to the object. Rate control is a first-order mapping: it maps the device location to the object \emph{velocity}, \ie~the derivative of position and orientation. While higher-order mappings such as acceleration control\footnote{The metaphor for a second-order mapping would be an impulse being applied to an object that causes it to continue moving until a reverse impulse is applied---similar to what happens on space vehicles~\citep{massimino89}.} are possible, they are known to perform worse than position control and rate control~\citep{massimino89,zhai95}. We thus do not consider them here.

\subsection{Control-display gain}
\label{sec:gain}

The mappings as they are formulated above do not change the scale of device movements which are applied to the object. Such mappings are called \emph{isomorphic} \citep{poupyrev00,zhai95}. However, we can easily extend them to amplify or reduce translations and rotations. The resulting mappings then become \emph{non-isomorphic}.

We thus introduce a~$\gain$ function to our framework that computes a scalar \emph{gain factor}~$k_t$ at each time~$t$:
\begin{equation}
  \begin{split}
    k_t &= \gain(t)
    \nolabel{eqn:gain-function}
  \end{split}
\end{equation}

This gain factor allows us to rescale the device translations and rotations before applying them to the object, so that the object on the screen can move faster or slower than the device itself. For a translation, the gain factor changes its length without altering its direction. This is accomplished by scaling the translation vector~$\vec{\Delta pc_t}$ by the gain factor~$k_t$, yielding a new translation vector $\vec{\Delta pc'_t}$:
\begin{equation}
  \begin{split}
    \vec{\Delta pc'_t} &= k_t\;\vec{\Delta pc_t}
    \nolabel{eqn:gain-translation}
  \end{split}
\end{equation}

In the case of a rotation of angle~$\theta$ around a given axis, the gain factor changes the angle without altering the axis. If the rotation $\quat{\Delta qc_t}$ is expressed as a quaternion, we can use a~$\slerp$ interpolation~\citep{shoemake85} from the identity quaternion~\textbf{1} to construct a new rotation $\quat{\Delta qc'_t}$ around the same axis but with an angle scaled by~$k_t$. We note this operation as $\quat{\Delta qc_t}^{k_t}$. If~$\quat{\Delta qc_t}$ is a non-null rotation, the new rotation $\quat{\Delta qc'_t}=\quat{\Delta qc_t}^{k_t}$ is given by:
\begin{equation}
  \begin{split}
    \quat{\Delta qc'_t} &= \slerp(\boldsymbol{\quat{1}}, \quat{\Delta qc_t}, k_t)\\
    &= \quat{\Delta qc_t}^{k_t}
    \nolabel{eqn:gain-rotation}
  \end{split}
\end{equation}

By substituting $\vec{\Delta pc_t}$ and $\quat{\Delta qc_t}$ with $k_t\:\vec{\Delta pc_t}$ and $\quat{\Delta qc_t}^{k_t}$ in the mappings presented above, it becomes possible to dynamically control the gain applied to translations and rotations.

The gain factor in our model is a function of the current time. Unlike some previous works (\eg,~\citet{teather14,poupyrev00,laviola07}) which only used static scaling coefficients, we emphasize that the gain may \emph{\mbox{dynamically} change} during manipulation. Such a variable gain factor is especially useful to increase the range and accuracy of manipulation as we show in \autoref{sec:motion-space-solutions}. We thus indicate below whether the properties of each mapping remain true even with a variable gain factor.

\section{Spatial feedback compliance}
\label{sec:spatial-compliances}

When users are moving a mobile device to control a 3D~object on the screen, they receive multiple forms of feedback. The first is ki\-nes\-the\-tic\discretionary{/}{}{/}proprioceptive feedback from translating and rotating the device itself. The second is visual feedback from the resulting object motion on the screen. To maintain user performance and comfort it is thus essential that the visual feedback matches the ki\-nes\-the\-tic\discretionary{/}{}{/}proprioceptive feedback~\citep{smith87}---a principle known as feedback compliance~\citep{bowman04}. Here, we focus specifically on \emph{spatial feedback compliance}, which refers to the motion of the virtual object and is thus especially relevant when designing mappings.

In this section we discuss the spatial compliance properties of each mapping, both for translations and for rotations. We begin with the two properties mentioned by \citet{bowman04}, directional and nulling compliance, along with the property of transitivity~\citep{bade05}. Finally, we address the question of the user's reference frame (allocentric or egocentric) and whether object motion matches the reference frame expected by the user.

\subsection{Directional compliance}

\begin{figure}[t]
  \centering
  \includegraphics[width=.9\linewidth]{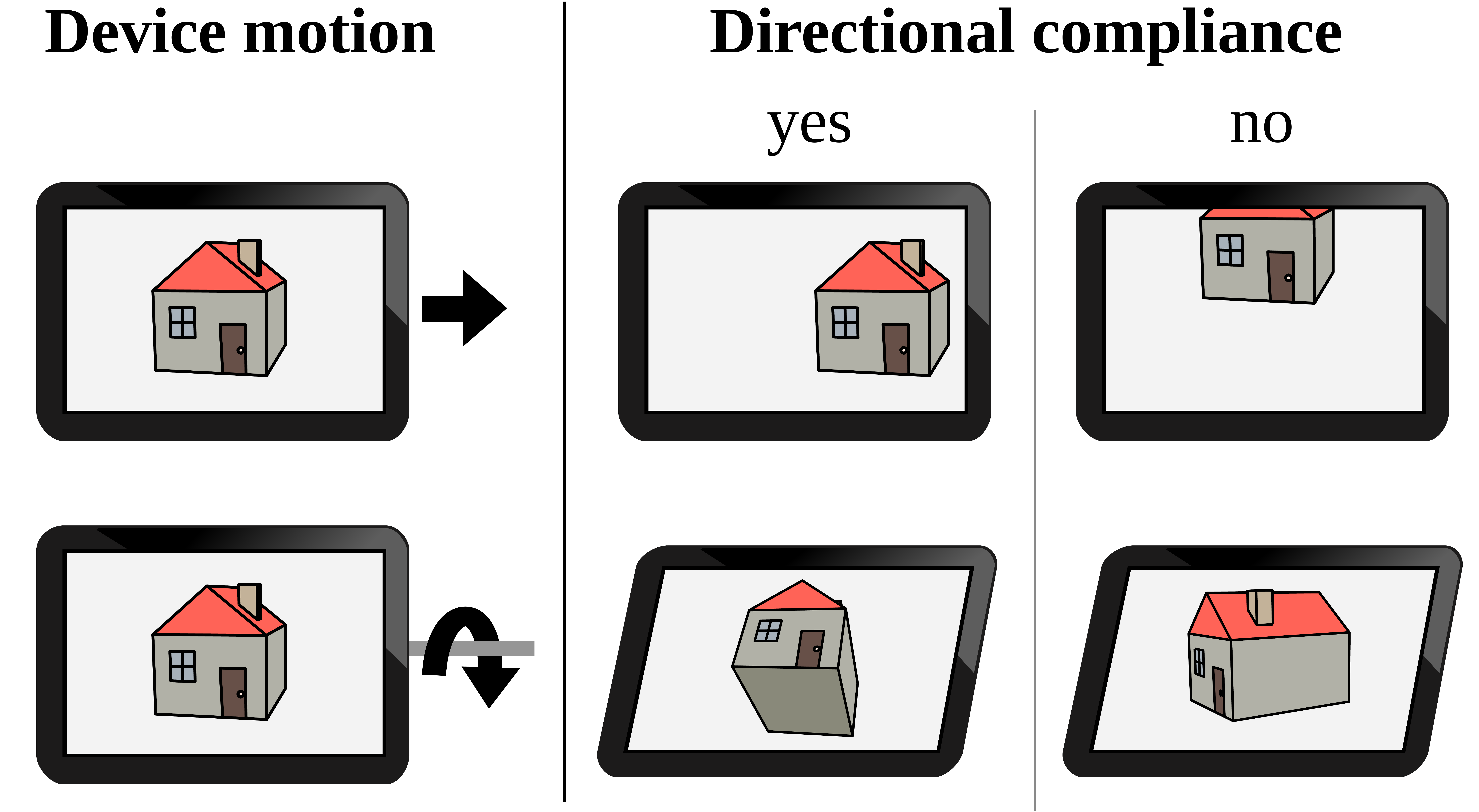}
  \caption[Directional compliance versus non-compliance.]{Directional compliance versus non-compliance, shown here on a tablet-shaped device.}
  \label{fig:dir-compl}
  \vspace{-3mm}
\end{figure}

Directional compliance~\citep{bowman04,poupyrev00}, also called ``kinesthetic correspondence''~\citep{britton78} or ``stimulus-response compatibility''~\citep{fitts53}, means that the manipulated object moves \emph{along the same direction} as the controlling device. In the configuration studied here, the object moves on the screen and is controlled by the mobile device's motion. The screen itself, however, is attached to the device and is also moving during manipulation. It is thus important to consider device motion relative to the screen (\ie,~in screen space). Directional compliance, therefore, means that the object is moving \emph{on the screen} along the same direction as the device is moving \emph{relative to the screen} (\autoref{fig:dir-compl}).

Note that the conversion to screen space described in \autoref{sec:tracker-to-screen-space} ensures that device motion is consistently aligned with screen space at~$t\,{=}\,0$, but does not guarantee directional compliance at any subsequent time during manipulation.

Object motion corresponds to the change of position and orientation on the screen between times~$t{-}1$ and~$t$: $\vec{pd_t}-\vec{pd_{t-1}}$ and $\quat{qd_t}\:\quat{qd_{t-1}}^{-1}$, resp. Mobile device motion corresponds to the change of position and orientation in tracking space between times~$t{-}1$ and~$t$: $\vec{pc_t}-\vec{pc_{t-1}}$ and $\quat{qc_t}\:\quat{qc_{t-1}}^{-1}$, resp. As before, a rotation of~$\quat{qc_{t-1}}^{-1}$ must be applied to this device motion to convert it to screen space. Formally stated, directional compliance means that object translations are collinear with device translations relative to the screen, and that object rotations have the same axis as device rotations relative to the screen. Thus, a mapping is directionally compliant at time~$t$ if it can be expressed~as:
\begin{multline}
  \exists\:(\alpha,\beta)\in\mathbb{R}^2:\\*
  \begin{aligned}
    \qquad\vec{pd_t}-\vec{pd_{t-1}} &= \alpha\;\left(\quat{qc_{t-1}}^{-1}\;(\vec{pc_t}-\vec{pc_{t-1}})\;\quat{qc_{t-1}}\right)\hspace{3cm}\\
    \qquad\quat{qd_t}\:\quat{qd_{t-1}}^{-1} &= \left(\quat{qc_{t-1}}^{-1}\;(\quat{qc_t}\:\quat{qc_{t-1}}^{-1})\;\quat{qc_{t-1}}\right)^{\beta}
  \end{aligned}
  \hspace{-2cm}
  \label{eqn:sr-compatibility}
\end{multline}
\vspace{2mm}

\textbf{Relative position control.}\quad
The relative position control mapping is always directionally compliant, for both translations and rotations. From the mapping formulation~(\autoref{eqn:relative-mapping}), and by taking into account the gain factor (\autoref{sec:gain}), we get $\vec{pd_t}-\vec{pd_{t-1}}=k_t\:\vec{\Delta pc_t}=k_t\:\left(\quat{qc_{t-1}}^{-1}\; (\vec{pc_t}-\vec{pc_{t-1}})\; \quat{qc_{t-1}}\right)$ and $\quat{qd_t}\:\quat{qd_{t-1}}^{-1}=\quat{\Delta qc_t}^{k_t}=\left(\quat{qc_{t-1}}^{-1}\; (\quat{qc_t}\:\quat{qc_{t-1}}^{-1})\; \quat{qc_{t-1}}\right)^{k_t}$---equivalent to the expression in \autoref{eqn:sr-compatibility}. Relative position control thus always guarantees directional compliance for both translations and rotations.

\textbf{Absolute position control.}\quad
The absolute position control mapping does \emph{not} guarantee directional compliance \emph{in the general case}. However, directional compliance can still be obtained under specific conditions. By taking into account a variable gain factor, we can express object translations between two times~$t{-}1$ and~$t$ as:
\begin{equation}
  \label{eqn:abs-delta-pos}
  \begin{split}
    &\vec{pd_t}-\vec{pd_{t-1}}\\
    &= (k_t\;\vec{\Delta pc_t} + \vec{pd_0}) - (k_{t-1}\;\vec{\Delta pc_{t-1}} + \vec{pd_0})\\
    &= k_t\;\vec{\Delta pc_t} - k_{t-1}\;\vec{\Delta pc_{t-1}}\\
    &= k_t\left(\quat{qc_0}^{-1}\,(\vec{pc_t}-\vec{pc_0})\;\quat{qc_0}\right) - k_{t-1}\left(\quat{qc_0}^{-1}\,(\vec{pc_{t-1}}-\vec{pc_0})\;\quat{qc_0}\right)\\
    &= \left(\quat{qc_0}^{-1}\,k_t\,(\vec{pc_t}-\vec{pc_0})\;\quat{qc_0}\right) - \left(\quat{qc_0}^{-1}\,k_{t-1}\,(\vec{pc_{t-1}}-\vec{pc_0})\;\quat{qc_0}\right)\\
    &= \quat{qc_0}^{-1}\;\big(k_t\;(\vec{pc_t}-\vec{pc_0})-k_{t-1}\;(\vec{pc_{t-1}}-\vec{pc_0})\big)\;\quat{qc_0}\\
  \end{split}
\end{equation}
Thus, in the general case, object translations do not correspond to expression~\autoref{eqn:sr-compatibility} and are not directionally compliant. For constant gain factors~$k_t$ ($k_t\,{=}\:\alpha\enskip\forall t{>}0$), however, \autoref{eqn:abs-delta-pos} can be reduced to:
\begin{flalign}
  \label{eqn:dir-compl-abs-alpha}
  &= \quat{qc_0}^{-1}\;\big(\alpha\;(\vec{pc_t}-\vec{pc_0})-\alpha\;(\vec{pc_{t-1}}-\vec{pc_0})\big)\;\quat{qc_0}\nonumber\\
  &= \quat{qc_0}^{-1}\;\alpha\;\big((\vec{pc_t}-\vec{pc_0})-(\vec{pc_{t-1}}-\vec{pc_0})\big)\;\quat{qc_0}\nonumber\\
  &= \quat{qc_0}^{-1}\;\alpha\;(\vec{pc_t}-\vec{pc_{t-1}})\;\quat{qc_0}\nonumber\hspace{4.7cm}\\
  &= \alpha\;\big(\quat{qc_0}^{-1}\;(\vec{pc_t}-\vec{pc_{t-1}})\;\,\quat{qc_0}\big)
\end{flalign}
Moreover, if at~$t{-}1$ the mobile device orientation~$\quat{qc_{t-1}}$ is equal to its initial orientation~$\quat{qc_0}$ then object translations can be written as:
\begin{flalign}
  \hspace{0.2cm}
  &= \alpha\;\big(\quat{qc_{t-1}}^{-1}\;(\vec{pc_t}-\vec{pc_{t-1}})\;\quat{qc_{t-1}}\big)&\nonumber
\end{flalign}
This corresponds to \autoref{eqn:sr-compatibility}. Translations in the absolute mapping are thus only directionally compliant if the gain factor remained constant between~$t{=}0$ and~$t{-}1$ \emph{and} if the mobile device orientation is the same as its initial orientation. Concerning rotations, incremental object motion can be written as:
\begin{equation}
  \label{eqn:abs-delta-rot}
  \begin{split}
    &\quat{qd_t}\:\quat{qd_{t-1}}^{-1}\\
    &= (\quat{\Delta qc_t}^{k_t}\;\quat{qd_0})\;(\quat{\Delta qc_{t-1}}^{k_{t-1}}\;\quat{qd_0})^{-1}\\
    &= (\quat{\Delta qc_t}^{k_t}\;\quat{qd_0})\;(\quat{qd_0}^{-1}\;\quat{\Delta qc_{t-1}}^{-k_{t-1}})\\
    &= \quat{\Delta qc_t}^{k_t}\;\quat{\Delta qc_{t-1}}^{-k_{t-1}}\hspace{0.95cm}\\
    &= \big(\quat{qc_0}^{-1}\;(\quat{qc_t}\:\quat{qc_0}^{-1})\;\quat{qc_0}\big)^{k_t}\;\big(\quat{qc_0}^{-1}\;(\quat{qc_{t-1}}\:\quat{qc_0}^{-1})\;\quat{qc_0}\big)^{-k_{t-1}}\\
  \end{split}
\end{equation}
In the general case, rotations do not correspond to the form stated in \autoref{eqn:sr-compatibility} and are not directionally compliant. However, if $\quat{qc_t}\:\quat{qc_0}^{-1}$ (device rotation from its initial orientation at time~$t$) and $\quat{qc_{t-1}}\:\quat{qc_0}^{-1}$ (same at time~$t{-}1$) have the same rotation axis then their difference $\quat{qc_t}\:\quat{qc_{t-1}}^{-1}$ also has the same rotation axis. There exist, therefore, two gain factors~$a_1$ and~$a_2$ such as $\quat{qc_t}\:\quat{qc_0}^{-1} = (\quat{qc_t}\:\quat{qc_{t-1}}^{-1})^{a_1}$ and $\quat{qc_{t-1}}\:\quat{qc_0}^{-1} = (\quat{qc_t}\:\quat{qc_{t-1}}^{-1})^{a_2}$ and \autoref{eqn:abs-delta-rot} can be rewritten as:
\begin{equation}
  \hspace{-0.5cm}
  \begin{split}
    &= \big(\quat{qc_0}^{-1}\;(\quat{qc_t}\:\quat{qc_{t-1}}^{-1})^{a_1}\;\quat{qc_0}\big)^{k_t}\;\big(\quat{qc_0}^{-1}\;(\quat{qc_t}\:\quat{qc_{t-1}}^{-1})^{a_2}\;\quat{qc_0}\big)^{-k_{t-1}}\\
    &= \big(\quat{qc_0}^{-1}\,(\quat{qc_t}\:\quat{qc_{t-1}}^{-1})\,\quat{qc_0}\big)^{a_1+k_t}\,\big(\quat{qc_0}^{-1}\,(\quat{qc_t}\:\quat{qc_{t-1}}^{-1})\,\quat{qc_0}\big)^{a_2-k_{t-1}}\\
    &= \big(\quat{qc_0}^{-1}\;(\quat{qc_t}\:\quat{qc_{t-1}}^{-1})\;\quat{qc_0}\big)^{a_1+k_t+a_2-k_{t-1}}\\
  \end{split}\nonumber
\end{equation}
\\
With $a_1+k_t+a_2-k_{t-1}\equiv\beta$, we can reduce the above to:\vspace{-1mm}
\\[-1mm]
\begin{equation}
  \label{eqn:dir-compl-abs-rot-beta}
  \hspace{-0.5cm}
  \begin{split}
    &= \big(\quat{qc_0}^{-1}\;(\quat{qc_t}\:\quat{qc_{t-1}}^{-1})\;\quat{qc_0}\big)^{\,\beta}\hspace{4.3cm}\\
  \end{split}
\end{equation}
\noindent Since $\quat{qc_{t-1}}\:\quat{qc_0}^{-1}=(\quat{qc_t}\:\quat{qc_{t-1}}^{-1})^{a_2}$, we have $\quat{qc_0}^{-1}=\quat{qc_{t-1}}^{-1}\;(\quat{qc_t}\:\quat{qc_{t-1}}^{-1})^{a_2}$. The above expression can thus be rewritten as:
\begin{equation}
  \hspace{-0.4cm}
  \begin{split}
    &= \Big(\quat{qc_{t\text{--}1}}^{-1}\,(\quat{qc_t}\:\quat{qc_{t\text{--}1}}^{-1})^{a_2}\,(\quat{qc_t}\,\quat{qc_{t\text{--}1}}^{-1})\,\big(\quat{qc_{t\text{--}1}}^{-1}\,(\quat{qc_t}\,\quat{qc_{t\text{--}1}}^{-1})^{a_2}\big)^{\text{--}1}\Big)^{\beta}\\
    &= \big(\quat{qc_{t\text{--}1}}^{-1}\:(\quat{qc_t}\:\quat{qc_{t\text{--}1}}^{-1})^{a_2}\;(\quat{qc_t}\:\quat{qc_{t\text{--}1}}^{-1})\;(\quat{qc_t}\:\quat{qc_{t\text{--}1}}^{-1})^{-a_2}\:\quat{qc_{t\text{--}1}}\big)^{\beta}\\
    &= \big(\quat{qc_{t\text{--}1}}^{-1}\;(\quat{qc_t}\:\quat{qc_{t\text{--}1}}^{-1})^{a_2+1-a_2}\;\quat{qc_{t\text{--}1}}\big)^{\beta}\\
    &= \big(\quat{qc_{t\text{--}1}}^{-1}\;(\quat{qc_t}\:\quat{qc_{t\text{--}1}}^{-1})\;\quat{qc_{t\text{--}1}}\big)^{\beta}\\
  \end{split}\nonumber\hspace{0.75cm}
\end{equation}
This corresponds to \autoref{eqn:sr-compatibility}. Hence, object rotations in the absolute mapping are only directionally compliant around the first, initial rotation axis. In practice, this means that only the first rotation step is directionally compliant: subsequent rotations are not if they happen on a different axis. Therefore, users who wish to rotate the object around another axis would have to return the device to its initial orientation in order to maintain directional compliance for further rotations.

\textbf{Rate control.}\quad
The rate control mapping does \emph{not} guarantee directio\-nal compliance in the general case, both for translations and for rotations. By taking into account a variable gain factor, we can express object translations between two times~$t{-}1$ and~$t$ as:
\begin{equation}
  \hspace{-1cm}
  \begin{split}
    \vec{pd_t}-\vec{pd_{t-1}} &= (\,k_t\:\vec{\Delta pc_t} + \vec{pd_{t-1}}) - \vec{pd_{t-1}}\\
    &= k_t\:\,\vec{\Delta pc_t}\\
    &= k_t\,\left(\quat{qc_0}^{-1}\,(\vec{pc_t}-\vec{pc_0})\;\quat{qc_0}\right)\\
  \end{split}
  \label{eqn:rate-delta-pos}
\end{equation}
\noindent Since this expression is not equivalent to \autoref{eqn:sr-compatibility}, translations in a rate control mapping are not directionally compliant in the general case.
However, if we assume that all device translations were performed along the same axis since the beginning of manipulation, then $\vec{pc_t}-\vec{pc_{t-1}}$ is collinear with $\vec{pc_t}-\vec{pc_0}$. There exist, therefore, a scalar~$a$ such as $\vec{pc_t}-\vec{pc_0}=a\:(\vec{pc_t}-\vec{pc_{t-1}})$ and \autoref{eqn:rate-delta-pos} can be rewritten as:
\begin{equation}
  \hspace{0.1cm}
  \begin{split}
    &= k_t\;\big(\quat{qc_0}^{-1}\,\left(a\:(\vec{pc_t}-\vec{pc_{t-1}})\right)\,\quat{qc_0}\big)\\
    &= (a\;k_t)\left(\quat{qc_0}^{-1}\,(\vec{pc_t}-\vec{pc_{t-1}})\;\kern.5pt\quat{qc_0}\right)\\
    &= \alpha\;\left(\quat{qc_0}^{-1}\,(\vec{pc_t}-\vec{pc_{t-1}})\;\quat{qc_0}\right)
  \end{split}\nonumber
\end{equation}
\noindent We find the same expression as \autoref{eqn:dir-compl-abs-alpha}, which leads to \autoref{eqn:sr-compatibility} if we make the second assumption that device orientation at~$t{-}1$ is equal to its initial orientation~$\quat{qc_0}$ (see the previous proof on the absolute mapping for more details).
Thus, translations in the rate control mapping are only directionally compliant along the first translation axis \emph{and} if the device orientation is equal to its initial orientation.

\noindent Rotations exhibit a similar behavior:
\begin{equation}
  \hspace{-.8cm}
  \begin{split}
    \quat{qd_t}\:\quat{qd_{t-1}}^{-1} &= (\quat{\Delta qc_t}^{k_t}\;\quat{qd_{t-1}})\;\quat{qd_{t-1}}^{-1}\\
    &= \quat{\Delta qc_t}^{k_t}\\
    &= \left(\quat{qc_0}^{-1}\;(\quat{qc_t}\:\quat{qc_0}^{-1})\;\quat{qc_0}\right)^{\,k_t}\\
  \end{split}\hspace{1.2cm}
  \label{eqn:rate-delta-rot}
\end{equation}
\noindent In the general case, this expression is not equivalent to \autoref{eqn:sr-compatibility}.
However, if we assume that all device rotations happened on the same axis since the beginning of manipulation, then there exists a scalar~$b$ such as $\quat{qc_t}\:\quat{qc_0}^{-1}=(\quat{qc_t}\:\quat{qc_{t-1}}^{-1})^{\:b}$ and \autoref{eqn:rate-delta-rot} can be rewritten as follows:
\begin{equation}
  \hspace{-.8cm}
  \begin{split}
    &= \left(\quat{qc_0}^{-1}\;(\quat{qc_t}\:\quat{qc_{t-1}}^{-1})^b\;\quat{qc_0}\right)^{\,k_t}\\
    &= \left(\quat{qc_0}^{-1}\;(\quat{qc_t}\:\quat{qc_{t-1}}^{-1})\;\quat{qc_0}\right)^{\,k_t+b}\\
    &= \left(\quat{qc_0}^{-1}\;(\quat{qc_t}\:\quat{qc_{t-1}}^{-1})\;\quat{qc_0}\right)^{\,\beta}\\
  \end{split}\nonumber\hspace{-0.4cm}
\end{equation}
\noindent We find the same expression as \autoref{eqn:dir-compl-abs-rot-beta}, which directly leads to \autoref{eqn:sr-compatibility} since we assumed that all device rotations were performed about the same axis (see the previous proof for more details).
As with the absolute mapping, rotations in the rate control mapping are thus only directionally compliant around the first rotation axis.

\subsection{Transitivity and nulling compliance}

\begin{figure}[t]
  \centering
  \includegraphics[width=1.0\linewidth]{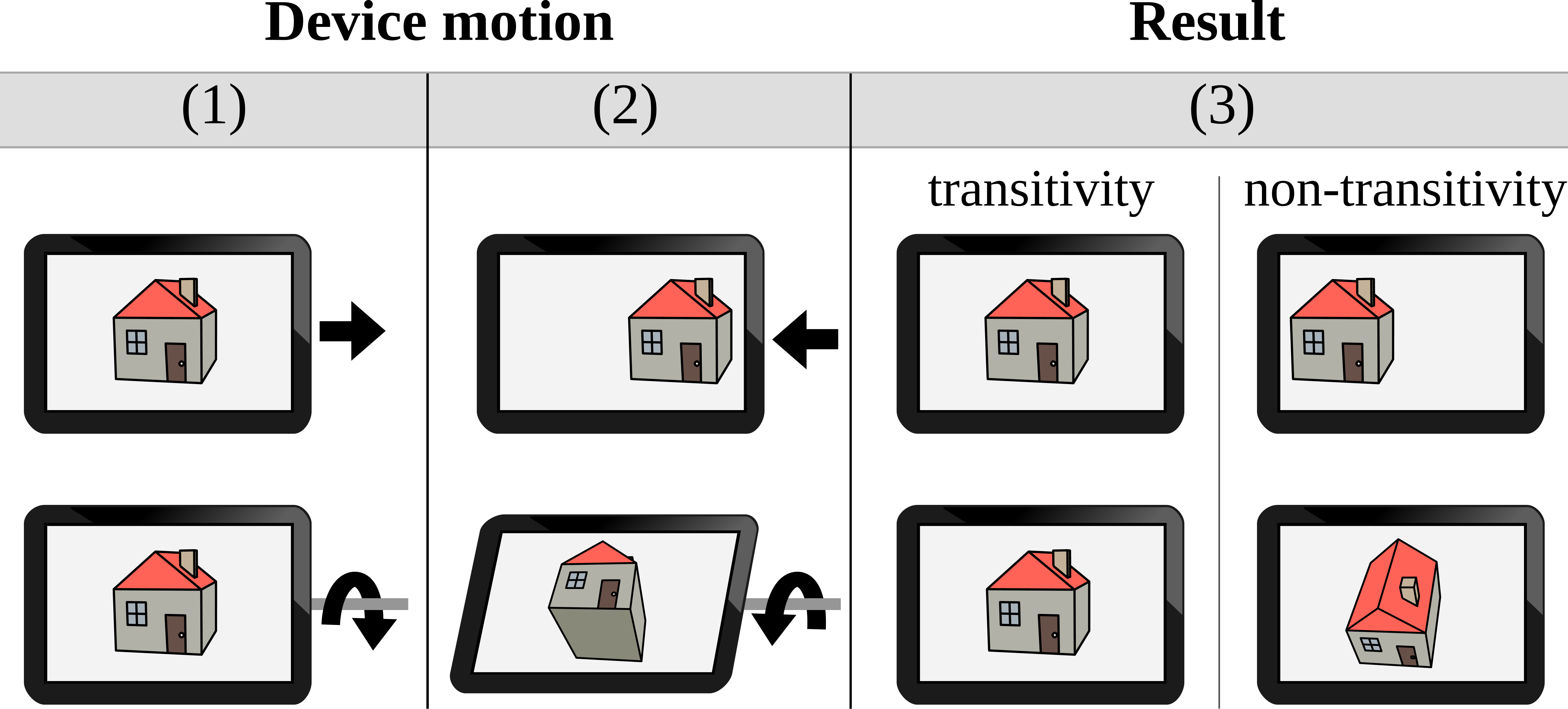}
  \caption[Transitivity versus non-transitivity.]{Transitivity versus non-transitivity (illustrating here the specific case of nulling compliance).}
  \label{fig:transitivity}
\end{figure}

Transitivity~\citep{bade05} refers to a property of the real world: moving an object from point~A to point~B then~C, or directly from~A to~C, results in the same final location for the object. According to this principle, translating and rotating the mobile device from~A to~B then~C, or directly from~A to~C, should bring the manipulated object to the same position and orientation (\autoref{fig:transitivity}). In particular, this property allows users to easily return the object to its initial location---which can be useful after a manipulation error---by simply returning the mobile device to its initial location. This specific case is known as \emph{nulling compliance}~\citep{poupyrev00}. Transitivity is a generalization of nulling compliance to any target location.

\textbf{Absolute position control.}\quad
The absolute position control mapping is \emph{generally} transitive, for both translations and rotations. The only terms in the mapping formulation (\autoref{eqn:absolute-mapping}) that are non-constant during manipulation are~$\vec{pc_t}$ and~$\quat{qc_t}$. The base formulation of the absolute mapping only depends on the \emph{current} position and orientation of the mobile device, regardless of the intermediate steps that led it there. An isomorphic absolute mapping is thus always transitive. For a non-isomorphic mapping, however, we must also take the gain function into account. A non-isomorphic absolute mapping is only transitive if the gain function itself does not depend on non-constant terms other than~$\vec{pc_t}$ and~$\quat{qc_t}$. This is obviously the case for any constant gain factor. But adding non-constant parameters to the gain function other than~$\vec{pc_t}$ and~$\quat{qc_t}$ breaks this guarantee. In any case, both isomorphic and non-isomorphic absolute mappings are always transitive in relation to the initial position and orientation $\vec{pc_0}$ and~$\quat{qc_0}$ since a null translation or rotation is unaffected by gain. They are thus \emph{always} nulling compliant. This property of absolute position control was identified by \citet{poupyrev00} for rotations, but it also holds for translations.

\textbf{Rate control.}\quad
The rate control mapping is \emph{not} transitive or nulling compliant, neither for translations nor rotation, because it is time-dependent by definition. The resulting object motion depends on the time spent by the mobile device between locations~A, B, and~C. There is thus no way to predict the final location of the manipulated object from a sequence of device locations alone.

\textbf{Relative position control.}\quad
The relative position control mapping is \emph{not} transitive or nulling compliant in the general case. But there are still specific conditions for which this mapping can be transitive. In order to demonstrate this, let $\vec{pd_t}^{\:\mathrm{step}}$ be the position of the manipulated object obtained after the device went through intermediate positions between $\vec{pc_0}$ and~$\vec{pc_t}$. Let $\vec{pd_t}^{\:\mathrm{direct}}$ be the object position obtained after the device moved \emph{directly} from $\vec{pc_0}$ to~$\vec{pc_1}\,{=}\:\vec{pc_t}$. Similarly, $\quat{qd_t}^{\mathrm{\:step}}$ is the object orientation obtained after the device rotated incrementally from~$\quat{qc_0}$ to~$\quat{qc_t}$, and $\quat{qd_t}^{\mathrm{\:direct}}$ is the object orientation obtained after the device directly rotated from~$\quat{qc_0}$ to~$\quat{qc_1}\,{=}\:\quat{qc_t}$. Concerning the object positions, we have:
\begin{equation}
  \begin{alignedat}{3}
    \vec{pd_t}^{\:\mathrm{step}} &= &\;&k_t\,\left(\quat{qc_{t-1}}^{-1}\;(\vec{pc_t}-\vec{pc_{t-1}})\;\quat{qc_{t-1}}\right) + \vec{pd_{t-1}}\\
    &= &&\left(\quat{qc_{t-1}}^{-1}\;k_t\,(\vec{pc_t}-\vec{pc_{t-1}})\;\quat{qc_{t-1}}\right) + \vec{pd_{t-1}}\\
    &= &&\left(\quat{qc_{t-1}}^{-1}\;k_t\,(\vec{pc_t}-\vec{pc_{t-1}})\;\quat{qc_{t-1}}\right)\\
    &  &      &+...+\left(\quat{qc_0}^{-1}\;k_1\;(\vec{pc_1}-\vec{pc_0})\;\quat{qc_0}\right) + \vec{pd_0}\\
    \vec{pd_t}^{\:\mathrm{direct}} &= &&k_t\,\left(\quat{qc_0}^{-1}\;(\vec{pc_t}-\vec{pc_0})\;\quat{qc_0}\right) + \vec{pd_0}\\
    &= &&\quat{qc_0}^{-1}\;k_t\;(\vec{pc_t}-\vec{pc_0})\;\quat{qc_0} + \vec{pd_0}
  \end{alignedat}\nonumber
\end{equation}
Due to the transformations (rotation to screen space and gain factor) applied to incremental $\vec{pc_i}-\vec{pc_{i-1}}$ vectors in $\vec{pd_t}^{\:\mathrm{step}}$, the result is generally not equivalent to $\vec{pd_t}^{\:\mathrm{direct}}$. This observation shows that translations are not transitive under this mapping in the general case. However, if mobile device orientation did not change so far ($\quat{qc_t}\,{=}\:\quat{qc_0}\enskip\forall t{>}0$) then $\vec{pd_t}^{\:\mathrm{step}}$ is reduced to:
\begin{equation}
  \label{eqn:rel-delta-pos}
  \begin{alignedat}{3}
    \vec{pd_t}^{\:\mathrm{step}} &= \quat{qc_0}^{-1}\;\big(&&\kern.5ptk_t\;(\vec{pc_t}-\vec{pc_{t-1}})\\
    &                    &&+...+ k_1\;(\vec{pc_1}-\vec{pc_0})\big)\;\quat{qc_0} + \vec{pd_0}
  \end{alignedat}
\end{equation}
\autoref{eqn:rel-delta-pos} applies arbitrary gain factors to each intermediate translation step. For \autoref{eqn:rel-delta-pos} to become equivalent to $\vec{pd_t}^{\:\mathrm{direct}}$, the gain factor must also have been constant during manipulation ($k_i\,{=}\,k_t\enskip\forall i{<}t$):
\begin{flalign}
  \quad\enskip\vec{pd_t}^{\:\mathrm{step}}
  &= \quat{qc_0}^{-1}\;k_t\,(\vec{pc_t}-\vec{pc_0})\;\quat{qc_0} + \vec{pd_0}\nonumber&\\
  \quad\enskip&= \vec{pd_t}^{\:\mathrm{direct}}\nonumber
\end{flalign}
Therefore, translations in the relative mapping are only transitive as long as the gain factor \emph{and} the mobile device orientation remained constant so far during manipulation. Concerning rotations we have:
\begin{equation}
  \begin{split}
    \quat{qd_t}^{\mathrm{\:step}} &= \left(\quat{qc_{t-1}}^{-1}\;(\quat{qc_t}\:\quat{qc_{t-1}}^{-1})\;\quat{qc_{t-1}}\right)^{k_t}\,\quat{qd_{t-1}}\\
    &= \left((\quat{qc_{t-1}}^{-1}\:\quat{qc_t})\;(\quat{qc_{t-1}}^{-1}\:\quat{qc_{t-1}})\right)^{k_t}\,\quat{qd_{t-1}}\\
    &= (\quat{qc_{t-1}}^{-1}\:\quat{qc_t})^{k_t}\;\quat{qd_{t-1}}\\
    &= (\quat{qc_{t-1}}^{-1}\:\quat{qc_t})^{k_t}\,(\quat{qc_{t-2}}^{-1}\:\quat{qc_{t-1}})^{k_{t-1}}...\,(\quat{qc_0}^{-1}\:\quat{qc_1})^{k_1}\,\quat{qd_0}\\
    \quat{qd_t}^{\mathrm{\:direct}} &= \left(\quat{qc_0}^{-1}\;(\quat{qc_t}\:\quat{qc_0}^{-1})\;\quat{qc_0}\right)^{k_t}\,\quat{qd_0}\\
    &= (\quat{qc_0}^{-1}\:\quat{qc_t})^{k_t}\:\quat{qd_0}
  \end{split}\nonumber
\end{equation}
Even if the gain factor remains constant, $\quat{qd_t}^{\mathrm{\:step}}$ is still not equivalent to $\quat{qd_t}^{\mathrm{\:direct}}$ in the general case. Rotations are thus not transitive for relative mappings (except when ${k_t\:{=}\:\text{\textminus1}}$; see below).

\subsection{Allocentric and egocentric manipulation}
\label{sec:allo-ego-manip}

Translating and rotating a 3D~object can be interpreted in two ways \citep{klatzky98}. One interpretation is that the object itself is moved relative to a stationary viewpoint. This type of transformation is called \emph{allocentric} (or exocentric). Another interpretation is that the object is stationary whereas the viewpoint is moved in an opposite way. This type of transformation is called \emph{egocentric}.

\begin{figure}[t]
  \centering
  \includegraphics[width=.9\linewidth]{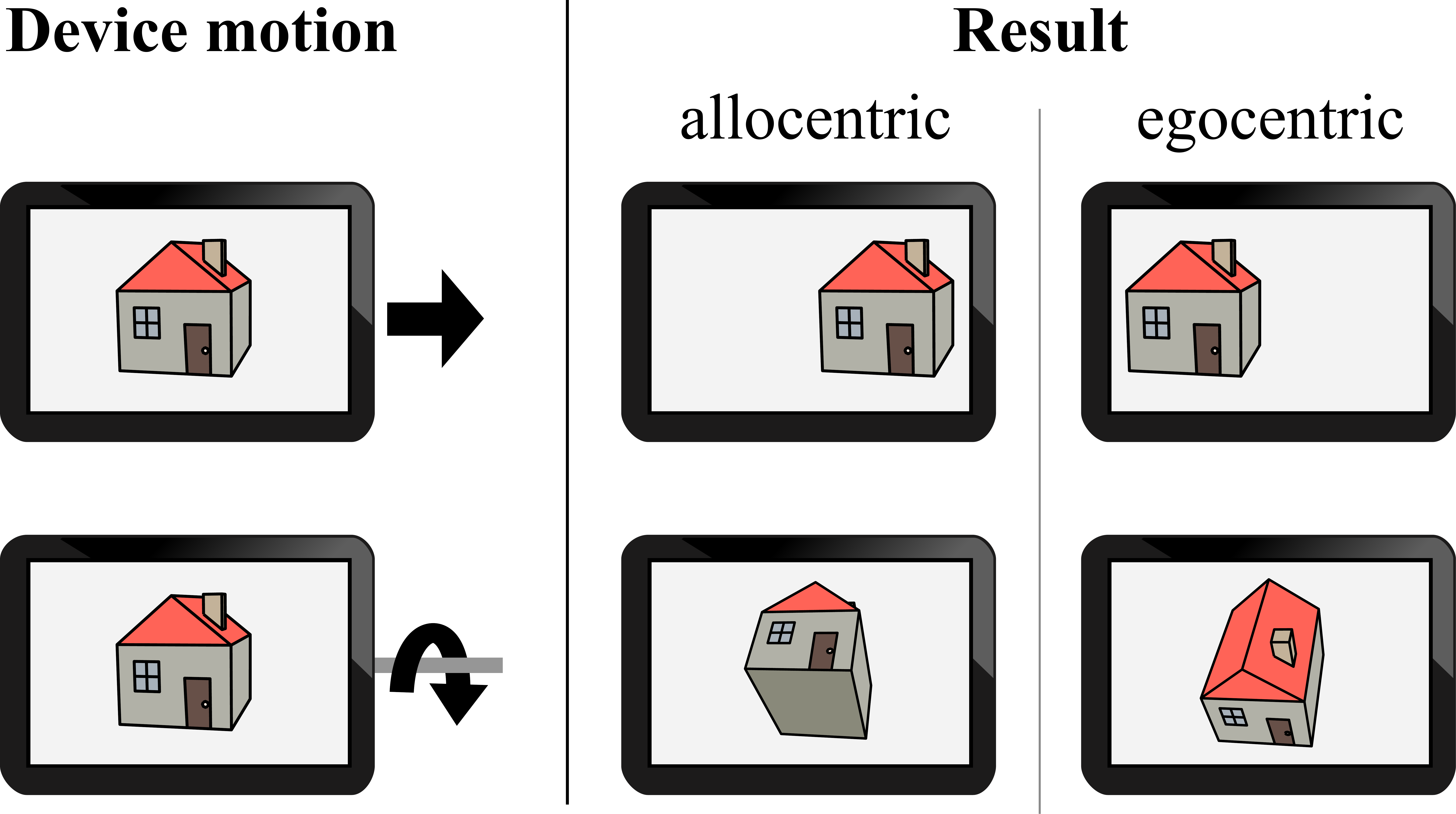}
  \caption[Difference between allocentric and egocentric manipulation.]{Difference between allocentric and egocentric manipulation.}
  \label{fig:allo-ego}
\end{figure}

In the configuration studied here, the object is displayed on the screen but the screen itself moves during manipulation. We thus do not consider whether the object appears to move from the viewpoint of the user, but how it moves \emph{on the screen} relative to the device. Allocentric mappings differ from egocentric manipulations based on whether the object moves on the screen in the same direction as the device (allocentric) or in the \emph{opposite} direction (egocentric), as illustrated in \autoref{fig:allo-ego}. This notion differs from directional compliance. Directional compliance means that the manipulated object moves along the same axis as the device, as opposed to another direction. Allocentric and egocentric manipulation, however, refer to the \emph{sense} of object motion along this axis. For example, if the object is moving along the same axis as the device but in an opposite way, then the mapping is directionally compliant and egocentric.

Using our formalism from \autoref{sec:formalization}, we can make any mapping either allocentric or egocentric by simply changing the gain factor~$k_t$. A positive gain factor, or no explicit gain (\ie,~a~gain factor of~1), makes mappings allocentric. Negative gain factors invert translations and rotations and the mapping becomes egocentric.

\textbf{Gain factor of~\textminus1.}\quad
This particular value is significant for position control mappings. It causes the direction of mobile device translations and rotations to be reversed, leaving their amplitude unchanged. In other words, a gain factor of~\textminus1 applies the exact opposite of the mobile device motion to the manipulated object. The mapping thus becomes egocentric, but remains isomorphic.

This has a notable effect on rotations. On a mobile device, the manipulated object is displayed on the screen, which is rotating along with the device during manipulation. By applying the exact opposite of the screen rotations to the object, these rotations are canceled from the user's point of view. The manipulated object will thus appear to have a \emph{fixed orientation} relative to the real world. Interestingly, the effect is identical in the absolute and relative mappings. For a constant gain factor ${k_t\,{=}\,\text{\textminus1}}\enskip\forall t{>}0$, the absolute mapping results in the following object orientations:
\begin{align}
  \begin{split}
    \quat{qd_t} &= \quat{\Delta qc_t}^{-1}\,\quat{qd_0}\\
    &= \left(\quat{qc_0}^{-1}\;(\quat{qc_t}\:\quat{qc_0}^{-1})\;\quat{qc_0}\right)^{-1}\,\quat{qd_0}\\
    &= \left(\quat{qc_0}^{-1}\;(\quat{qc_t}\:\quat{qc_0}^{-1})^{-1}\;\quat{qc_0}\right)\,\quat{qd_0}\\
    &= \left(\quat{qc_0}^{-1}\;(\quat{qc_0}\:\quat{qc_t}^{-1})\;\quat{qc_0}\right)\,\quat{qd_0}\\
    &= (\quat{qc_t}^{-1}\:\quat{qc_0})\; \quat{qd_0}
  \end{split}\nonumber
  \intertext{And the relative mapping results in:}
  \begin{split}
    \quat{qd_t} &= \quat{\Delta qc_t}^{-1}\,\quat{qd_{t-1}}\\
    &= \left(\quat{qc_{t-1}}^{-1}\;(\quat{qc_t}\:\quat{qc_{t-1}}^{-1})\;\quat{qc_{t-1}}\right)^{-1}\,\quat{qd_{t-1}}\\
    &= (\quat{qc_{t-1}}^{-1}\:\quat{qc_t})^{-1}\:\quat{qd_{t-1}}\\
    &= (\quat{qc_{t-1}}^{-1}\:\quat{qc_t})^{-1}\:(\quat{qc_{t-2}}^{-1}\:\quat{qc_{t-1}})^{-1}...\:(\quat{qc_0}^{-1}\:\quat{qc_1})^{-1}\:\quat{qd_0}\\
    &= (\quat{qc_t}^{-1}\:\quat{qc_{t-1}})\;(\quat{qc_{t-1}}^{-1}\:\quat{qc_{t-2}})\;...\;(\quat{qc_1}^{-1}\:\quat{qc_0})\;\quat{qd_0}\\
    &= (\quat{qc_t}^{-1}\:\quat{qc_0})\;\quat{qd_0}
  \end{split}
  \nolabel{eqn:equiv-rotations}
\end{align}
Hence, rotations become strictly equivalent in absolute and relative mappings for a constant gain factor~${k_t\:{=}\:\text{\textminus1}}$. This implies that they now  share the \emph{same spatial compliances}. Rotations in the absolute mapping become directionally compliant, as in the equivalent relative mapping. Rotations in the relative mapping become transitive, as in the equivalent absolute mapping. A gain factor of~\textminus1 is thus the only way to have \emph{both} directional compliance and transitivity for rotations in position control mappings.

The same equivalence, however, is not guaranteed for translations. As we showed before, spatial compliances of translations not only depend on the device's position but also on its orientation. In the absolute mapping, a constant gain factor~${k_t\:{=}\:\text{\textminus1}}$ results in the following object positions:
\begin{equation}
  \hspace{-1cm}
  \begin{split}
    \vec{pd_t} &= -\vec{\Delta pc_t} + \vec{pd_0}\\
    &= -\left(\quat{qc_0}^{-1}\;(\vec{pc_t}-\vec{pc_0})\;\quat{qc_0}\right) + \vec{pd_0}\\
    &= \left(\quat{qc_0}^{-1}\;(\vec{pc_0}-\vec{pc_t})\;\quat{qc_0}\right) + \vec{pd_0}
  \end{split}\nonumber
\end{equation}
The vector $\vec{pc_0}-\vec{pc_t}$ is the opposite of the total device translation. One might expect that applying it to the object would compensate for the device translation, making the object appear to have a fixed position in the real world. This vector, however, has to be converted to screen space. In this mapping, the conversion is done according to the initial device orientation~$\quat{qc_0}$. The object can thus only appear to be stationary if device orientation remains equal to~$\quat{qc_0}$.
In the relative mapping, a constant gain factor of~\textminus1 yields:
\begin{equation}
  \begin{alignedat}{2}
    \vec{pd_t} &= &\;&-\vec{\Delta pc_t} + \vec{pd_{t-1}}\\
    &= &\;&-\left(\quat{qc_{t-1}}^{-1}\;(\vec{pc_t}-\vec{pc_{t-1}})\;\quat{qc_{t-1}}\right) + \vec{pd_{t-1}}\\
    &= &  &\left(\quat{qc_{t-1}}^{-1}\;(\vec{pc_{t-1}}-\vec{pc_t})\;\quat{qc_{t-1}}\right) + \vec{pd_{t-1}}\\
    &= &  &\left(\quat{qc_{t-1}}^{-1}\;(\vec{pc_{t-1}}-\vec{pc_t})\;\quat{qc_{t-1}}\right)\\
    &  &        &+...+\left(\quat{qc_0}^{-1}\;(\vec{pc_0}-\vec{pc_1})\;\quat{qc_0}\right) + \vec{pd_0}
  \end{alignedat}\nonumber
\end{equation}
Even though the opposite of each device translation substep $\vec{pc_t}-\vec{pc_{t-1}}$ is applied to the manipulated object, each substep is first converted to screen space according to the \emph{intermediate} device orientation. The total object translation thus depends on intermediate device orientations. The object position will only appear to be fixed relative to the real world if device orientation does not change during manipulation, and thus remains equal to~$\quat{qc_0}$.

Note that simultaneously performing translations and rotations with a gain factor of~\textminus1 would therefore \emph{not} result in an \mbox{AR-like} mapping---\ie, the object having both a fixed position and a fixed orientation relative to the real world---at least under the above formulation of the position-control mappings.

\subsection{Summary of spatial compliances}

\autoref{tab:compliances} summarizes spatial compliance properties of each mapping, both for translations and rotations. With $k_t\,{=}\,\text{\textminus1}$ we indicate that the property is only guaranteed when the gain factor~$k_t$ (see \autoref{sec:gain}) remained constant and equal to~\textminus1 since the start of manipulation.

\newcommand\cellyes{yes}
\newcommand\cellno{no}
\newcommand\cellcond[1]{#1}
\begin{table}[ht]
  \small
  \centering
  \caption{Spatial compliances of each mapping.}
  \label{tab:compliances}
  \newcolumntype{Y}{>{\centering\arraybackslash}X}
  \begin{minipage*}{.95\linewidth}
  \renewcommand\footnoterule{}
  \begin{tabularx}{\linewidth}{rYYYY}
    \toprule
    & \multicolumn{2}{c}{\textbf{directional compliance}}
    & \multicolumn{2}{c}{\textbf{transitivity}}
    \\\cmidrule(lr){2-3}\cmidrule(lr){4-5}
    & translation & rotation
    & translation & rotation
    \\\midrule
    \textbf{absolute}
      & \hspace{.5mm}\cellno{}\footnotemark[a]
      & \cellcond{$k_t\,{=}\,\text{\textminus1}$}\footnotemark[b]
      & \hspace{-.5mm}\cellyes{}
      & \cellyes{}
    \\\textbf{relative}
      & \hspace{-.5mm}\cellyes{}
      & \hspace{-.5mm}\cellyes{}
      & \hspace{.5mm}\cellno{}\footnotemark[c]
      & \cellcond{$k_t\,{=}\,\text{\textminus1}$}
    \\\textbf{rate}
      & \hspace{.5mm}\cellno{}\footnotemark[d]
      & \hspace{.3mm}\cellno{}\footnotemark[b]
      & \hspace{-.5mm}\cellno{}
      & \cellno{}
    \\\bottomrule
  \end{tabularx}
  \vspace{2mm}\\
  \footnotetext[a]{%
    \begin{minipage}[t]{.93\linewidth}
      unless the gain factor is constant \emph{and} the device orientation is equal to its initial orientation
    \end{minipage}
  }\\[2mm]
  \footnotetext[b]{%
    \begin{minipage}[t]{.93\linewidth}%
      unless the device only rotates about a single axis,~in which case rotations remain directionally compliant
    \end{minipage}
  }\\[2mm]
  \footnotetext[c]{%
    \begin{minipage}[t]{.93\linewidth}%
      unless the gain factor is constant \emph{and} there is no device rotation (either its rotation is ignored, or its orientation does not change)
    \end{minipage}
  }\\[2mm]
  \footnotetext[d]{%
    \begin{minipage}[t]{.93\linewidth}%
      unless all translations occur along a single axis \emph{and} the device orientation is equal to its initial orientation
    \end{minipage}\vspace{-1mm}
  }
\end{minipage*}
\end{table}

\subsection{Choosing between spatial compliances}

From the results shown in \autoref{tab:compliances}, we see that none of the three mappings provide both directional compliance and transitivity in all cases. Furthermore, a choice must be made between allocentric and egocentric manipulation. In this section, we present arguments and evidence to help designers choose between these alternatives.

\subsubsection{Directional compliance versus transitivity}

We demonstrated above that none of the three main mappings can generally guarantee simultaneous directional compliance and transitivity. According to \autoref{tab:compliances}, rotations only exhibit both compliances when the mapping is position-controlled, isomorphic, and egocentric ($k_t\,{=}\,\text{\textminus1}$). Translations only exhibit both compliances when the mapping is position-controlled, without simultaneous rotations, and with a constant gain factor. Those are substantial constraints which may not be acceptable in many applications. In practice, the choice of the right mapping for a given use case will thus depend on which of the two spatial compliances is the most important.

Directional compliance ensures that the motion of the manipulated object matches the motion applied to the interaction device. According to \citet{bowman04}, this helps the user anticipate object motion and plan its trajectory during manipulation. There is indeed evidence that directional compliance plays a role in usability and user performance. \citet{fitts53} showed that user responses were slower and less accurate when visual stimuli and user input were not spatially aligned. Furthermore, the difference in performance could not be fully eliminated by training. \citet{ware04} studied the effect of a rotation between input and display reference frames---\ie,~a~lack of directional compliance---on a 3D~object rotation task. Their results showed an strong reduction of performance with large angles of mismatch, though the effect was more limited with smaller mismatches. \Citet{vanrhijn06} showed that performance in a 3D~manipulation task (translation and rotation) was best when object motion matches device motion relative to the object reference frame, which corresponds to our description of directional compliance. Otherwise, completion time increased significantly. Directional compliance thus appears to be essential for effective manipulation---unless the device is not rotated much during manipulation so that the misalignment between input and display reference frames would remain small~\citep{poupyrev00}.

Transitivity, or nulling compliance, is desirable in some situations. As previously mentioned, transitivity is useful to recover from manipulation errors. It allows users to exploit muscle memory~\citep{bowman04} to reliably return the object to its initial location (nulling compliance), or any valid intermediate location. Transitivity is also useful when the manipulated object has a meaningful upright orientation such as a human head~\citep{buda12}, a building, or a landscape since the object can be easily and predictably returned to an upright orientation. According to \citet{poupyrev00}, the shape of the interaction device is also important. If the device has a perceivable ``base'' orientation, a lack of nulling compliance in rotations will be noticed by the user, and may impact usability. Here, the interaction device is the mobile device itself. Most current mobile devices have a planar shape, designed to be facing the user. Therefore, they have a base orientation that can be perceived visually and haptically by the user. The lack of nulling compliance (hence the lack of rotation transitivity) can thus be noticeable during manipulation. Other devices such as cube displays~\citep{stavness10} do not have a single preferred orientation. The absence of rotation transitivity might be less noticeable on such devices.

Despite its situational usefulness, transitivity is unfortunately incompatible with directional compliance in most cases as we demonstrated above. At least one study comparing rotation mappings~\citep{buda12} reported that directionally compliant mappings performed better and were rated higher than transitive mappings. This suggests that transitive mappings should be preferred over directionally compliant mappings only for specific applications.

\subsubsection{Allocentric versus egocentric mappings}

Another key factor for usability is the choice between allocentric and egocentric manipulation. Both alternatives are functionally equivalent: they preserve other spatial compliances and produce equivalent---albeit mirrored---object movements in response to the same device movements. Therefore, a choice cannot be made on this basis. As mentioned in \autoref{sec:allo-ego-manip}, the difference between allocentric and egocentric manipulation is primarily a question of interpretation \citep[\eg,][]{lopez15}. When users manipulate the mobile device, are they expecting to manipulate the object, or the viewpoint on the object? It is important that the mapping matches user expectations~\citep{chan03} to reduce manipulation errors and improve usability. We thus need to determine what should be the default setting.

\citet{fitts51} introduced the concept of \emph{population stereotype} which refers to the option preferred (or expected) by a majority of users in a given population when faced with an arbitrary choice. A typical example is the fact a pushbutton is expected to be activated when depressed. It should be noted that population stereotypes are defined in relation to a given population, and do not necessarily generalize to users from different cultures or even professions~\citep{wiebe09}. Still, a number of stereotypes were found to be sufficiently prevalent to become design guidelines. For instance, Warrick's principle~\citep{warrick47,wiebe09} states that a controlled object should move in the same direction as the side of the control device closest to it. The clockwise-to-increase principle~\citep{wiebe09} states that a controlled value indicator should increase when the control device is turned in a clockwise direction. However, these guidelines were established for 1D~translation tasks with separate control devices. They are thus difficult to apply to 3D~manipulation in a locally coupled configuration.

\begin{figure*}
  \renewcommand\thesubfigure{\hspace{2mm}\normalsize\bfseries\protect\raisebox{1mm}{Scene~\arabic{subfigure}}}
  \setlength{\subfigcapskip}{-3.5ex}%
  \textcolor{white}{%
    \subfigure[\hspace{.21\textwidth}]{\label{fig:scene1}\includegraphics[scale=.065]{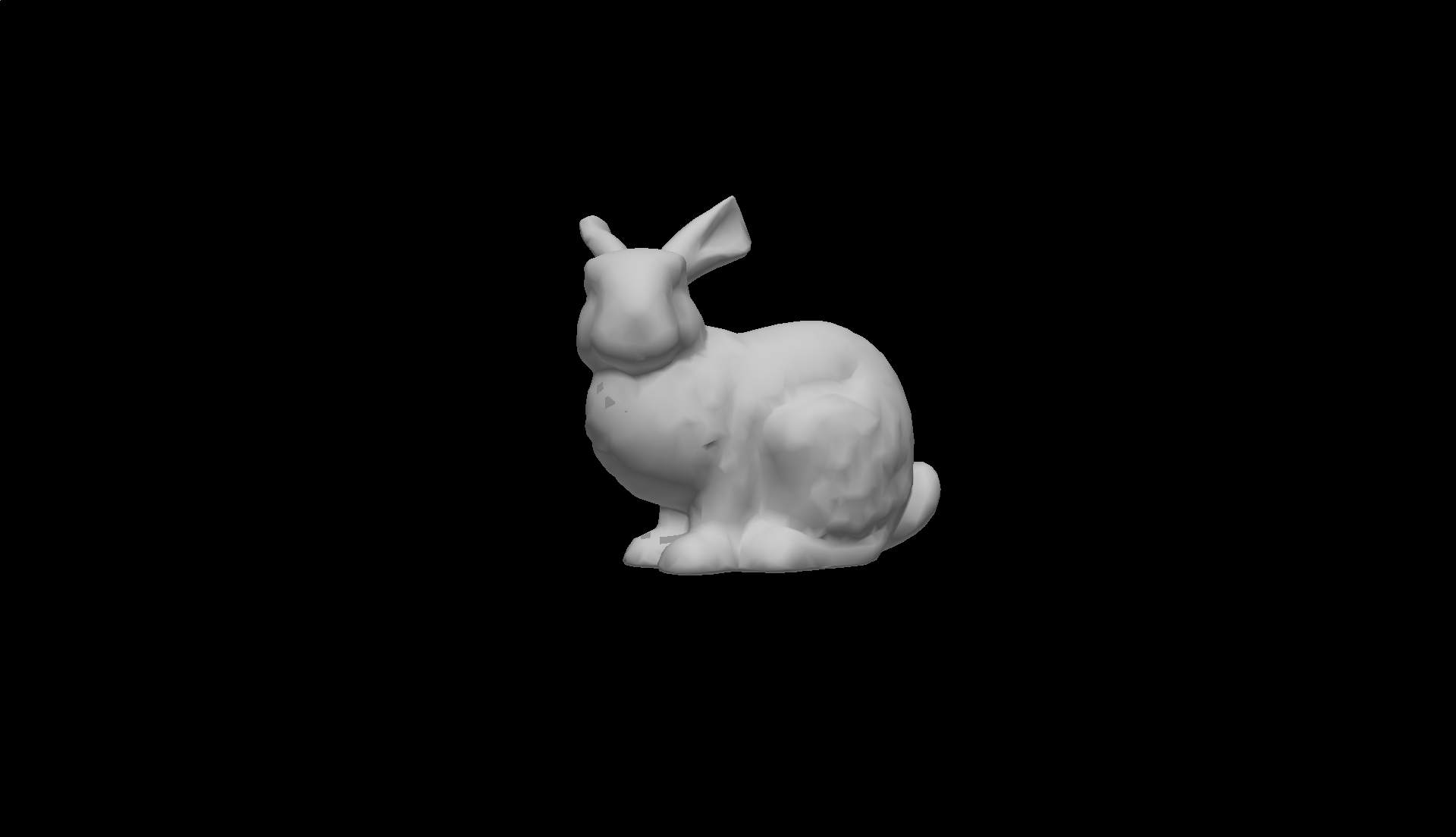}}\hfill%
    \subfigure[\hspace{.21\textwidth}]{\label{fig:scene2}\includegraphics[scale=.065]{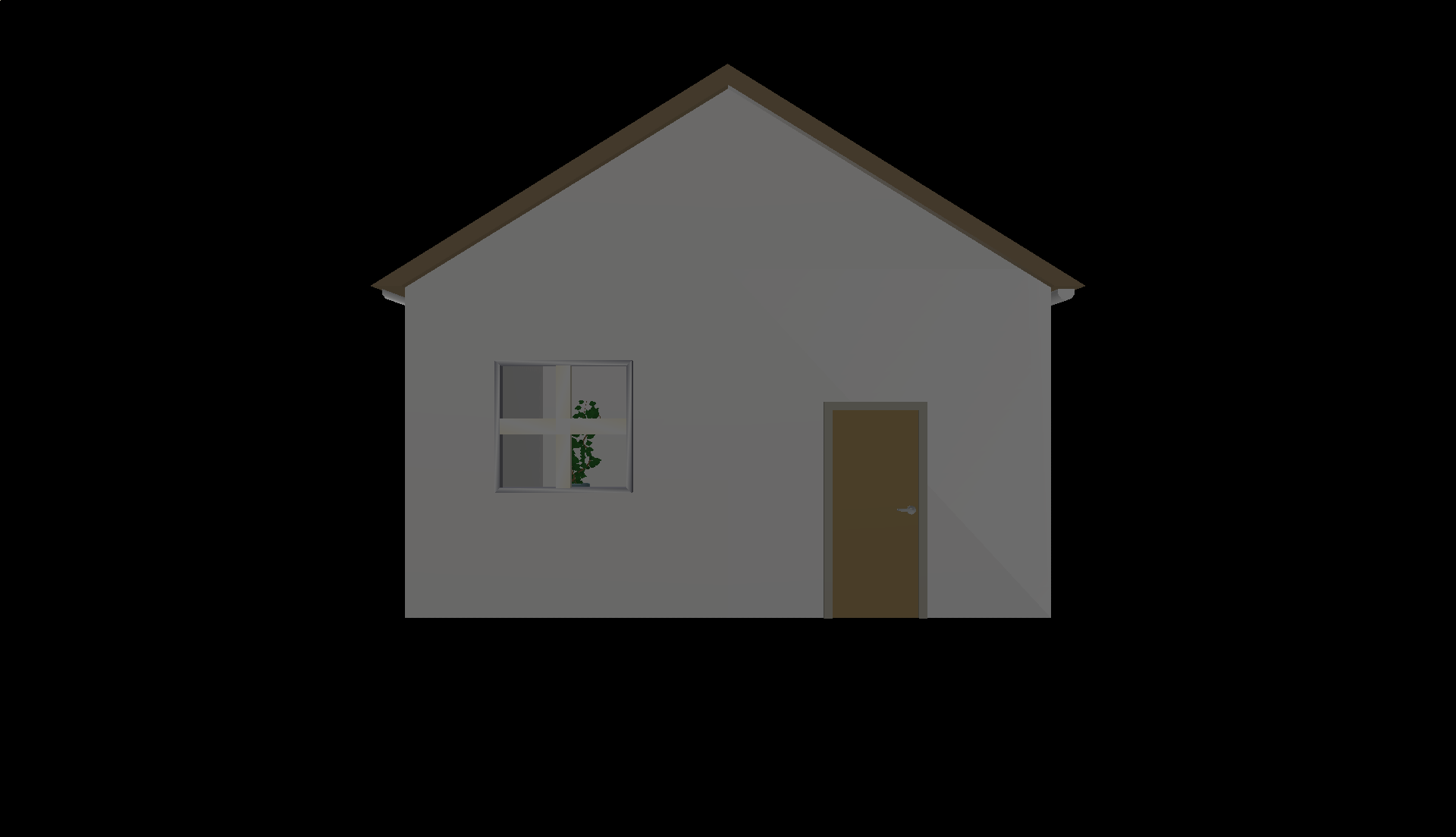}}\hfill%
    \subfigure[\hspace{.21\textwidth}]{\label{fig:scene3}\includegraphics[scale=.065]{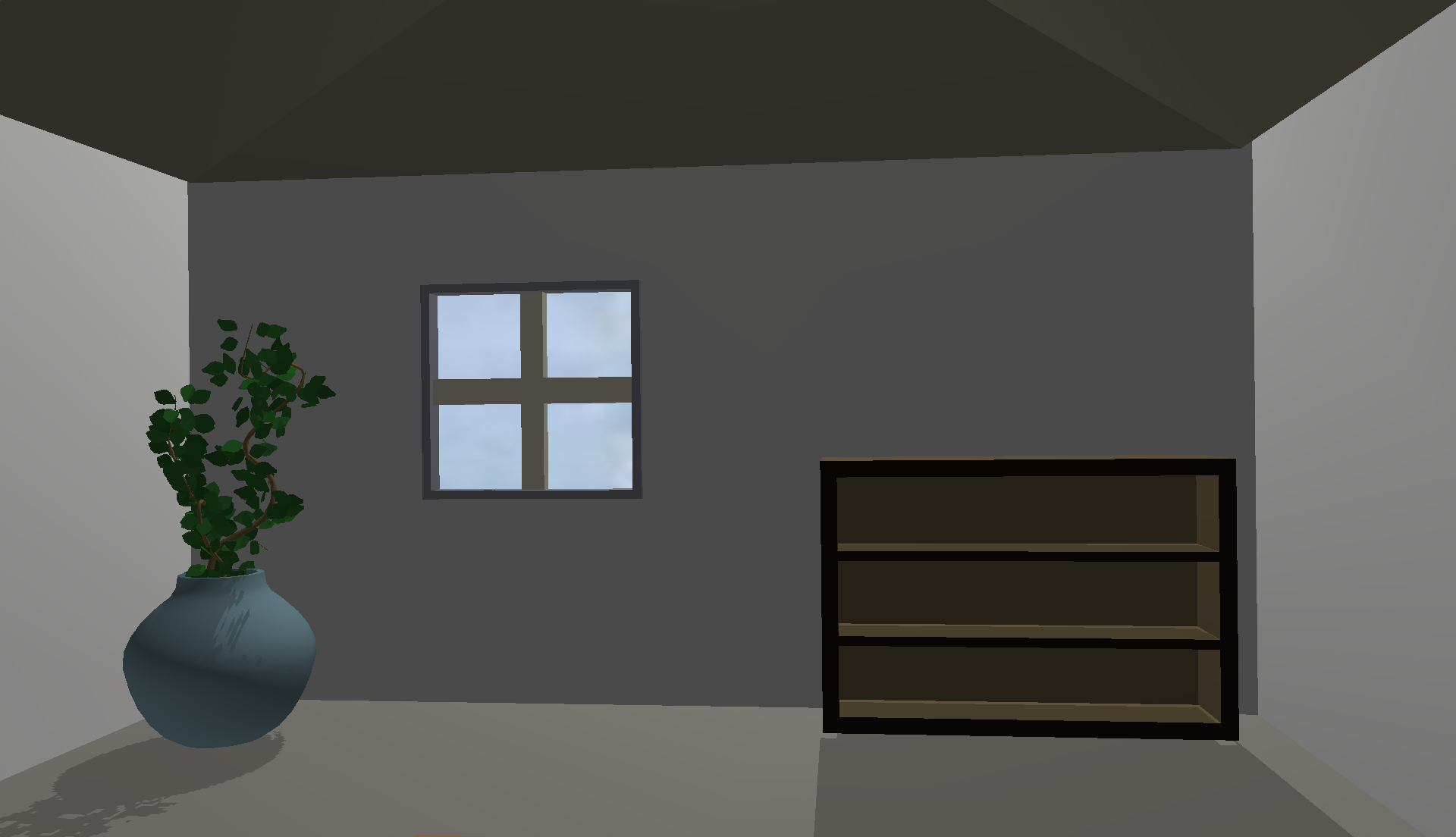}}\hfill%
    \subfigure[\hspace{.21\textwidth}]{\label{fig:scene4}\includegraphics[scale=.065]{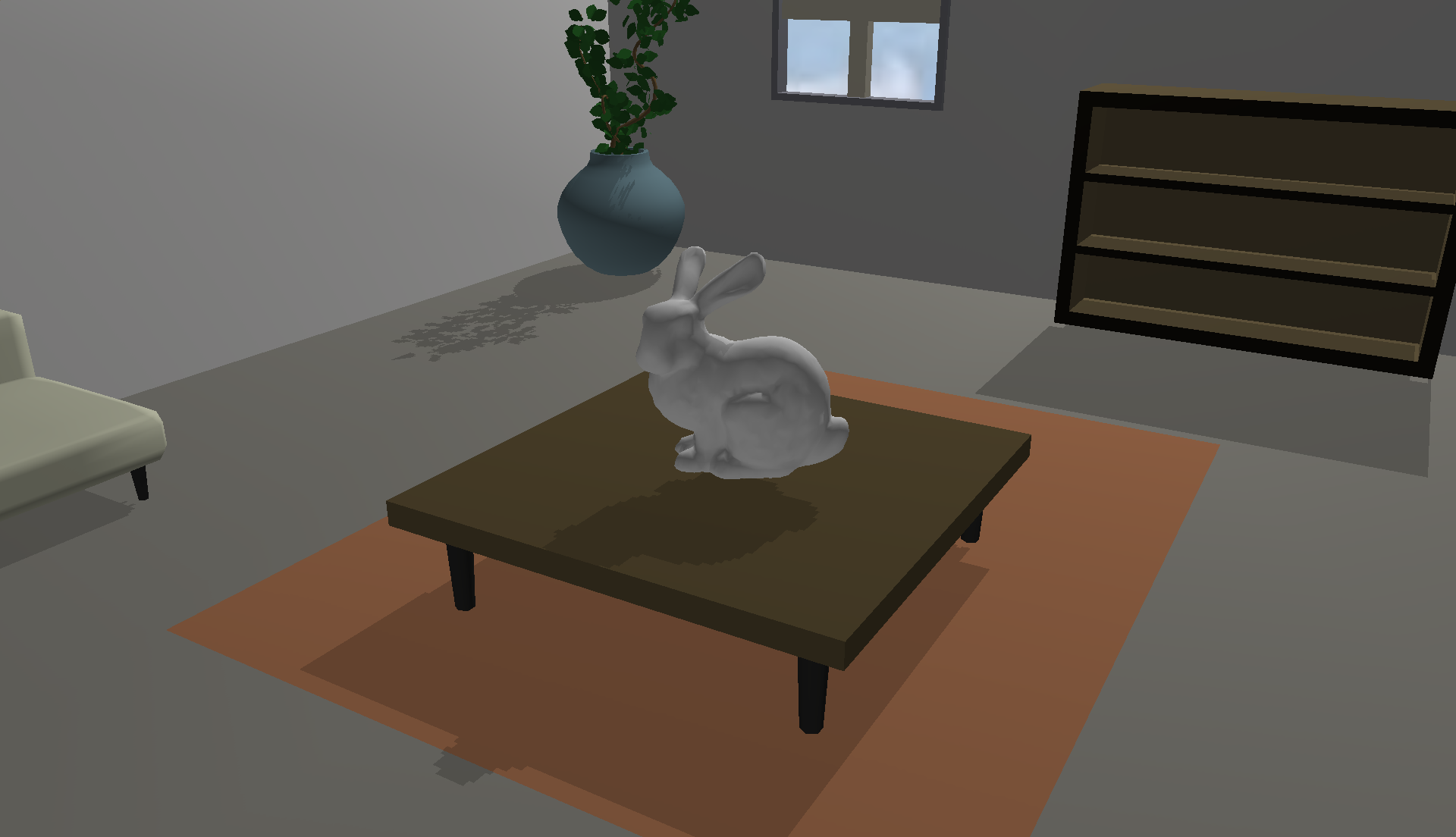}}%
 }%
 \caption[The four scenes used in the experiment]{%
    Illustrations of the four scenes used in the experiment. Participants were asked to perform translations or rotations, starting from a random location, in order to obtain the result shown above. In~scene~4, everything was fixed in screen space apart from the object on the table.%
    \vspace{3mm}
  }%
  \label{fig:scenes}
\end{figure*}

Fewer works have focused on population stereotypes for actual 3D~manipulation tasks. \citet{kaminaka85} studied stereotypes for translations and rotations of a cube about each axis. The cube was controlled by a lever which could be pushed forward or pulled backward. Their results suggest that allocentric manipulation might be preferred for some axes, but there were no significant difference in others. Besides, a lever is only a 1D~control device whereas a mobile device can be freely moved in 3D~space. In another study, \citet{diaz05} investigated ``accidental inversions'' of rotations, \ie~when a user mistakenly rotates the manipulated 3D~object in a direction opposite than intended. The goal was to reveal inconsistencies between what the user was expecting (either allocentric or egocentric manipulation) and the actual mapping encountered. The results revealed two types of user expectations. One group of participants strongly expected a direction that ``matched'' the motion of the input device, while other participants were more uncertain. At first glance, these results seem to discourage inverted mappings. The meaning of ``matching'' and ``inverted,'' however, cannot be easily transposed to our case because in this experiment 3D~rotations were controlled by a separate 2D~input device.

Due to the specifics of the locally coupled configuration it is especially difficult to apply previous results and guidelines to our case. Experiments are generally performed in front of a fixed screen that shows the manipulated object, with a separate input device to control it. In the locally coupled configuration studied here, however, the screen is attached to the input device. The mobile device thus serves a dual function: as a handle to control the object and as a way to depict this object. It can be seen both as an input device controlling the object and as a ``window'' controlling the viewpoint---the chosen interpretation solely depending on the user.

Therefore, the only results applicable here would be those obtained in a locally coupled configuration, so that users face the two possible interpretations. Unfortunately, previous works on locally coupled mappings do not provide sufficient evidence to decide between allocentric and egocentric mappings. In a tilt-based menu selection interface, \citet{rekimoto96} chose to move the menu instead of the cursor (egocentric manipulation) to avoid cases where the cursor would be clamped by the device's screen. \citet{weberg01} preferred allocentric cursor manipulation as it ``felt very intuitive and natural.'' In a tilt-based scrolling technique, \citet{bartlett00} reported that a number of users expected the image to scroll in one direction and a number of others expected the opposite. In a a tilt-based zoom mapping, \citet{hinckley11} merely mentioned that ``some users with a different mental model preferred the opposite mapping.''

Since there do not appear to be sufficient results applicable to 3D~manipulation in a locally coupled configuration, we conducted an exploratory experiment\footnote{A follow-up study with a larger number of participants and an extended analysis can be found at \citep{issartel2016pref}.}. to get more insight into which alternative should be preferred. 

\vspace{2mm}
\textbf{Experiment.}\quad
The goal of this experiment was to identify if there exists a preference for either allocentric of egocentric manipulation---\ie,~whether a user expects that the mobile device would control the object or would be moving relative to this object. We assumed that a mapping matching this expectation would be perceived as more \emph{natural}. Therefore, in this experiment we chose a subjective assessment of the ``naturalness'' of each mode in interactive 3D~manipulation tasks.

In addition to this primary question, we also wanted to investigate whether the type of 3D~object or the type of virtual scene can influence the preference between allocentric and egocentric manipulation. Whether the object is perceived as ``movable'' or ``static'' may influence the preference since it depends on interpretation. Any hint that suggests the viewpoint is moving may also have an influence. We will refer to these cues as \emph{contextual cues}. We thus stated the following hypotheses.

\vspace{1mm}

\textbf{H1}: the preferred mode is likely to be the egocentric one when the manipulated object is viewed from inside, since motion of the surrounding environment suggests that the viewpoint is moving;

\textbf{H2}: the preferred mode is likely to be the allocentric one when an object is manipulated relative to a fixed environment (\ie,~not moving in screen space) since this would suggest that the viewpoint itself is not moving; and

\textbf{H3}: the preferred mode is likely to be the egocentric one when the manipulated object such as a house is perceived as unmovable since in the real world people tend to move around a house rather than moving the house itself.

\vspace{1mm}

To test these hypotheses, we conducted this experiment on four 3D~scenes with different contextual cues (\autoref{fig:scenes}):
\begin{enumerate}[topsep=1mm]
\item\label{item:scene1} a generic object on an empty background,
\item\label{item:scene2} a house model (less likely to be perceived as ``movable'') on an empty background,
\item\label{item:scene3} the same house model viewed from inside, and
\item\label{item:scene4} a generic object inside the house model.
\end{enumerate}
\vspace{5mm}

\textbf{Procedure and tasks.}\quad
The experiment proceeded as follows. Participants were first told they would have to ``perform translations/rotations by moving/tilting the device''. We carefully avoided instructions such as ``manipulating the object'' or ``moving the scene'' which could have biased the results. The first two trials were performed with \sceneref{item:scene1}, the most generic case. One comprised translation tasks, and the other one rotation tasks, presented in random order. Thus, potential learning effects were minimal for this scene. Participants then performed similar translations and rotation trials with the other scenes. All these remaining trials were presented in random order.

A trial consisted of a series of manipulation tasks, either translations or rotations, in both allocentric and egocentric modes. The main purpose of these tasks was to encourage the use of the interface so that participants could rate each mode in realistic conditions. The canonical task for object manipulation is the \emph{docking task} in which an object must be translated and\discretionary{/}{}{/}or rotated from an initial location to a target location. The object was initially put at a random position or orientation (depending on the type of trial) within the scene. Generally, the target location is visually represented by a ``ghost'' copy of the object on the screen. However, we could not display the docking target on the screen as we hypothesized (H2) that a fixed object in screen space could influence the results. Instead, we printed on paper an image of the target position/orientation for each scene~(the same images as in \autoref{fig:scenes}) and we asked participants to ``try to obtain the same result''. On success, the object was translated or rotated to a different location for them to try again, as many times as a participant needed to form an opinion on each mode.

There were two unlabeled, randomly assigned buttons that allowed participants to switch freely between allocentric and egocentric modes. In each trial, participants were asked to perform the manipulation task a few times with each mode before rating them. Ratings were given on a Likert-type scale with 4~points (to avoid neutral answers) going from ``unnatural'' to ``natural''. The meaning of ``natural'' was explained as ``whether your actions produce a translation or a rotation in the direction you expected''.

\textbf{Apparatus.}\quad
The mobile device was a 7''~tactile tablet, held horizontally with two hands. Participants were seated during the experiment. We chose to implement a relative position control mapping, as described above, with a constant gain factor~$k_t\,{=}\,1$ for translations and~$k_t\,{=}\,3$ for rotations. The direction of translations and rotations was inverted in the egocentric condition. Tracking was accomplished with the tablet's integrated gyroscopes and magnetic sensors for orientation as well as with an optically tracked marker for position. The marker was a large textured board (420$\times$297\;mm), placed on a table in front of the user and tracked by the tablet's rear camera using the Vuforia framework\footnote{\url{{http://www.vuforia.com/}}}.
We do not expect the chosen mapping or device to have an influence on the answers since the preference between allocentric and egocentric manipulation relates to the user's own mental model. The results should thus remain applicable to other mappings and mobile devices.

\textbf{Participants.}\quad
Ten unpaid participants (2~female, 8~male) took part in this experiment. Four of them were familiar with 3D~manipulation techniques---though not in this particular configuration---and the others were novices.

The purpose of this study was exploratory: obtaining a first insight into the preferred mode of manipulation in a locally coupled configuration. Although the limited number of participants might not be enough to draw definite conclusions, it should nevertheless allow us to observe sufficiently strong effects.

\begin{figure}[t]
  \centering
  \includegraphics[width=\linewidth]{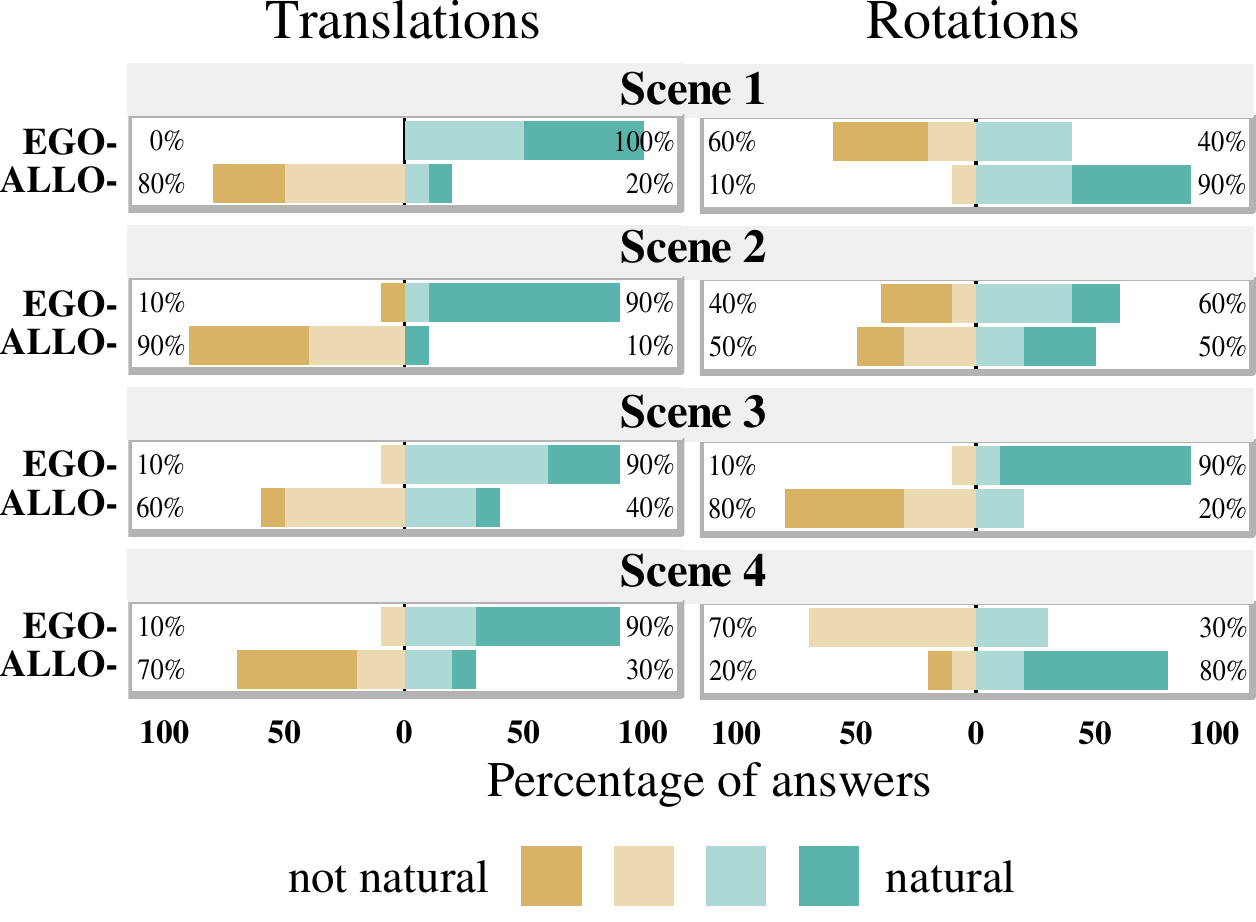}\vspace{.5mm}
  \caption[``Naturalness'' ratings in each of the four scenes]{%
    Subjective ``naturalness'' ratings for \mbox{allocentric} and \mbox{egocentric} mappings in each of the four scenes.%
    \vspace{1mm}
  }
  \label{fig:exp1-results}
\end{figure}

\textbf{Results and discussion.}\quad
The ratings given by participants to each condition are shown in \autoref{fig:exp1-results}. In addition, we used the Wilcoxon signed-rank test to quantify the difference in ratings between the allocentric and egocentric modes. The effect size~$r$ was computed from the~$z$ statistic as described by \citet{fritz12}. Guidelines are that~$r\,{>}\,0.5$ is a large effect, $r\,{>}\,0.3$ is a medium effect and $r\,{>}\,0.1$ is a small effect~\citep{fritz12}. We also report a bootstrapped standard error~$\sigma$ of the effect size from which confidence intervals can be derived for a meta-analysis (bias was~${<}0.01$ in all cases).

In \sceneref{item:scene1} which was presented first to minimize learning effects and which did not contain any contextual cues, we had no reason to expect that participants would prefer either allocentric or egocentric manipulation. Yet, the results reveal a noticeable difference between the two modes. With translations, there was a strong preference ($|r|{=}0.57$, $\sigma{=}0.08$) for egocentric manipulation. With rotations, there was a slightly less contrasted but still strong preference ($|r|{=}0.54$, $\sigma{=}0.09$) for allocentric manipulation.

The three other scenes were designed to study the influence of contextual cues. \sceneref{item:scene2} was almost identical to \sceneref{item:scene1}, except for the manipulated object: a house, \ie~an object that is generally \emph{unmovable} in reality as well as in many virtual environments. This scene was designed to test our hypothesis \textbf{H3} that egocentric manipulation is preferred when the manipulated object is perceived as unmovable. With translations, there was a distinct preference for egocentric manipulation ($|r|{=}0.47$, $\sigma{=}0.16$). The ratings also seem more contrasted than in \sceneref{item:scene1}, which may support our hypothesis. With rotations, both allocentric and egocentric manipulation received a similar proportion of positive and negative ratings ($|r|{=}0.02$, $\sigma{=}0.22$). Compared to \sceneref{item:scene1}, however, allocentric rotation was rated lower and egocentric rotation was rated higher. The preference for egocentric rotation was thus higher than that for \sceneref{item:scene1}, supporting hypothesis \textbf{H3}. Allocentric rotation may also be a sensible choice in this case since the ratings were similar.

\sceneref{item:scene3} was designed to test our hypothesis \textbf{H1} that egocentric manipulation is preferred when the manipulated object is viewed from inside. Indeed, the results show a preference for egocentric manipulation with both translations and rotations. This preference was strong for rotations ($|r|{=}0.64$, $\sigma{=}0.007$). It was less marked but still noticeable for translations ($|r|{=}0.42$, $\sigma{=}0.15$). These results thus support our hypothesis~\textbf{H1}, although we cannot explain why this effect was weaker for translations.

\sceneref{item:scene4} was designed to test our hypothesis \textbf{H2} that allocentric manipulation is preferred when an object is manipulated within a surrounding environment. Concerning rotations, allocentric manipulation was widely considered as natural which tends to support our hypothesis, although egocentric manipulation was neither truly preferred nor truly rejected by participants. The difference in ratings between the two conditions was $|r|{=}0.48$ ($\sigma{=}0.13$). Surprisingly, preferences for translations did not match our \textbf{H2} hypothesis. Egocentric translations were perceived as more natural than allocentric translations ($|r|{=}0.51$, $\sigma{=}0.13$). At this point, we are left with two possible explanations. Either the contextual cues of the 3D~scene actually have little or no influence on translations and thus any scene would lead to the same (egocentric) preference as \sceneref{item:scene1} (some participants commented that they did not pay attention to the surrounding environment in \sceneref{item:scene4} and just performed the task as if it were \sceneref{item:scene1}). Or there could be an unexpected bias in the experiment that led users toward the egocentric interpretation. One possible cause might be the marker used for position tracking. Some participants reported that they were consciously moving the device relative to the marker, which may reveal a potential bias. There are also arguments, however, against the existence of such a bias: If one scene was truly encouraging allocentric translation, while a bias in the experiment caused a shift toward egocentric translation, then the results should be closer to a tie between the two modes (\ie,~similar to the rotation ratings in \sceneref{item:scene2}).

To summarize, for translations the preference was clearly toward egocentric manipulation in every case. It is unknown whether this was caused by an experimental bias or just a coincidence. But if we were to draw guidelines from these preliminary results, we would recommend to make \emph{egocentric translations} the default setting. Concerning rotations, they were apparently affected by the contextual cues in different types of scenes. The preferences were not always as marked as for translations but we can still derive recommendations from these results. Allocentric rotations seem to be a safe choice in most cases, except in \sceneref{item:scene3}. We would thus tend to recommend making \emph{allocentric rotations} the default setting, \emph{except} when the manipulated object is actually viewed from inside.

\section{Motion space}
\label{sec:motion-space}

Using the motion of a handheld device as input modality is subject to a limiting factor: the space in which the device can be translated and rotated by the user---which we refer to as \emph{motion space}. Depending on the application, the manipulated 3D~object may have to be moved across arbitrarily large virtual distances or rotated by arbitrarily large angles. The user, however, cannot apply arbitrarily large displacements to the mobile device itself. In addition, the device itself may not provide enough visual feedback in some positions or orientations. Therefore, a manipulation mapping must provide a way to address these limitations.

In this section, we first present the factors that influence the motion space. We then present several solutions to overcome this limited range in 3D~manipulation mappings, with arguments for and against each alternative.

\subsection{Factors influencing the motion space}

\subsubsection{Anatomical constraints}

Since the user is holding the mobile device during manipulation, device motion is primarily limited to the space reachable by the user's hands. Translations are limited by arm length\footnote{Unless the user moves within tracking space (\eg, by walking while holding the mobile device), a case which we choose not to consider here since the manipulated objects are situated in screen space rather than in an external reference frame.}, and rotations are limited by wrist constraint~\citep{rahman09}. The exact limitations, however, depend on how the mobile device is held. First, a handheld object can be held with either one or two hands. Two types of grasps can also be distinguished: a \emph{power grasp} (or force grasp) in which the device is held firmly in the hand(s), and a \emph{precision grasp} in which the device is held by the fingers~\citep{fitzmaurice96}.

In a power grasp, device movements are more constrained. When the device is held with one hand, translations and rotations are limited by the space reachable by the hand and orientations reachable by the wrist. With two hands, device movements become subject to the limitations of both hands and wrists at the same time, although the range of rotations may increase by using the hands as pivots. In a precision grasp, fingers can compensate for these constraints and greatly enlarge the motion space. With two hands, the freedom of manipulation increases even more.

\subsubsection{Visual feedback}

For devices whose display surface does not cover all directions, a second restriction to motion space is screen visibility. Manipulation of a 3D~virtual object normally requires continuous visual feedback to control the path of the object, so the screen has to remain visible during manipulation. This further reduces the motion space to device configurations that maintain visual feedback. Since many current mobile devices only have a single flat front-facing screen, this restricts the range of possible orientations to those where the screen is facing the user~\citep{rahman09}. Recent panel technologies (such as~IPS or~OLED) provide close to~180\textdegree{} viewing angles. Still, a flat one-sided screen---regardless of its technology---necessarily becomes unreadable when facing away from the user.

\subsubsection{Comfortable motion space}

Even though device motion space can theoretically extend up to arm's length and to wide viewing angles, such positions and orientations are not always practical for prolonged manipulation. Holding the device at arm's length causes fatigue and reduces manipulation accuracy. The screen may also be visible but not necessarily legible. When the device is translated away from the user's eyes, the screen contents appear smaller which reduces legibility. When the device is rotated, perspective distortion (for flat screens) or less defined/darker areas (for some spherical and volumetric displays) can make the screen contents unreadable, effectively interrupting visual feedback.
Therefore, we could further distinguish the \emph{full motion space}, theoretically reachable by the user, from the smaller \emph{comfortable motion space} in which the device can be handled comfortably. A good mapping should not require the user to move the device outside the comfortable motion space. Still, the additional accessible space can be exploited for short-term or less frequent manipulation tasks.

The extents of the full motion space and the comfortable motion space may be determined experimentally. However, the results would be strongly device-specific. Such an experiment thus falls outside the scope of our theoretical framework. A future, dedicated study could be conducted on the most commonly available devices, in order to better adjust the mappings to current mobile platforms.

\subsection{Overcoming motion space limitations}
\label{sec:motion-space-solutions}

There are different ways to work around motion space limitations in a manipulation mapping. Although this remains an open problem, we present here some of the traditional solutions. We then demonstrate how more advanced techniques can be implemented by extending the mappings we presented before.

\subsubsection{Rate control}

One way to circumvent almost entirely this issue is to use a rate control mapping. Under a rate control mapping, the manipulated object moves continuously as long as the mobile device is away from the starting position and orientation, with a speed that depends on the distance from this point. Thus, with a reasonable gain factor and/or given enough time, unlimited translations and rotations can be applied to the manipulated object while keeping the device well inside its motion space.

Unlimited movement is undoubtedly a significant advantage of rate control mappings. However, such mappings are not without drawbacks compared to position control. For instance, changing the direction of motion \emph{first} requires to return the device to its initial location, \emph{then} to move it toward the new direction. This is slower than position control, in which such change of direction would happen instantly. Stopping the manipulated object at a given position or orientation is also difficult with rate control. This requires to bring back the mobile device to its initial location with a precise timing. Since a mobile device is untethered and freely manipulable, \ie,~isotonic, there is no self-centering mechanism and no feedback to help the user return to the initial location~\citep{zhai98}. The numerous degrees of freedom in 3D~space make it particularly difficult. This is supported by experimental results: \citet{zhai98-2} showed that rate control mappings are less efficient than position control mappings, when isotonic input devices are used for a 3D~docking task.

The difficulty of zeroing the device in mid air can be alleviated by adding a ``deadband'' to the mapping. In our formalism, this can be accomplished by setting the gain factor to~0 when the distance (or angle) between the current device location and its initial location is below a given threshold. We first introduce a~$\mathit{dist}$ function to compute the distance or angle between two device locations at times~$t_1$ and~$t_2$ (here, $(\quat{q})_w$~denotes the real part of a quaternion~$\quat{q}$):
\begin{flalign}
  \quad
  \dist(t_1,t_2)
  &= \left\{\begin{array}{ll}
      \|\vec{pc_{t_1}}-\vec{pc_{t_2}}\|&\mathrm{(translations)}\\
      2\:\arccos\!\big((\quat{qc_{t_1}}\:\quat{qc_{t_2}}^{-1})_w\big)&\mathrm{(rotations)}
    \end{array}\right.
  \label{eqn:dist-func}
\end{flalign}
The distance between the current and initial device locations is given by~$\dist(t,0)$. A deadband of radius $\mathrm{threshold}$ can thus be obtained with the following gain function:
\begin{flalign}
  \quad
  d &= \dist(t,0)&\nonumber\\
  \quad
  \gain(t)
  &= \left\{\begin{array}{ll}
      0&\mathrm{when}~\mathrm{d}<\mathrm{threshold}\\
      (d-\mathrm{threshold})\;/\;d&\mathrm{otherwise}
    \end{array}\right.\nonumber
  \nolabel{eqn:gain-deadband}
\end{flalign}
A deadband, however, uses up a valuable part of the motion space, forcing the device into less comfortable positions and orientations. Also a change of motion direction becomes even less efficient.

Therefore, the use of rate control mappings should be considered carefully. Whether unlimited object motion is worth the loss of accuracy and efficiency in manipulation depends on the application, and especially the frequency of long-distance object manipulation.

\subsubsection{Clutching}
\label{sec:clutching}

Position control mappings do not allow unlimited object movement, but there are still ways to overcome the motion space limitation. \emph{Clutching} consists in temporarily disengaging the mapping between device and object motion, returning the device to a more convenient location in the motion space, then resuming manipulation from there. This enables arbitrarily large translations and rotations to be applied to the manipulated object, by decomposing them into smaller steps that fit within the motion space. Clutching requires a way to explicitly engage or disengage manipulation, such as a button on the mobile device. In any case, such a mechanism is recommended since it lets users move the device for other purposes than object manipulation.

However, clutching is best avoided \emph{during} manipulation. By interrupting the interaction, it slows down object manipulation and reduces efficiency~\citep{jellinek90}. The further the object has to be moved or rotated, the more clutching must be used and the less efficient manipulation becomes. Clutching also causes ``wasted'' user movement~\citep{zhai98}. Moving a handheld device in 3D~space is demanding, and frequent clutching can quickly lead to user fatigue.

Clutching is generally required for position control mappings to support exceptionally large translations and rotations. Occasional clutching is a normal occurrence in a position control mapping. But a good mapping should be designed to reduce it as much as possible for normal use cases.

\subsubsection{Larger gain factor}

The need for clutching can be reduced by mapping device movements to larger object movements. In the formalism we presented before, this is easily accomplished by increasing the gain factor~$k_t$ (see \autoref{sec:gain}) to a value greater than~1. With an increased gain factor, the same device movements will result in larger object translations or rotations. The larger the gain factor, the less likely the user will be limited by motion space during manipulation.

The gain factor, on the other hand, cannot be infinite. The need for clutching can thus never be completely eliminated with this method. Furthermore, human motor resolution is also limited~\citep{berard11}. There is a limit to how much the gain can be increased before it exceeds a user's ability to control the manipulated object, after which accuracy begins to drop. The exact threshold depends both on the user's own motor skills and the manipulation task to be carried out. Conversely, reducing the gain factor to less than~1 could artificially increase manipulation accuracy beyond the human motor abilities (within the limits of the tracking system). But this would also require larger device movements to translate or rotate the object along the same distance, and cause more frequent clutching.

Increasing the gain factor is thus an effective solution against clutching in position control mappings, but leads to a trade-off between motion space and accuracy. To avoid this trade-off, more advanced strategies must be employed for managing the gain factor.

\subsubsection{Dynamic gain factor}
\label{sec:dynamic-gain}

The problem with a static gain factor is that it affects every manipulation task performed with the interface. Ideally, the gain factor should be dynamically adapted to the type of manipulation: larger for coarse long-distance object manipulation, and smaller for precise short-distance manipulation \citep{frees07,kopper11}. However, in order to do that, we need to know what the user intends to perform next.

One solution is to give the user explicit control on the gain factor~\citep[\eg,][]{ware89}. However, this would require an additional input modality, since all the degrees of freedom of the mobile device are already used for manipulation. This would also cause additional cognitive load during interaction (the need for a specific learning period is mentioned by \citet{ware89}). Another solution is to give the user \emph{implicit} control on the gain factor: by varying the gain factor according to patterns in device movement. We present here two methods to implicitly control the gain factor.

\textbf{Gain based on distance.}\quad
This method consists in increasing the gain when the mobile device moves or rotates away from the location where manipulation was initiated. It is motivated by some observations and assumptions. First, users generally begin manipulation from a comfortable device position and orientation. If users only move or rotate the device by a small amount around this starting location, we can assume they are trying to perform precise manipulation of the virtual object. We thus keep the gain factor low in this area. This is safe because clutching is unlikely to be needed when the device is still inside the comfortable motion space. On the other hand, if users move or rotate the device far away from its initial location, we can assume they are attempting to translate or rotate the virtual object by a large amount. We thus increase the gain factor to support long-distance object manipulation and reduce the risk of reaching the limits of the motion space. Since there is no way to know what the user intends to do before the actual movement occurred, we can transition smoothly between the two situations by making the gain a function of the distance from the starting location. This behavior is implemented by the following gain function, based on the~$\mathit{dist}$ function we introduced before (\autoref{eqn:dist-func}):
\begin{flalign}
  \gain(t)
  &= a + b\:\dist(t,0)^c
  \nolabel{eqn:gain-dist}
\end{flalign}
Previous works have proposed similar techniques that increase the gain with distance, such as the mapping proposed by \citet{poupyrev99} for rotations, or the non-linear part of the the Go-Go Interaction Technique~\citep{poupyrev96} for translations. In the gain function we formalize here, the terms~$a$ (minimum gain factor), $b$~(scaling coefficient) and~$c$ (exponent) can be adjusted to recreate these mappings.

\textbf{Gain based on speed.}\quad
A different approach, inspired by the ``pointer acceleration'' technique for computer mice and the ``PRISM'' technique for 3D~manipulation \citep{frees07}, is to base the gain factor on the mobile device's \emph{speed}. In other words, the faster the mobile device is moved\discretionary{/}{}{/}rotated by the user, the faster the ma\-nipulated object will move\discretionary{/}{}{/}rotate. In the real world, precise object manipulation has to be performed slowly and carefully, whereas coarser and larger-scale manipulation can be performed with faster movements. We can thus exploit this metaphor to let users implicitly control the amount of gain they expect during manipulation.

The speed of the mobile device corresponds to the distance or angle crossed by the device between times~$t{-}1$ and~$t$, divided by the interval of time~$\Delta{\kern.1em}t$ elapsed between the two steps. Note that this method relies on incremental device translations and rotations, it is thus \emph{only meaningful for relative position control mappings}. On such mappings, the speed-based gain can be implemented with the following function:
\begin{flalign}
  \gain(t)
  &= a + b\;\bigg(\frac{\dist(t,\,t{-}1)}{\Delta{\kern.1em}t}\bigg)^{\kern-2ptc}
  \nolabel{eqn:gain-speed}
\end{flalign}
As in the previous method, the~$a$, $b$ and~$c$ terms make it possible to fine-tune the shape of the gain function, for instance to match existing pointer acceleration schemes.

\section{Conclusion}

In this work, we investigated a specific mode of interaction: using the motion of a mobile device to manipulate a 3D~object displayed on the device itself. More precisely, we conducted the first in-depth analysis of the possible \emph{mappings} between device motion and object motion in this particular configuration. We introduced a formalization of the main mappings, which---unlike many previous works---covers both device translations and device rotations and supports a variable control-display gain. We performed a theoretical analysis of the spatial compliance properties of each mapping. We then reviewed the arguments for each property, contributing new results when needed, allowing implementors to choose which mapping is best suited to their needs. We conducted a first study on user preference between allocentric and egocentric manipulation, in order to help implementors select a sensible default setting. We finally presented a theoretical analysis of the motion space available for manipulation while holding a mobile device, as well as several solutions including non-isomorphic gain functions to overcome this limitation.

By presenting the main mappings in a consistent notation and addressing important questions often overlooked by previous work, our framework provides implementors with readily available advice to design such manipulation techniques. Furthermore, our framework is applicable to any handheld device capable of displaying a virtual object (according to the definition given in \autoref{sec:mobile-device-definition}), regardless of its shape or the technologies used. Since mobile devices are going to become even more ubiquitous, our work may serve a basis for 3D~manipulation techniques on future~devices.

To better understand the further implications of this mode of interaction, a number of user studies could be conducted in the \mbox{future}. One study could compare the usability of the non-isomorphic gain functions presented in \autoref{sec:motion-space}. Although they are based on real-world metaphors, it is unknown whether they can be easily understood in the specific configuration studied here. Another question of interest is the usability of integrated (\ie,~simultaneous) translations and rotations. Our formalized mappings cover both translations and rotations, and our analysis of spatial compliances did take into account the effects of simultaneously rotating and translating the \emph{device}. However, the usability consequences of simultaneously translating and rotating the \emph{manipulated object} remain to be studied. In addition, different mappings may be preferable for translations than for rotations (\eg,~relative position control for translations and rate control for rotations). It might also be preferable to use different gain functions. Further work is thus needed to evaluate the effects of having different mappings for translations and rotations, and to identify potential usability challenges arising in this new situation.

\def\url#1{}
\makeatletter
\renewcommand\@biblabel[1]{}
\makeatother
\bibliographystyle{apalike-doi-hyperref}
\bibliography{article}

\begin{thebibliography}{}

\bibitem[Alexander et~al., 2012]{alexander12}
\href{http://dx.doi.org/10.1145/2371574.2371600}{Alexander, J., Lucero, A., and
  Subramanian, S.} (2012).
\newblock \href{http://dx.doi.org/10.1145/2371574.2371600}{Tilt displays:
  Designing display surfaces with multi-axis tilting and actuation}.
\newblock \href{http://dx.doi.org/10.1145/2371574.2371600}{In {\em Proc.\
  MobileHCI}}, \href{http://dx.doi.org/10.1145/2371574.2371600}{pages
  161--170}. \href{http://dx.doi.org/10.1145/2371574.2371600}{ACM},
  \href{http://dx.doi.org/10.1145/2371574.2371600}{New York}.
  \href{http://dx.doi.org/10.1145/2371574.2371600}
{doi: {{%
10\hspace{.1pt}\discretionary{.}{%
}{.}\hspace{.4pt}1145\discretionary{/}{%
}{/}2371574\hspace{.1pt}\discretionary{.}{%
}{.}\hspace{.4pt}2371600}}}


\bibitem[Bade et~al., 2005]{bade05}
\href{http://dx.doi.org/10.1007/11536482_12}{Bade, R., Ritter, F., and Preim,
  B.} (2005).
\newblock \href{http://dx.doi.org/10.1007/11536482_12}{Usability comparison of
  mouse-based interaction techniques for predictable {3D} rotation}.
\newblock \href{http://dx.doi.org/10.1007/11536482_12}{In {\em Proc.\ Smart
  Graphics}}, \href{http://dx.doi.org/10.1007/11536482_12}{pages 138--150}.
  \href{http://dx.doi.org/10.1007/11536482_12}{Springer},
  \href{http://dx.doi.org/10.1007/11536482_12}{Berlin\discretionary{/}{}{/}Heidelberg}.
  \href{http://dx.doi.org/10.1007/11536482_12}
{doi: {{%
10\hspace{.1pt}\discretionary{.}{%
}{.}\hspace{.4pt}1007\discretionary{/}{%
}{/}11536482\_12}}}


\bibitem[Bartlett, 2000]{bartlett00}
\href{http://dx.doi.org/10.1109/38.844371}{Bartlett, J.} (2000).
\newblock \href{http://dx.doi.org/10.1109/38.844371}{{R}ock 'n' {S}croll is
  here to stay}.
\newblock \href{http://dx.doi.org/10.1109/38.844371}{{\em {IEEE} Computer
  Graphics and Applications}},
  \href{http://dx.doi.org/10.1109/38.844371}{20(3):40--45}.
  \href{http://dx.doi.org/10.1109/38.844371}
{doi: {{%
10\hspace{.1pt}\discretionary{.}{%
}{.}\hspace{.4pt}1109\discretionary{/}{%
}{/}38\hspace{.1pt}\discretionary{.}{%
}{.}\hspace{.4pt}844371}}}


\bibitem[Benko et~al., 2008]{benko08}
\href{http://dx.doi.org/10.1145/1449715.1449729}{Benko, H., Wilson, A.~D., and
  Balakrishnan, R.} (2008).
\newblock \href{http://dx.doi.org/10.1145/1449715.1449729}{Sphere: Multi-touch
  interactions on a spherical display}.
\newblock \href{http://dx.doi.org/10.1145/1449715.1449729}{In {\em Proc.\
  UIST}}, \href{http://dx.doi.org/10.1145/1449715.1449729}{pages 77--86}.
  \href{http://dx.doi.org/10.1145/1449715.1449729}{ACM},
  \href{http://dx.doi.org/10.1145/1449715.1449729}{New York}.
  \href{http://dx.doi.org/10.1145/1449715.1449729}
{doi: {{%
10\hspace{.1pt}\discretionary{.}{%
}{.}\hspace{.4pt}1145\discretionary{/}{%
}{/}1449715\hspace{.1pt}\discretionary{.}{%
}{.}\hspace{.4pt}1449729}}}


\bibitem[Benzina et~al., 2012]{benzina12}
\href{http://dx.doi.org/10.1145/2350046.2350078}{Benzina, A., Dey, A., Tönnis,
  M., and Klinker, G.} (2012).
\newblock \href{http://dx.doi.org/10.1145/2350046.2350078}{Empirical evaluation
  of mapping functions for navigation in virtual reality using phones with
  integrated sensors}.
\newblock \href{http://dx.doi.org/10.1145/2350046.2350078}{In {\em Proc.\
  APCHI}}, \href{http://dx.doi.org/10.1145/2350046.2350078}{pages 149--158}.
  \href{http://dx.doi.org/10.1145/2350046.2350078}{ACM},
  \href{http://dx.doi.org/10.1145/2350046.2350078}{New York}.
  \href{http://dx.doi.org/10.1145/2350046.2350078}
{doi: {{%
10\hspace{.1pt}\discretionary{.}{%
}{.}\hspace{.4pt}1145\discretionary{/}{%
}{/}2350046\hspace{.1pt}\discretionary{.}{%
}{.}\hspace{.4pt}2350078}}}


\bibitem[B{\'e}rard et~al., 2011]{berard11}
\href{http://dx.doi.org/10.1007/978-3-642-23771-3_10}{B{\'e}rard, F., Wang, G.,
  and Cooperstock, J.~R.} (2011).
\newblock \href{http://dx.doi.org/10.1007/978-3-642-23771-3_10}{On the limits
  of the human motor control precision: The search for a device’s human
  resolution}.
\newblock \href{http://dx.doi.org/10.1007/978-3-642-23771-3_10}{In {\em Proc.\
  INTERACT}}, \href{http://dx.doi.org/10.1007/978-3-642-23771-3_10}{pages
  107--122}. \href{http://dx.doi.org/10.1007/978-3-642-23771-3_10}{Springer},
  \href{http://dx.doi.org/10.1007/978-3-642-23771-3_10}{Berlin\discretionary{/}{}{/}Heidelberg}.
  \href{http://dx.doi.org/10.1007/978-3-642-23771-3_10}
{doi: {{%
10\hspace{.1pt}\discretionary{.}{%
}{.}\hspace{.4pt}1007\discretionary{/}{%
}{/}978\discretionary{%
}{-}{-}3\discretionary{%
}{-}{-}642\discretionary{%
}{-}{-}23771\discretionary{%
}{-}{-}3\_10}}}


\bibitem[Bowman et~al., 2004]{bowman04}
\href{http://people.cs.vt.edu/~bowman/3dui.org/3D_UI_Book.html}{Bowman, D.~A.,
  Kruijff, E., LaViola, Jr., J.~J., and Poupyrev, I.} (2004).
\newblock \href{http://people.cs.vt.edu/~bowman/3dui.org/3D_UI_Book.html}{{\em
  {3D} User Interfaces: Theory and Practice}}.
\newblock
  \href{http://people.cs.vt.edu/~bowman/3dui.org/3D_UI_Book.html}{Addison-Wesley},
  \href{http://people.cs.vt.edu/~bowman/3dui.org/3D_UI_Book.html}{Boston}.

\bibitem[Britton et~al., 1978]{britton78}
\href{http://dx.doi.org/10.1145/800248.807394}{Britton, E.~G., Lipscomb, J.~S.,
  and Pique, M.~E.} (1978).
\newblock \href{http://dx.doi.org/10.1145/800248.807394}{Making nested
  rotations convenient for the user}.
\newblock \href{http://dx.doi.org/10.1145/800248.807394}{{\em ACM SIGGRAPH
  Computer Graphics}},
  \href{http://dx.doi.org/10.1145/800248.807394}{12(3):222--227}.
  \href{http://dx.doi.org/10.1145/800248.807394}
{doi: {{%
10\hspace{.1pt}\discretionary{.}{%
}{.}\hspace{.4pt}1145\discretionary{/}{%
}{/}800248\hspace{.1pt}\discretionary{.}{%
}{.}\hspace{.4pt}807394}}}


\bibitem[Buda, 2012]{buda12}
\href{http://www.staff.science.uu.nl/~hurst101/temp/MSc_Vlad.pdf}{Buda, V.}
  (2012).
\newblock
  \href{http://www.staff.science.uu.nl/~hurst101/temp/MSc_Vlad.pdf}{Rotation
  techniques for {3D} object interaction on mobile devices}.
\newblock
  \href{http://www.staff.science.uu.nl/~hurst101/temp/MSc_Vlad.pdf}{Master's
  thesis},
  \href{http://www.staff.science.uu.nl/~hurst101/temp/MSc_Vlad.pdf}{Utrecht
  University},
  \href{http://www.staff.science.uu.nl/~hurst101/temp/MSc_Vlad.pdf}{the
  Netherlands}.

\bibitem[Chan et~al., 2003]{chan03}
\href{http://dx.doi.org/10.1007/s00170-002-1491-z}{Chan, A. H.~S., Shum, V.
  W.~Y., Law, H.~W., and Hui, I.~K.} (2003).
\newblock \href{http://dx.doi.org/10.1007/s00170-002-1491-z}{Precise effects of
  control position, indicator type, and scale side on human performance}.
\newblock \href{http://dx.doi.org/10.1007/s00170-002-1491-z}{{\em The
  International Journal of Advanced Manufacturing Technology}},
  \href{http://dx.doi.org/10.1007/s00170-002-1491-z}{22(5--6):380--386}.
  \href{http://dx.doi.org/10.1007/s00170-002-1491-z}
{doi: {{%
10\hspace{.1pt}\discretionary{.}{%
}{.}\hspace{.4pt}1007\discretionary{/}{%
}{/}s00170\discretionary{%
}{-}{-}002\discretionary{%
}{-}{-}1491\discretionary{%
}{-}{-}z}}}


\bibitem[Cho et~al., 2007]{cho07}
\href{http://dx.doi.org/10.1145/1240866.1240930}{Cho, S.-J., Choi, C., Sung,
  Y., Lee, K., Kim, Y.-B., and Murray-Smith, R.} (2007).
\newblock \href{http://dx.doi.org/10.1145/1240866.1240930}{Dynamics of
  tilt-based browsing on mobile devices}.
\newblock \href{http://dx.doi.org/10.1145/1240866.1240930}{In {\em CHI Extended
  Abstracts}}, \href{http://dx.doi.org/10.1145/1240866.1240930}{pages
  1947--1952}. \href{http://dx.doi.org/10.1145/1240866.1240930}{ACM},
  \href{http://dx.doi.org/10.1145/1240866.1240930}{New York}.
  \href{http://dx.doi.org/10.1145/1240866.1240930}
{doi: {{%
10\hspace{.1pt}\discretionary{.}{%
}{.}\hspace{.4pt}1145\discretionary{/}{%
}{/}1240866\hspace{.1pt}\discretionary{.}{%
}{.}\hspace{.4pt}1240930}}}


\bibitem[Daiber et~al., 2012]{daiber12}
\href{http://dx.doi.org/10.1145/2406367.2406371}{Daiber, F., Li, L., and
  Krüger, A.} (2012).
\newblock \href{http://dx.doi.org/10.1145/2406367.2406371}{Designing gestures
  for mobile {3D} gaming}.
\newblock \href{http://dx.doi.org/10.1145/2406367.2406371}{In {\em Proc.\
  MUM}}, \href{http://dx.doi.org/10.1145/2406367.2406371}{pages 3:1--3:4}.
  \href{http://dx.doi.org/10.1145/2406367.2406371}{ACM},
  \href{http://dx.doi.org/10.1145/2406367.2406371}{New York}.
  \href{http://dx.doi.org/10.1145/2406367.2406371}
{doi: {{%
10\hspace{.1pt}\discretionary{.}{%
}{.}\hspace{.4pt}1145\discretionary{/}{%
}{/}2406367\hspace{.1pt}\discretionary{.}{%
}{.}\hspace{.4pt}2406371}}}


\bibitem[Diaz and Sims, 2005]{diaz05}
\href{http://dx.doi.org/10.1177/154193120504901307}{Diaz, D.~D. and Sims,
  V.~K.} (2005).
\newblock \href{http://dx.doi.org/10.1177/154193120504901307}{Accidental
  inversion during three-dimensional orientational control}.
\newblock \href{http://dx.doi.org/10.1177/154193120504901307}{{\em Proceedings
  of the Human Factors and Ergonomics Society Annual Meeting}},
  \href{http://dx.doi.org/10.1177/154193120504901307}{49(13):1248--1250}.
  \href{http://dx.doi.org/10.1177/154193120504901307}
{doi: {{%
10\hspace{.1pt}\discretionary{.}{%
}{.}\hspace{.4pt}1177\discretionary{/}{%
}{/}154193120504901307}}}


\bibitem[Du et~al., 2011]{du11}
\href{http://dx.doi.org/10.1145/2030112.2030183}{Du, Y., Ren, H., Pan, G., and
  Li, S.} (2011).
\newblock \href{http://dx.doi.org/10.1145/2030112.2030183}{Tilt \& touch:
  Mobile phone for {3D} interaction}.
\newblock \href{http://dx.doi.org/10.1145/2030112.2030183}{In {\em Proc.\
  UbiComp}}, \href{http://dx.doi.org/10.1145/2030112.2030183}{pages 485--486}.
  \href{http://dx.doi.org/10.1145/2030112.2030183}{ACM},
  \href{http://dx.doi.org/10.1145/2030112.2030183}{New York}.
  \href{http://dx.doi.org/10.1145/2030112.2030183}
{doi: {{%
10\hspace{.1pt}\discretionary{.}{%
}{.}\hspace{.4pt}1145\discretionary{/}{%
}{/}2030112\hspace{.1pt}\discretionary{.}{%
}{.}\hspace{.4pt}2030183}}}


\bibitem[Eslambolchilar and Murray-Smith, 2008]{eslambolchilar08}
\href{http://dx.doi.org/10.1016/j.ijhcs.2008.07.005}{Eslambolchilar, P. and
  Murray-Smith, R.} (2008).
\newblock \href{http://dx.doi.org/10.1016/j.ijhcs.2008.07.005}{Control centric
  approach in designing scrolling and zooming user interfaces}.
\newblock \href{http://dx.doi.org/10.1016/j.ijhcs.2008.07.005}{{\em
  International Journal of Human-Computer Studies}},
  \href{http://dx.doi.org/10.1016/j.ijhcs.2008.07.005}{66(12):838--856}.
  \href{http://dx.doi.org/10.1016/j.ijhcs.2008.07.005}
{doi: {{%
10\hspace{.1pt}\discretionary{.}{%
}{.}\hspace{.4pt}1016\discretionary{/}{%
}{/}j\hspace{.1pt}\discretionary{.}{%
}{.}\hspace{.4pt}ijhcs\hspace{.1pt}\discretionary{.}{%
}{.}\hspace{.4pt}2008\hspace{.1pt}\discretionary{.}{%
}{.}\hspace{.4pt}07\hspace{.1pt}\discretionary{.}{%
}{.}\hspace{.4pt}005}}}


\bibitem[Feinstein, 2002]{feinstein02}
\href{http://www.google.com/patents/US6466198}{Feinstein, D.} (2002).
\newblock \href{http://www.google.com/patents/US6466198}{View navigation and
  magnification of a hand-held device with a display}.
\newblock \href{http://www.google.com/patents/US6466198}{{US} Patent No.\
  6,466,198}.

\bibitem[Fitts, 1951]{fitts51}
\href{http://psycnet.apa.org/psycinfo/1951-07758-000}{Fitts, P.~M.} (1951).
\newblock \href{http://psycnet.apa.org/psycinfo/1951-07758-000}{Engineering
  psychology and equipment design}.
\newblock \href{http://psycnet.apa.org/psycinfo/1951-07758-000}{In Stevens,
  S.~S., editor, {\em Handbook of Experimental Psychology}},
  \href{http://psycnet.apa.org/psycinfo/1951-07758-000}{pages 1287--1340}.
  \href{http://psycnet.apa.org/psycinfo/1951-07758-000}{Wiley}.

\bibitem[Fitts and Seeger, 1953]{fitts53}
\href{http://dx.doi.org/10.1037/h0062827}{Fitts, P.~M. and Seeger, C.~M.}
  (1953).
\newblock \href{http://dx.doi.org/10.1037/h0062827}{{SR} compatibility: Spatial
  characteristics of stimulus and response codes}.
\newblock \href{http://dx.doi.org/10.1037/h0062827}{{\em Journal of
  Experimental Psychology}},
  \href{http://dx.doi.org/10.1037/h0062827}{46(3):199--210}.
  \href{http://dx.doi.org/10.1037/h0062827}
{doi: {{%
10\hspace{.1pt}\discretionary{.}{%
}{.}\hspace{.4pt}1037\discretionary{/}{%
}{/}h0062827}}}


\bibitem[Fitzmaurice, 1996]{fitzmaurice96}
\href{http://www.dgp.toronto.edu/~gf/papers/PhD%20-%20Graspable%20UIs/Thesis.gf.html}{Fitzmaurice,
  G.~W.} (1996).
\newblock
  \href{http://www.dgp.toronto.edu/~gf/papers/PhD%20-%20Graspable%20UIs/Thesis.gf.html}{{\em
  Graspable user interfaces}}.
\newblock
  \href{http://www.dgp.toronto.edu/~gf/papers/PhD%20-%20Graspable%20UIs/Thesis.gf.html}{PhD
  thesis},
  \href{http://www.dgp.toronto.edu/~gf/papers/PhD%20-%20Graspable%20UIs/Thesis.gf.html}{University
  of Toronto},
  \href{http://www.dgp.toronto.edu/~gf/papers/PhD%20-%20Graspable%20UIs/Thesis.gf.html}{Canada}.

\bibitem[Fitzmaurice et~al., 1993]{fitzmaurice93}
\href{http://dx.doi.org/10.1145/159161.159160}{Fitzmaurice, G.~W., Zhai, S.,
  and Chignell, M.~H.} (1993).
\newblock \href{http://dx.doi.org/10.1145/159161.159160}{Virtual reality for
  palmtop computers}.
\newblock \href{http://dx.doi.org/10.1145/159161.159160}{{\em {ACM}
  Transactions on Information Systems}},
  \href{http://dx.doi.org/10.1145/159161.159160}{11(3):197--218}.
  \href{http://dx.doi.org/10.1145/159161.159160}
{doi: {{%
10\hspace{.1pt}\discretionary{.}{%
}{.}\hspace{.4pt}1145\discretionary{/}{%
}{/}159161\hspace{.1pt}\discretionary{.}{%
}{.}\hspace{.4pt}159160}}}


\bibitem[Frees et~al., 2007]{frees07}
\href{http://dx.doi.org/10.1145/1229855.1229857}{Frees, S., Kessler, G.~D., and
  Kay, E.} (2007).
\newblock \href{http://dx.doi.org/10.1145/1229855.1229857}{{PRISM} interaction
  for enhancing control in immersive virtual environments}.
\newblock \href{http://dx.doi.org/10.1145/1229855.1229857}{{\em ACM Trans.
  Comput.-Hum. Interact.}},
  \href{http://dx.doi.org/10.1145/1229855.1229857}{14(1)}.
  \href{http://dx.doi.org/10.1145/1229855.1229857}
{doi: {{%
10\hspace{.1pt}\discretionary{.}{%
}{.}\hspace{.4pt}1145\discretionary{/}{%
}{/}1229855\hspace{.1pt}\discretionary{.}{%
}{.}\hspace{.4pt}1229857}}}


\bibitem[Fritz et~al., 2012]{fritz12}
\href{http://dx.doi.org/10.1037/a0024338}{Fritz, C.~O., Morris, P.~E., and
  Richler, J.~J.} (2012).
\newblock \href{http://dx.doi.org/10.1037/a0024338}{Effect size estimates:
  Current use, calculations, and interpretation}.
\newblock \href{http://dx.doi.org/10.1037/a0024338}{{\em Journal of
  Experimental Psychology: General}},
  \href{http://dx.doi.org/10.1037/a0024338}{141(1):2--18}.
  \href{http://dx.doi.org/10.1037/a0024338}
{doi: {{%
10\hspace{.1pt}\discretionary{.}{%
}{.}\hspace{.4pt}1037\discretionary{/}{%
}{/}a0024338}}}


\bibitem[Fr{\"o}hlich et~al., 2000]{frohlich00}
\href{http://dx.doi.org/10.1109/38.851743}{Fr{\"o}hlich, B., Plate, J., Wind,
  J., Wesche, G., and G{\"o}bel, M.} (2000).
\newblock \href{http://dx.doi.org/10.1109/38.851743}{{C}ubic-{M}ouse-based
  interaction in virtual environments}.
\newblock \href{http://dx.doi.org/10.1109/38.851743}{{\em {IEEE} Computer
  Graphics and Applications}},
  \href{http://dx.doi.org/10.1109/38.851743}{20(4):12--15}.
  \href{http://dx.doi.org/10.1109/38.851743}
{doi: {{%
10\hspace{.1pt}\discretionary{.}{%
}{.}\hspace{.4pt}1109\discretionary{/}{%
}{/}38\hspace{.1pt}\discretionary{.}{%
}{.}\hspace{.4pt}851743}}}


\bibitem[Grossman and Balakrishnan, 2006]{grossman06}
\href{http://dx.doi.org/10.1145/1166253.1166257}{Grossman, T. and Balakrishnan,
  R.} (2006).
\newblock \href{http://dx.doi.org/10.1145/1166253.1166257}{The design and
  evaluation of selection techniques for {3D} volumetric displays}.
\newblock \href{http://dx.doi.org/10.1145/1166253.1166257}{In {\em Proc.\
  UIST}}, \href{http://dx.doi.org/10.1145/1166253.1166257}{pages 3--12}.
  \href{http://dx.doi.org/10.1145/1166253.1166257}{ACM},
  \href{http://dx.doi.org/10.1145/1166253.1166257}{New York}.
  \href{http://dx.doi.org/10.1145/1166253.1166257}
{doi: {{%
10\hspace{.1pt}\discretionary{.}{%
}{.}\hspace{.4pt}1145\discretionary{/}{%
}{/}1166253\hspace{.1pt}\discretionary{.}{%
}{.}\hspace{.4pt}1166257}}}


\bibitem[Ha and Woo, 2011]{ha11}
\href{http://dx.doi.org//10.1109/ISUVR.2011.14}{Ha, T. and Woo, W.} (2011).
\newblock \href{http://dx.doi.org//10.1109/ISUVR.2011.14}{{ARWand}: Phone-based
  {3D} object manipulation in augmented reality environment}.
\newblock \href{http://dx.doi.org//10.1109/ISUVR.2011.14}{In {\em Proc.\
  ISUVR}}, \href{http://dx.doi.org//10.1109/ISUVR.2011.14}{pages 44--47}.
  \href{http://dx.doi.org//10.1109/ISUVR.2011.14}{IEEE Computer Society},
  \href{http://dx.doi.org//10.1109/ISUVR.2011.14}{Los Alamitos}.
  \href{http://dx.doi.org//10.1109/ISUVR.2011.14}
{doi: {{%
\discretionary{/}{%
}{/}10\hspace{.1pt}\discretionary{.}{%
}{.}\hspace{.4pt}1109\discretionary{/}{%
}{/}ISUVR\hspace{.1pt}\discretionary{.}{%
}{.}\hspace{.4pt}2011\hspace{.1pt}\discretionary{.}{%
}{.}\hspace{.4pt}14}}}


\bibitem[Ha and Woo, 2013]{ha13}
\href{http://dx.doi.org/10.1109/3DUI.2013.6550216}{Ha, T. and Woo, W.} (2013).
\newblock \href{http://dx.doi.org/10.1109/3DUI.2013.6550216}{Poster: A pilot
  study on stepwise 6-{DoF} manipulation of virtual {3D} objects using
  smartphone in wearable augmented reality environment}.
\newblock \href{http://dx.doi.org/10.1109/3DUI.2013.6550216}{In {\em Proc.\
  3DUI}}, \href{http://dx.doi.org/10.1109/3DUI.2013.6550216}{pages 137--138}.
  \href{http://dx.doi.org/10.1109/3DUI.2013.6550216}{IEEE Computer Society},
  \href{http://dx.doi.org/10.1109/3DUI.2013.6550216}{Los Alamitos}.
  \href{http://dx.doi.org/10.1109/3DUI.2013.6550216}
{doi: {{%
10\hspace{.1pt}\discretionary{.}{%
}{.}\hspace{.4pt}1109\discretionary{/}{%
}{/}3DUI\hspace{.1pt}\discretionary{.}{%
}{.}\hspace{.4pt}2013\hspace{.1pt}\discretionary{.}{%
}{.}\hspace{.4pt}6550216}}}


\bibitem[Hansen et~al., 2006]{eriksson06}
\href{http://dx.doi.org/10.1007/1-84628-249-7_23}{Hansen, T.~R., Eriksson, E.,
  and Lykke-Olesen, A.} (2006).
\newblock \href{http://dx.doi.org/10.1007/1-84628-249-7_23}{Mixed interaction
  space -- {E}xpanding the interaction space with mobile devices}.
\newblock \href{http://dx.doi.org/10.1007/1-84628-249-7_23}{In {\em People and
  Computers {XIX} — The Bigger Picture}},
  \href{http://dx.doi.org/10.1007/1-84628-249-7_23}{pages 365--380}.
  \href{http://dx.doi.org/10.1007/1-84628-249-7_23}{Springer},
  \href{http://dx.doi.org/10.1007/1-84628-249-7_23}{London}.
  \href{http://dx.doi.org/10.1007/1-84628-249-7_23}
{doi: {{%
10\hspace{.1pt}\discretionary{.}{%
}{.}\hspace{.4pt}1007\discretionary{/}{%
}{/}1\discretionary{%
}{-}{-}84628\discretionary{%
}{-}{-}249\discretionary{%
}{-}{-}7\_23}}}


\bibitem[Harrison et~al., 1998]{harrison98}
\href{http://dx.doi.org/10.1145/274644.274647}{Harrison, B.~L., Fishkin, K.~P.,
  Gujar, A., Mochon, C., and Want, R.} (1998).
\newblock \href{http://dx.doi.org/10.1145/274644.274647}{Squeeze me, hold me,
  tilt me! {A}n exploration of manipulative user interfaces}.
\newblock \href{http://dx.doi.org/10.1145/274644.274647}{In {\em Proc.\ CHI}},
  \href{http://dx.doi.org/10.1145/274644.274647}{pages 17--24}.
  \href{http://dx.doi.org/10.1145/274644.274647}{ACM},
  \href{http://dx.doi.org/10.1145/274644.274647}{New York}.
  \href{http://dx.doi.org/10.1145/274644.274647}
{doi: {{%
10\hspace{.1pt}\discretionary{.}{%
}{.}\hspace{.4pt}1145\discretionary{/}{%
}{/}274644\hspace{.1pt}\discretionary{.}{%
}{.}\hspace{.4pt}274647}}}


\bibitem[Henrysson et~al., 2005]{henrysson05}
\href{http://dx.doi.org/10.1145/1152399.1152430}{Henrysson, A., Billinghurst,
  M., and Ollila, M.} (2005).
\newblock \href{http://dx.doi.org/10.1145/1152399.1152430}{Virtual object
  manipulation using a mobile phone}.
\newblock \href{http://dx.doi.org/10.1145/1152399.1152430}{In {\em Proc.\
  ICAT}}, \href{http://dx.doi.org/10.1145/1152399.1152430}{pages 164--171}.
  \href{http://dx.doi.org/10.1145/1152399.1152430}{ACM},
  \href{http://dx.doi.org/10.1145/1152399.1152430}{New York}.
  \href{http://dx.doi.org/10.1145/1152399.1152430}
{doi: {{%
10\hspace{.1pt}\discretionary{.}{%
}{.}\hspace{.4pt}1145\discretionary{/}{%
}{/}1152399\hspace{.1pt}\discretionary{.}{%
}{.}\hspace{.4pt}1152430}}}


\bibitem[Hinckley et~al., 1994]{hinckley94}
\href{http://dx.doi.org/10.1145/191666.191821}{Hinckley, K., Pausch, R., Goble,
  J.~C., and Kassell, N.~F.} (1994).
\newblock \href{http://dx.doi.org/10.1145/191666.191821}{Passive real-world
  interface props for neurosurgical visualization}.
\newblock \href{http://dx.doi.org/10.1145/191666.191821}{In {\em Proc.\ CHI}},
  \href{http://dx.doi.org/10.1145/191666.191821}{pages 452--458}.
  \href{http://dx.doi.org/10.1145/191666.191821}{ACM},
  \href{http://dx.doi.org/10.1145/191666.191821}{New York}.
  \href{http://dx.doi.org/10.1145/191666.191821}
{doi: {{%
10\hspace{.1pt}\discretionary{.}{%
}{.}\hspace{.4pt}1145\discretionary{/}{%
}{/}191666\hspace{.1pt}\discretionary{.}{%
}{.}\hspace{.4pt}191821}}}


\bibitem[Hinckley et~al., 2000]{hinckley00}
\href{http://dx.doi.org/10.1145/354401.354417}{Hinckley, K., Pierce, J.,
  Sinclair, M., and Horvitz, E.} (2000).
\newblock \href{http://dx.doi.org/10.1145/354401.354417}{Sensing techniques for
  mobile interaction}.
\newblock \href{http://dx.doi.org/10.1145/354401.354417}{In {\em Proc.\ UIST}},
  \href{http://dx.doi.org/10.1145/354401.354417}{pages 91--100}.
  \href{http://dx.doi.org/10.1145/354401.354417}{ACM},
  \href{http://dx.doi.org/10.1145/354401.354417}{New York}.
  \href{http://dx.doi.org/10.1145/354401.354417}
{doi: {{%
10\hspace{.1pt}\discretionary{.}{%
}{.}\hspace{.4pt}1145\discretionary{/}{%
}{/}354401\hspace{.1pt}\discretionary{.}{%
}{.}\hspace{.4pt}354417}}}


\bibitem[Hinckley and Song, 2011]{hinckley11}
\href{http://dx.doi.org/10.1145/1978942.1979059}{Hinckley, K. and Song, H.}
  (2011).
\newblock \href{http://dx.doi.org/10.1145/1978942.1979059}{Sensor synaesthesia:
  Touch in motion, and motion in touch}.
\newblock \href{http://dx.doi.org/10.1145/1978942.1979059}{In {\em Proc.\
  CHI}}, \href{http://dx.doi.org/10.1145/1978942.1979059}{pages 801--810}.
  \href{http://dx.doi.org/10.1145/1978942.1979059}{ACM},
  \href{http://dx.doi.org/10.1145/1978942.1979059}{New York}.
  \href{http://dx.doi.org/10.1145/1978942.1979059}
{doi: {{%
10\hspace{.1pt}\discretionary{.}{%
}{.}\hspace{.4pt}1145\discretionary{/}{%
}{/}1978942\hspace{.1pt}\discretionary{.}{%
}{.}\hspace{.4pt}1979059}}}


\bibitem[Ishii, 2008]{ishii08}
\href{http://dx.doi.org/10.1145/1349026.1349034}{Ishii, H.} (2008).
\newblock \href{http://dx.doi.org/10.1145/1349026.1349034}{The tangible user
  interface and its evolution}.
\newblock \href{http://dx.doi.org/10.1145/1349026.1349034}{{\em Communications
  of the ACM}}, \href{http://dx.doi.org/10.1145/1349026.1349034}{51(6):32--36}.
  \href{http://dx.doi.org/10.1145/1349026.1349034}
{doi: {{%
10\hspace{.1pt}\discretionary{.}{%
}{.}\hspace{.4pt}1145\discretionary{/}{%
}{/}1349026\hspace{.1pt}\discretionary{.}{%
}{.}\hspace{.4pt}1349034}}}


\bibitem[Issartel et~al., 2016]{issartel2016pref}
\href{http://dx.doi.org/10.1145/2983310.2985750}{Issartel, P., Besan{\c c}on,
  L., Gu{\'e}niat, F., Isenberg, T., and Ammi, M.} (2016).
\newblock \href{http://dx.doi.org/10.1145/2983310.2985750}{Preference between
  allocentric and egocentric {3D} manipulation in a locally coupled
  configuration}.
\newblock \href{http://dx.doi.org/10.1145/2983310.2985750}{In {\em Proceedings
  of the 2016 Symposium on Spatial User Interaction}},
  \href{http://dx.doi.org/10.1145/2983310.2985750}{{SUI} '16},
  \href{http://dx.doi.org/10.1145/2983310.2985750}{pages 79--88}.
  \href{http://dx.doi.org/10.1145/2983310.2985750}{{ACM}}.
  \href{http://dx.doi.org/10.1145/2983310.2985750}
{doi: {{%
10\hspace{.1pt}\discretionary{.}{%
}{.}\hspace{.4pt}1145\discretionary{/}{%
}{/}2983310\hspace{.1pt}\discretionary{.}{%
}{.}\hspace{.4pt}2985750}}}


\bibitem[Issartel et~al., 2014]{issartel14}
\href{http://dx.doi.org/10.1145/2671015.2671130}{Issartel, P., Guéniat, F.,
  and Ammi, M.} (2014).
\newblock \href{http://dx.doi.org/10.1145/2671015.2671130}{A portable interface
  for tangible exploration of volumetric data}.
\newblock \href{http://dx.doi.org/10.1145/2671015.2671130}{In {\em Proc.\
  VRST}}, \href{http://dx.doi.org/10.1145/2671015.2671130}{pages 209--210}.
  \href{http://dx.doi.org/10.1145/2671015.2671130}{ACM},
  \href{http://dx.doi.org/10.1145/2671015.2671130}{New York}.
  \href{http://dx.doi.org/10.1145/2671015.2671130}
{doi: {{%
10\hspace{.1pt}\discretionary{.}{%
}{.}\hspace{.4pt}1145\discretionary{/}{%
}{/}2671015\hspace{.1pt}\discretionary{.}{%
}{.}\hspace{.4pt}2671130}}}


\bibitem[Jellinek and Card, 1990]{jellinek90}
\href{http://dx.doi.org/10.1145/97243.97276}{Jellinek, H.~D. and Card, S.~K.}
  (1990).
\newblock \href{http://dx.doi.org/10.1145/97243.97276}{Powermice and user
  performance}.
\newblock \href{http://dx.doi.org/10.1145/97243.97276}{In {\em Proc.\ CHI}},
  \href{http://dx.doi.org/10.1145/97243.97276}{pages 213--220}.
  \href{http://dx.doi.org/10.1145/97243.97276}{ACM},
  \href{http://dx.doi.org/10.1145/97243.97276}{New York}.
  \href{http://dx.doi.org/10.1145/97243.97276}
{doi: {{%
10\hspace{.1pt}\discretionary{.}{%
}{.}\hspace{.4pt}1145\discretionary{/}{%
}{/}97243\hspace{.1pt}\discretionary{.}{%
}{.}\hspace{.4pt}97276}}}


\bibitem[Joshi et~al., 2012]{joshi12}
\href{http://dx.doi.org/10.1145/2207676.2208375}{Joshi, N., Kar, A., and Cohen,
  M.} (2012).
\newblock \href{http://dx.doi.org/10.1145/2207676.2208375}{Looking at you:
  Fused gyro and face tracking for viewing large imagery on mobile devices}.
\newblock \href{http://dx.doi.org/10.1145/2207676.2208375}{In {\em Proc.\
  CHI}}, \href{http://dx.doi.org/10.1145/2207676.2208375}{pages 2211--2220}.
  \href{http://dx.doi.org/10.1145/2207676.2208375}{ACM},
  \href{http://dx.doi.org/10.1145/2207676.2208375}{New York}.
  \href{http://dx.doi.org/10.1145/2207676.2208375}
{doi: {{%
10\hspace{.1pt}\discretionary{.}{%
}{.}\hspace{.4pt}1145\discretionary{/}{%
}{/}2207676\hspace{.1pt}\discretionary{.}{%
}{.}\hspace{.4pt}2208375}}}


\bibitem[Kaminaka and Egli, 1985]{kaminaka85}
\href{http://dx.doi.org/10.1016/0003-6870(85)90008-0}{Kaminaka, M.~S. and Egli,
  E.~A.} (1985).
\newblock \href{http://dx.doi.org/10.1016/0003-6870(85)90008-0}{Lever controls
  on specialised farm equipment: Some control/response stereotypes}.
\newblock \href{http://dx.doi.org/10.1016/0003-6870(85)90008-0}{{\em Applied
  Ergonomics}},
  \href{http://dx.doi.org/10.1016/0003-6870(85)90008-0}{16(3):193--199}.
  \href{http://dx.doi.org/10.1016/0003-6870(85)90008-0}
{doi: {{%
10\hspace{.1pt}\discretionary{.}{%
}{.}\hspace{.4pt}1016\discretionary{/}{%
}{/}0003\discretionary{%
}{-}{-}6870\discretionary{%
}{(}{(}85\discretionary{)}{%
}{)}90008\discretionary{%
}{-}{-}0}}}


\bibitem[Katzakis and Hori, 2009]{katzakis09}
\href{http://dx.doi.org/10.1109/DASC.2009.76}{Katzakis, N. and Hori, M.}
  (2009).
\newblock \href{http://dx.doi.org/10.1109/DASC.2009.76}{Mobile phones as
  3-{DOF} controllers: A comparative study}.
\newblock \href{http://dx.doi.org/10.1109/DASC.2009.76}{In {\em Proc.\ DASC}},
  \href{http://dx.doi.org/10.1109/DASC.2009.76}{pages 345--349}.
  \href{http://dx.doi.org/10.1109/DASC.2009.76}{IEEE Computer Society},
  \href{http://dx.doi.org/10.1109/DASC.2009.76}{Los Alamitos}.
  \href{http://dx.doi.org/10.1109/DASC.2009.76}
{doi: {{%
10\hspace{.1pt}\discretionary{.}{%
}{.}\hspace{.4pt}1109\discretionary{/}{%
}{/}DASC\hspace{.1pt}\discretionary{.}{%
}{.}\hspace{.4pt}2009\hspace{.1pt}\discretionary{.}{%
}{.}\hspace{.4pt}76}}}


\bibitem[Klatzky, 1998]{klatzky98}
\href{http://dx.doi.org/10.1007/3-540-69342-4_1}{Klatzky, R.~L.} (1998).
\newblock \href{http://dx.doi.org/10.1007/3-540-69342-4_1}{Allocentric and
  egocentric spatial representations: Definitions, distinctions, and
  interconnections}.
\newblock \href{http://dx.doi.org/10.1007/3-540-69342-4_1}{In {\em Spatial
  Cognition}}, \href{http://dx.doi.org/10.1007/3-540-69342-4_1}{pages 1--17}.
  \href{http://dx.doi.org/10.1007/3-540-69342-4_1}{Springer},
  \href{http://dx.doi.org/10.1007/3-540-69342-4_1}{Berlin\discretionary{/}{}{/}Heidelberg}.
  \href{http://dx.doi.org/10.1007/3-540-69342-4_1}
{doi: {{%
10\hspace{.1pt}\discretionary{.}{%
}{.}\hspace{.4pt}1007\discretionary{/}{%
}{/}3\discretionary{%
}{-}{-}540\discretionary{%
}{-}{-}69342\discretionary{%
}{-}{-}4\_1}}}


\bibitem[Kopper, 2011]{kopper11}
Kopper, R. A.~P. (2011).
\newblock {\em Understanding and Improving Distal Pointing Interaction}.
\newblock PhD thesis, Virginia Tech.

\bibitem[Kratz and Rohs, 2010]{kratz10}
\href{http://dx.doi.org/10.1109/3DUI.2010.5444712}{Kratz, S. and Rohs, M.}
  (2010).
\newblock \href{http://dx.doi.org/10.1109/3DUI.2010.5444712}{Extending the
  virtual trackball metaphor to rear touch input}.
\newblock \href{http://dx.doi.org/10.1109/3DUI.2010.5444712}{In {\em Proc.\
  3DUI}}, \href{http://dx.doi.org/10.1109/3DUI.2010.5444712}{pages 111--114}.
  \href{http://dx.doi.org/10.1109/3DUI.2010.5444712}{IEEE Computer Society},
  \href{http://dx.doi.org/10.1109/3DUI.2010.5444712}{Los Alamitos}.
  \href{http://dx.doi.org/10.1109/3DUI.2010.5444712}
{doi: {{%
10\hspace{.1pt}\discretionary{.}{%
}{.}\hspace{.4pt}1109\discretionary{/}{%
}{/}3DUI\hspace{.1pt}\discretionary{.}{%
}{.}\hspace{.4pt}2010\hspace{.1pt}\discretionary{.}{%
}{.}\hspace{.4pt}5444712}}}


\bibitem[LaViola and Katzourin, 2007]{laviola07}
\href{http://dx.doi.org/10.1109/3DUI.2007.340774}{LaViola, J.~J. and Katzourin,
  M.} (2007).
\newblock \href{http://dx.doi.org/10.1109/3DUI.2007.340774}{An exploration of
  non-isomorphic {3D} rotation in surround screen virtual environments}.
\newblock \href{http://dx.doi.org/10.1109/3DUI.2007.340774}{In {\em Proc.\
  3DUI}}, \href{http://dx.doi.org/10.1109/3DUI.2007.340774}{pages 49--54}.
  \href{http://dx.doi.org/10.1109/3DUI.2007.340774}{IEEE Computer Society},
  \href{http://dx.doi.org/10.1109/3DUI.2007.340774}{Los Alamitos}.
  \href{http://dx.doi.org/10.1109/3DUI.2007.340774}
{doi: {{%
10\hspace{.1pt}\discretionary{.}{%
}{.}\hspace{.4pt}1109\discretionary{/}{%
}{/}3DUI\hspace{.1pt}\discretionary{.}{%
}{.}\hspace{.4pt}2007\hspace{.1pt}\discretionary{.}{%
}{.}\hspace{.4pt}340774}}}


\bibitem[Liang, 2013]{liang13}
\href{http://dx.doi.org/10.1145/2508468.2508470}{Liang, R.-H.} (2013).
\newblock \href{http://dx.doi.org/10.1145/2508468.2508470}{Augmenting the input
  space of portable displays using add-on {H}all-sensor grid}.
\newblock \href{http://dx.doi.org/10.1145/2508468.2508470}{In {\em UIST Adjunct
  Proc.}}, \href{http://dx.doi.org/10.1145/2508468.2508470}{pages 33--36}.
  \href{http://dx.doi.org/10.1145/2508468.2508470}{ACM},
  \href{http://dx.doi.org/10.1145/2508468.2508470}{New York}.
  \href{http://dx.doi.org/10.1145/2508468.2508470}
{doi: {{%
10\hspace{.1pt}\discretionary{.}{%
}{.}\hspace{.4pt}1145\discretionary{/}{%
}{/}2508468\hspace{.1pt}\discretionary{.}{%
}{.}\hspace{.4pt}2508470}}}


\bibitem[L{\'o}pez et~al., 2016]{lopez15}
\href{http://dx.doi.org/10.1109/TVCG.2015.2440233}{L{\'o}pez, D., Oehlberg, L.,
  Doger, C., and Isenberg, T.} (2016).
\newblock \href{http://dx.doi.org/10.1109/TVCG.2015.2440233}{Towards an
  understanding of mobile touch navigation in a stereoscopic viewing
  environment for {3D} data exploration}.
\newblock \href{http://dx.doi.org/10.1109/TVCG.2015.2440233}{{\em IEEE
  Transactions on Visualization and Computer Graphics}},
  \href{http://dx.doi.org/10.1109/TVCG.2015.2440233}{22}.
\newblock \href{http://dx.doi.org/10.1109/TVCG.2015.2440233}{To appear}.
  \href{http://dx.doi.org/10.1109/TVCG.2015.2440233}
{doi: {{%
10\hspace{.1pt}\discretionary{.}{%
}{.}\hspace{.4pt}1109\discretionary{/}{%
}{/}TVCG\hspace{.1pt}\discretionary{.}{%
}{.}\hspace{.4pt}2015\hspace{.1pt}\discretionary{.}{%
}{.}\hspace{.4pt}2440233}}}


\bibitem[Lopez-Gulliver et~al., 2009]{lopez-gulliver09}
\href{http://dx.doi.org/10.1145/1597956.1597967}{Lopez-Gulliver, R., Yoshida,
  S., Yano, S., and Inoue, N.} (2009).
\newblock \href{http://dx.doi.org/10.1145/1597956.1597967}{{gCubik}: Real-time
  integral image rendering for a cubic {3D} display}.
\newblock \href{http://dx.doi.org/10.1145/1597956.1597967}{In {\em SIGGRAPH
  Emerging Technologies}},
  \href{http://dx.doi.org/10.1145/1597956.1597967}{page 11:1}.
  \href{http://dx.doi.org/10.1145/1597956.1597967}{ACM},
  \href{http://dx.doi.org/10.1145/1597956.1597967}{New York}.
  \href{http://dx.doi.org/10.1145/1597956.1597967}
{doi: {{%
10\hspace{.1pt}\discretionary{.}{%
}{.}\hspace{.4pt}1145\discretionary{/}{%
}{/}1597956\hspace{.1pt}\discretionary{.}{%
}{.}\hspace{.4pt}1597967}}}


\bibitem[Lyons et~al., 2012]{lyons12}
\href{http://dx.doi.org/10.1145/2380116.2380134}{Lyons, K., Nguyen, D.,
  Ashbrook, D., and White, S.} (2012).
\newblock \href{http://dx.doi.org/10.1145/2380116.2380134}{{F}acet: A
  multi-segment wrist worn system}.
\newblock \href{http://dx.doi.org/10.1145/2380116.2380134}{In {\em Proc.\
  UIST}}, \href{http://dx.doi.org/10.1145/2380116.2380134}{pages 123--130}.
  \href{http://dx.doi.org/10.1145/2380116.2380134}{ACM},
  \href{http://dx.doi.org/10.1145/2380116.2380134}{New York}.
  \href{http://dx.doi.org/10.1145/2380116.2380134}
{doi: {{%
10\hspace{.1pt}\discretionary{.}{%
}{.}\hspace{.4pt}1145\discretionary{/}{%
}{/}2380116\hspace{.1pt}\discretionary{.}{%
}{.}\hspace{.4pt}2380134}}}


\bibitem[Marzo et~al., 2014]{marzo14}
\href{http://dx.doi.org/10.1145/2659766.2659775}{Marzo, A., Bossavit, B., and
  Hachet, M.} (2014).
\newblock \href{http://dx.doi.org/10.1145/2659766.2659775}{Combining
  multi-touch input and device movement for {3D} manipulations in mobile
  augmented reality environments}.
\newblock \href{http://dx.doi.org/10.1145/2659766.2659775}{In {\em Proc.\
  SUI}}, \href{http://dx.doi.org/10.1145/2659766.2659775}{pages 13--16}.
  \href{http://dx.doi.org/10.1145/2659766.2659775}{ACM},
  \href{http://dx.doi.org/10.1145/2659766.2659775}{New York}.
  \href{http://dx.doi.org/10.1145/2659766.2659775}
{doi: {{%
10\hspace{.1pt}\discretionary{.}{%
}{.}\hspace{.4pt}1145\discretionary{/}{%
}{/}2659766\hspace{.1pt}\discretionary{.}{%
}{.}\hspace{.4pt}2659775}}}


\bibitem[Massimino et~al., 1989]{massimino89}
\href{http://dx.doi.org/10.1109/ICSMC.1989.71346}{Massimino, M.~J., Sheridan,
  T.~B., and Roseborough, J.~B.} (1989).
\newblock \href{http://dx.doi.org/10.1109/ICSMC.1989.71346}{One handed tracking
  in six degrees of freedom}.
\newblock \href{http://dx.doi.org/10.1109/ICSMC.1989.71346}{In {\em Proc.\
  Systems, Man and Cybernetics}},
  \href{http://dx.doi.org/10.1109/ICSMC.1989.71346}{volume~2},
  \href{http://dx.doi.org/10.1109/ICSMC.1989.71346}{pages 498--503}.
  \href{http://dx.doi.org/10.1109/ICSMC.1989.71346}{IEEE},
  \href{http://dx.doi.org/10.1109/ICSMC.1989.71346}{New York}.
  \href{http://dx.doi.org/10.1109/ICSMC.1989.71346}
{doi: {{%
10\hspace{.1pt}\discretionary{.}{%
}{.}\hspace{.4pt}1109\discretionary{/}{%
}{/}ICSMC\hspace{.1pt}\discretionary{.}{%
}{.}\hspace{.4pt}1989\hspace{.1pt}\discretionary{.}{%
}{.}\hspace{.4pt}71346}}}


\bibitem[Mossel et~al., 2013]{mossel13}
\href{http://dx.doi.org/10.1145/2466816.2466829}{Mossel, A., Venditti, B., and
  Kaufmann, H.} (2013).
\newblock \href{http://dx.doi.org/10.1145/2466816.2466829}{{3DTouch} and
  {HOMER-S}: Intuitive manipulation techniques for one-handed handheld
  augmented reality}.
\newblock \href{http://dx.doi.org/10.1145/2466816.2466829}{In {\em Proc.\
  VRIC}}, \href{http://dx.doi.org/10.1145/2466816.2466829}{pages 12:1--12:10}.
  \href{http://dx.doi.org/10.1145/2466816.2466829}{ACM},
  \href{http://dx.doi.org/10.1145/2466816.2466829}{New York}.
  \href{http://dx.doi.org/10.1145/2466816.2466829}
{doi: {{%
10\hspace{.1pt}\discretionary{.}{%
}{.}\hspace{.4pt}1145\discretionary{/}{%
}{/}2466816\hspace{.1pt}\discretionary{.}{%
}{.}\hspace{.4pt}2466829}}}


\bibitem[Neale et~al., 2013]{neale13}
\href{http://dx.doi.org/10.1007/978-3-642-40483-2_7}{Neale, S., Chinthammit,
  W., Lueg, C., and Nixon, P.} (2013).
\newblock \href{http://dx.doi.org/10.1007/978-3-642-40483-2_7}{{RelicPad}: A
  hands-on, mobile approach to collaborative exploration of virtual museum
  artifacts}.
\newblock \href{http://dx.doi.org/10.1007/978-3-642-40483-2_7}{In {\em Proc.\
  INTERACT}}, \href{http://dx.doi.org/10.1007/978-3-642-40483-2_7}{pages
  86--103}. \href{http://dx.doi.org/10.1007/978-3-642-40483-2_7}{Springer},
  \href{http://dx.doi.org/10.1007/978-3-642-40483-2_7}{Berlin\discretionary{/}{}{/}Heidelberg}.
  \href{http://dx.doi.org/10.1007/978-3-642-40483-2_7}
{doi: {{%
10\hspace{.1pt}\discretionary{.}{%
}{.}\hspace{.4pt}1007\discretionary{/}{%
}{/}978\discretionary{%
}{-}{-}3\discretionary{%
}{-}{-}642\discretionary{%
}{-}{-}40483\discretionary{%
}{-}{-}2\_7}}}


\bibitem[Noma et~al., 1996]{noma96}
\href{http://dx.doi.org/10.1145/238386.238454}{Noma, H., Miyasato, T., and
  Kishino, F.} (1996).
\newblock \href{http://dx.doi.org/10.1145/238386.238454}{A palmtop display for
  dextrous manipulation with haptic sensation}.
\newblock \href{http://dx.doi.org/10.1145/238386.238454}{In {\em Proc.\ CHI}},
  \href{http://dx.doi.org/10.1145/238386.238454}{pages 126--133}.
  \href{http://dx.doi.org/10.1145/238386.238454}{ACM},
  \href{http://dx.doi.org/10.1145/238386.238454}{New York}.
  \href{http://dx.doi.org/10.1145/238386.238454}
{doi: {{%
10\hspace{.1pt}\discretionary{.}{%
}{.}\hspace{.4pt}1145\discretionary{/}{%
}{/}238386\hspace{.1pt}\discretionary{.}{%
}{.}\hspace{.4pt}238454}}}


\bibitem[Oakley and O'Modhrain, 2005]{oakley05}
\href{http://dx.doi.org/10.1109/WHC.2005.138}{Oakley, I. and O'Modhrain, S.}
  (2005).
\newblock \href{http://dx.doi.org/10.1109/WHC.2005.138}{Tilt to scroll:
  Evaluating a motion based vibrotactile mobile interface}.
\newblock \href{http://dx.doi.org/10.1109/WHC.2005.138}{In {\em Proc.\ World
  Haptics}}, \href{http://dx.doi.org/10.1109/WHC.2005.138}{pages 40--49}.
  \href{http://dx.doi.org/10.1109/WHC.2005.138}{IEEE Computer Society},
  \href{http://dx.doi.org/10.1109/WHC.2005.138}{Los Alamitos}.
  \href{http://dx.doi.org/10.1109/WHC.2005.138}
{doi: {{%
10\hspace{.1pt}\discretionary{.}{%
}{.}\hspace{.4pt}1109\discretionary{/}{%
}{/}WHC\hspace{.1pt}\discretionary{.}{%
}{.}\hspace{.4pt}2005\hspace{.1pt}\discretionary{.}{%
}{.}\hspace{.4pt}138}}}


\bibitem[Poupyrev et~al., 1996]{poupyrev96}
\href{http://dx.doi.org/10.1145/237091.237102}{Poupyrev, I., Billinghurst, M.,
  Weghorst, S., and Ichikawa, T.} (1996).
\newblock \href{http://dx.doi.org/10.1145/237091.237102}{The {Go}-{Go}
  interaction technique: Non-linear mapping for direct manipulation in {VR}}.
\newblock \href{http://dx.doi.org/10.1145/237091.237102}{In {\em Proc.\ UIST}},
  \href{http://dx.doi.org/10.1145/237091.237102}{pages 79--80}.
  \href{http://dx.doi.org/10.1145/237091.237102}{ACM},
  \href{http://dx.doi.org/10.1145/237091.237102}{New York}.
  \href{http://dx.doi.org/10.1145/237091.237102}
{doi: {{%
10\hspace{.1pt}\discretionary{.}{%
}{.}\hspace{.4pt}1145\discretionary{/}{%
}{/}237091\hspace{.1pt}\discretionary{.}{%
}{.}\hspace{.4pt}237102}}}


\bibitem[Poupyrev et~al., 2000]{poupyrev00}
\href{http://dx.doi.org/10.1145/332040.332497}{Poupyrev, I., Weghorst, S., and
  Fels, S.} (2000).
\newblock \href{http://dx.doi.org/10.1145/332040.332497}{Non-isomorphic {3D}
  rotational techniques}.
\newblock \href{http://dx.doi.org/10.1145/332040.332497}{In {\em Proc.\ CHI}},
  \href{http://dx.doi.org/10.1145/332040.332497}{pages 540--547}.
  \href{http://dx.doi.org/10.1145/332040.332497}{ACM},
  \href{http://dx.doi.org/10.1145/332040.332497}{New York}.
  \href{http://dx.doi.org/10.1145/332040.332497}
{doi: {{%
10\hspace{.1pt}\discretionary{.}{%
}{.}\hspace{.4pt}1145\discretionary{/}{%
}{/}332040\hspace{.1pt}\discretionary{.}{%
}{.}\hspace{.4pt}332497}}}


\bibitem[Poupyrev et~al., 1999]{poupyrev99}
\href{http://dx.doi.org/10.1145/632716.632874}{Poupyrev, I., Weghorst, S.,
  Otsuka, T., and Ichikawa, T.} (1999).
\newblock \href{http://dx.doi.org/10.1145/632716.632874}{Amplifying spatial
  rotations in {3D} interfaces}.
\newblock \href{http://dx.doi.org/10.1145/632716.632874}{In {\em CHI Extended
  Abstracts}}, \href{http://dx.doi.org/10.1145/632716.632874}{pages 256--257}.
  \href{http://dx.doi.org/10.1145/632716.632874}{ACM},
  \href{http://dx.doi.org/10.1145/632716.632874}{New York}.
  \href{http://dx.doi.org/10.1145/632716.632874}
{doi: {{%
10\hspace{.1pt}\discretionary{.}{%
}{.}\hspace{.4pt}1145\discretionary{/}{%
}{/}632716\hspace{.1pt}\discretionary{.}{%
}{.}\hspace{.4pt}632874}}}


\bibitem[Rahman et~al., 2009]{rahman09}
\href{http://dx.doi.org/10.1145/1518701.1518997}{Rahman, M., Gustafson, S.,
  Irani, P., and Subramanian, S.} (2009).
\newblock \href{http://dx.doi.org/10.1145/1518701.1518997}{Tilt techniques:
  Investigating the dexterity of wrist-based input}.
\newblock \href{http://dx.doi.org/10.1145/1518701.1518997}{In {\em Proc.\
  CHI}}, \href{http://dx.doi.org/10.1145/1518701.1518997}{pages 1943--1952}.
  \href{http://dx.doi.org/10.1145/1518701.1518997}{ACM},
  \href{http://dx.doi.org/10.1145/1518701.1518997}{New York}.
  \href{http://dx.doi.org/10.1145/1518701.1518997}
{doi: {{%
10\hspace{.1pt}\discretionary{.}{%
}{.}\hspace{.4pt}1145\discretionary{/}{%
}{/}1518701\hspace{.1pt}\discretionary{.}{%
}{.}\hspace{.4pt}1518997}}}


\bibitem[Ramakers et~al., 2014]{ramakers14}
\href{http://dx.doi.org/10.1145/2556288.2557340}{Ramakers, R., Sch\"{o}ning,
  J., and Luyten, K.} (2014).
\newblock \href{http://dx.doi.org/10.1145/2556288.2557340}{{P}addle: Highly
  deformable mobile devices with physical controls}.
\newblock \href{http://dx.doi.org/10.1145/2556288.2557340}{In {\em Proc.\
  CHI}}, \href{http://dx.doi.org/10.1145/2556288.2557340}{pages 2569--2578}.
  \href{http://dx.doi.org/10.1145/2556288.2557340}{ACM},
  \href{http://dx.doi.org/10.1145/2556288.2557340}{New York}.
  \href{http://dx.doi.org/10.1145/2556288.2557340}
{doi: {{%
10\hspace{.1pt}\discretionary{.}{%
}{.}\hspace{.4pt}1145\discretionary{/}{%
}{/}2556288\hspace{.1pt}\discretionary{.}{%
}{.}\hspace{.4pt}2557340}}}


\bibitem[Rekimoto, 1996]{rekimoto96}
\href{http://dx.doi.org/10.1145/237091.237115}{Rekimoto, J.} (1996).
\newblock \href{http://dx.doi.org/10.1145/237091.237115}{Tilting operations for
  small screen interfaces}.
\newblock \href{http://dx.doi.org/10.1145/237091.237115}{In {\em Proc.\ UIST}},
  \href{http://dx.doi.org/10.1145/237091.237115}{pages 167--168}.
  \href{http://dx.doi.org/10.1145/237091.237115}{ACM},
  \href{http://dx.doi.org/10.1145/237091.237115}{New York}.
  \href{http://dx.doi.org/10.1145/237091.237115}
{doi: {{%
10\hspace{.1pt}\discretionary{.}{%
}{.}\hspace{.4pt}1145\discretionary{/}{%
}{/}237091\hspace{.1pt}\discretionary{.}{%
}{.}\hspace{.4pt}237115}}}


\bibitem[Shoemake, 1985]{shoemake85}
\href{http://dx.doi.org/10.1145/325165.325242}{Shoemake, K.} (1985).
\newblock \href{http://dx.doi.org/10.1145/325165.325242}{Animating rotation
  with quaternion curves}.
\newblock \href{http://dx.doi.org/10.1145/325165.325242}{{\em ACM SIGGRAPH
  Computer Graphics}},
  \href{http://dx.doi.org/10.1145/325165.325242}{19(3):245--254}.
  \href{http://dx.doi.org/10.1145/325165.325242}
{doi: {{%
10\hspace{.1pt}\discretionary{.}{%
}{.}\hspace{.4pt}1145\discretionary{/}{%
}{/}325165\hspace{.1pt}\discretionary{.}{%
}{.}\hspace{.4pt}325242}}}


\bibitem[Small and Ishii, 1997]{small97}
\href{http://dx.doi.org/10.1145/1120212.1120437}{Small, D. and Ishii, H.}
  (1997).
\newblock \href{http://dx.doi.org/10.1145/1120212.1120437}{Design of spatially
  aware graspable displays}.
\newblock \href{http://dx.doi.org/10.1145/1120212.1120437}{In {\em CHI Extended
  Abstracts}}, \href{http://dx.doi.org/10.1145/1120212.1120437}{pages
  367--368}. \href{http://dx.doi.org/10.1145/1120212.1120437}{ACM},
  \href{http://dx.doi.org/10.1145/1120212.1120437}{New York}.
  \href{http://dx.doi.org/10.1145/1120212.1120437}
{doi: {{%
10\hspace{.1pt}\discretionary{.}{%
}{.}\hspace{.4pt}1145\discretionary{/}{%
}{/}1120212\hspace{.1pt}\discretionary{.}{%
}{.}\hspace{.4pt}1120437}}}


\bibitem[Smith and Smith, 1987]{smith87}
Smith, T.~J. and Smith, K.~U. (1987).
\newblock Feedback-control mechanisms of human behavior.
\newblock In Salvendy, G., editor, {\em Handbook of Human Factors}, pages
  251--293. Wiley, New York.

\bibitem[Song et~al., 2011]{song11}
\href{http://dx.doi.org/10.1145/1978942.1979140}{Song, P., Goh, W.~B., Fu,
  C.-W., Meng, Q., and Heng, P.-A.} (2011).
\newblock \href{http://dx.doi.org/10.1145/1978942.1979140}{{WYSIWYF}: Exploring
  and annotating volume data with a tangible handheld device}.
\newblock \href{http://dx.doi.org/10.1145/1978942.1979140}{In {\em Proc.\
  CHI}}, \href{http://dx.doi.org/10.1145/1978942.1979140}{pages 1333--1342}.
  \href{http://dx.doi.org/10.1145/1978942.1979140}{ACM},
  \href{http://dx.doi.org/10.1145/1978942.1979140}{New York}.
  \href{http://dx.doi.org/10.1145/1978942.1979140}
{doi: {{%
10\hspace{.1pt}\discretionary{.}{%
}{.}\hspace{.4pt}1145\discretionary{/}{%
}{/}1978942\hspace{.1pt}\discretionary{.}{%
}{.}\hspace{.4pt}1979140}}}


\bibitem[Spindler et~al., 2012]{spindler12}
\href{http://dx.doi.org/10.1145/2396636.2396674}{Spindler, M., Büschel, W.,
  and Dachselt, R.} (2012).
\newblock \href{http://dx.doi.org/10.1145/2396636.2396674}{Use your head:
  Tangible windows for {3D} information spaces in a tabletop environment}.
\newblock \href{http://dx.doi.org/10.1145/2396636.2396674}{In {\em Proceedings
  of the 2012 {ACM} International Conference on Interactive Tabletops and
  Surfaces}}, \href{http://dx.doi.org/10.1145/2396636.2396674}{{ITS} '12},
  \href{http://dx.doi.org/10.1145/2396636.2396674}{page 245–254}.
  \href{http://dx.doi.org/10.1145/2396636.2396674}{{ACM}},
  \href{http://dx.doi.org/10.1145/2396636.2396674}{New York, {NY}, {USA}}.
  \href{http://dx.doi.org/10.1145/2396636.2396674}
{doi: {{%
10\hspace{.1pt}\discretionary{.}{%
}{.}\hspace{.4pt}1145\discretionary{/}{%
}{/}2396636\hspace{.1pt}\discretionary{.}{%
}{.}\hspace{.4pt}2396674}}}


\bibitem[Spindler et~al., 2014]{spindler14}
\href{http://dx.doi.org/10.1145/2556288.2557028}{Spindler, M., Schuessler, M.,
  Martsch, M., and Dachselt, R.} (2014).
\newblock \href{http://dx.doi.org/10.1145/2556288.2557028}{Pinch-drag-flick vs.
  spatial input: Rethinking zoom \& pan on mobile displays}.
\newblock \href{http://dx.doi.org/10.1145/2556288.2557028}{In {\em Proceedings
  of the {SIGCHI} Conference on Human Factors in Computing Systems}},
  \href{http://dx.doi.org/10.1145/2556288.2557028}{{CHI} '14},
  \href{http://dx.doi.org/10.1145/2556288.2557028}{page 1113–1122}.
  \href{http://dx.doi.org/10.1145/2556288.2557028}{{ACM}},
  \href{http://dx.doi.org/10.1145/2556288.2557028}{New York, {NY}, {USA}}.
  \href{http://dx.doi.org/10.1145/2556288.2557028}
{doi: {{%
10\hspace{.1pt}\discretionary{.}{%
}{.}\hspace{.4pt}1145\discretionary{/}{%
}{/}2556288\hspace{.1pt}\discretionary{.}{%
}{.}\hspace{.4pt}2557028}}}


\bibitem[Stavness et~al., 2010]{stavness10}
\href{http://dx.doi.org/10.1145/1753326.1753535}{Stavness, I., Lam, B., and
  Fels, S.} (2010).
\newblock \href{http://dx.doi.org/10.1145/1753326.1753535}{{pCubee}: A
  perspective-corrected handheld cubic display}.
\newblock \href{http://dx.doi.org/10.1145/1753326.1753535}{In {\em Proc.\
  CHI}}, \href{http://dx.doi.org/10.1145/1753326.1753535}{pages 1381--1390}.
  \href{http://dx.doi.org/10.1145/1753326.1753535}{ACM},
  \href{http://dx.doi.org/10.1145/1753326.1753535}{New York}.
  \href{http://dx.doi.org/10.1145/1753326.1753535}
{doi: {{%
10\hspace{.1pt}\discretionary{.}{%
}{.}\hspace{.4pt}1145\discretionary{/}{%
}{/}1753326\hspace{.1pt}\discretionary{.}{%
}{.}\hspace{.4pt}1753535}}}


\bibitem[Teather and {MacKenzie}, 2014]{teather14}
\href{http://dl.acm.org/citation.cfm?id=2619658}{Teather, R.~J. and
  {MacKenzie}, I.~S.} (2014).
\newblock \href{http://dl.acm.org/citation.cfm?id=2619658}{Position
  vs.~velocity control for tilt-based interaction}.
\newblock \href{http://dl.acm.org/citation.cfm?id=2619658}{In {\em Proc.\ GI}},
  \href{http://dl.acm.org/citation.cfm?id=2619658}{pages 51--58}.
  \href{http://dl.acm.org/citation.cfm?id=2619658}{CIPS},
  \href{http://dl.acm.org/citation.cfm?id=2619658}{Toronto}.

\bibitem[Tsandilas et~al., 2013]{tsandilas13}
\href{http://dx.doi.org/10.1007/978-3-642-40477-1_31}{Tsandilas, T., Dubois,
  E., and Raynal, M.} (2013).
\newblock \href{http://dx.doi.org/10.1007/978-3-642-40477-1_31}{Modeless
  pointing with low-precision wrist movements}.
\newblock \href{http://dx.doi.org/10.1007/978-3-642-40477-1_31}{In {\em Proc.\
  INTERACT}}, \href{http://dx.doi.org/10.1007/978-3-642-40477-1_31}{pages
  494--511}. \href{http://dx.doi.org/10.1007/978-3-642-40477-1_31}{Springer},
  \href{http://dx.doi.org/10.1007/978-3-642-40477-1_31}{Berlin\discretionary{/}{}{/}Heidelberg}.
  \href{http://dx.doi.org/10.1007/978-3-642-40477-1_31}
{doi: {{%
10\hspace{.1pt}\discretionary{.}{%
}{.}\hspace{.4pt}1007\discretionary{/}{%
}{/}978\discretionary{%
}{-}{-}3\discretionary{%
}{-}{-}642\discretionary{%
}{-}{-}40477\discretionary{%
}{-}{-}1\_31}}}


\bibitem[Tsang et~al., 2002]{tsang02}
\href{http://dx.doi.org/10.1145/571985.572001}{Tsang, M., Fitzmaurice, G.~W.,
  Kurtenbach, G., Khan, A., and Buxton, B.} (2002).
\newblock \href{http://dx.doi.org/10.1145/571985.572001}{Boom chameleon:
  Simultaneous capture of {3D} viewpoint, voice and gesture annotations on a
  spatially-aware display}.
\newblock \href{http://dx.doi.org/10.1145/571985.572001}{In {\em Proc.\ UIST}},
  \href{http://dx.doi.org/10.1145/571985.572001}{pages 111--120}.
  \href{http://dx.doi.org/10.1145/571985.572001}{ACM},
  \href{http://dx.doi.org/10.1145/571985.572001}{New York}.
  \href{http://dx.doi.org/10.1145/571985.572001}
{doi: {{%
10\hspace{.1pt}\discretionary{.}{%
}{.}\hspace{.4pt}1145\discretionary{/}{%
}{/}571985\hspace{.1pt}\discretionary{.}{%
}{.}\hspace{.4pt}572001}}}


\bibitem[van Rhijn and Mulder, 2006]{vanrhijn06}
\href{http://dx.doi.org/10.1145/1180495.1180507}{van Rhijn, A. and Mulder,
  J.~D.} (2006).
\newblock \href{http://dx.doi.org/10.1145/1180495.1180507}{Spatial input device
  structure and bimanual object manipulation in virtual environments}.
\newblock \href{http://dx.doi.org/10.1145/1180495.1180507}{In {\em Proc.\
  VRST}}, \href{http://dx.doi.org/10.1145/1180495.1180507}{pages 51--60}.
  \href{http://dx.doi.org/10.1145/1180495.1180507}{ACM},
  \href{http://dx.doi.org/10.1145/1180495.1180507}{New York}.
  \href{http://dx.doi.org/10.1145/1180495.1180507}
{doi: {{%
10\hspace{.1pt}\discretionary{.}{%
}{.}\hspace{.4pt}1145\discretionary{/}{%
}{/}1180495\hspace{.1pt}\discretionary{.}{%
}{.}\hspace{.4pt}1180507}}}


\bibitem[Wang et~al., 2006]{wang06}
\href{http://dx.doi.org/10.1145/1166253.1166270}{Wang, J., Zhai, S., and Canny,
  J.} (2006).
\newblock \href{http://dx.doi.org/10.1145/1166253.1166270}{Camera phone based
  motion sensing: Interaction techniques, applications and performance study}.
\newblock \href{http://dx.doi.org/10.1145/1166253.1166270}{In {\em Proc.\
  UIST}}, \href{http://dx.doi.org/10.1145/1166253.1166270}{pages 101--110}.
  \href{http://dx.doi.org/10.1145/1166253.1166270}{ACM},
  \href{http://dx.doi.org/10.1145/1166253.1166270}{New York}.
  \href{http://dx.doi.org/10.1145/1166253.1166270}
{doi: {{%
10\hspace{.1pt}\discretionary{.}{%
}{.}\hspace{.4pt}1145\discretionary{/}{%
}{/}1166253\hspace{.1pt}\discretionary{.}{%
}{.}\hspace{.4pt}1166270}}}


\bibitem[Ware and Arsenault, 2004]{ware04}
\href{http://dx.doi.org/10.1145/1012551.1012576}{Ware, C. and Arsenault, R.}
  (2004).
\newblock \href{http://dx.doi.org/10.1145/1012551.1012576}{Frames of reference
  in virtual object rotation}.
\newblock \href{http://dx.doi.org/10.1145/1012551.1012576}{In {\em Proc.\
  APGV}}, \href{http://dx.doi.org/10.1145/1012551.1012576}{pages 135--141}.
  \href{http://dx.doi.org/10.1145/1012551.1012576}{ACM},
  \href{http://dx.doi.org/10.1145/1012551.1012576}{New York}.
  \href{http://dx.doi.org/10.1145/1012551.1012576}
{doi: {{%
10\hspace{.1pt}\discretionary{.}{%
}{.}\hspace{.4pt}1145\discretionary{/}{%
}{/}1012551\hspace{.1pt}\discretionary{.}{%
}{.}\hspace{.4pt}1012576}}}


\bibitem[Ware and Baxter, 1989]{ware89}
\href{http://dx.doi.org/10.1145/67449.67482}{Ware, C. and Baxter, C.} (1989).
\newblock \href{http://dx.doi.org/10.1145/67449.67482}{{B}at {B}rushes: On the
  uses of six position and orientation parameters in a paint program}.
\newblock \href{http://dx.doi.org/10.1145/67449.67482}{In {\em Proc.\ CHI}},
  \href{http://dx.doi.org/10.1145/67449.67482}{pages 155--160}.
  \href{http://dx.doi.org/10.1145/67449.67482}{ACM},
  \href{http://dx.doi.org/10.1145/67449.67482}{New York}.
  \href{http://dx.doi.org/10.1145/67449.67482}
{doi: {{%
10\hspace{.1pt}\discretionary{.}{%
}{.}\hspace{.4pt}1145\discretionary{/}{%
}{/}67449\hspace{.1pt}\discretionary{.}{%
}{.}\hspace{.4pt}67482}}}


\bibitem[Warrick, 1947]{warrick47}
Warrick, M.~J. (1947).
\newblock Direction of movement in the use of control knobs to position visual
  indicators.
\newblock {\em Psychological Research on Equipment Design}, pages 137--146.

\bibitem[Weberg et~al., 2001]{weberg01}
\href{http://dx.doi.org/10.1145/634067.634320}{Weberg, L., Brange, T., and
  Wendelbo-Hansson, {\AA}.} (2001).
\newblock \href{http://dx.doi.org/10.1145/634067.634320}{A piece of butter on
  the {PDA} display}.
\newblock \href{http://dx.doi.org/10.1145/634067.634320}{In {\em CHI Extended
  Abstracts}}, \href{http://dx.doi.org/10.1145/634067.634320}{pages 435--436}.
  \href{http://dx.doi.org/10.1145/634067.634320}{ACM},
  \href{http://dx.doi.org/10.1145/634067.634320}{New York}.
  \href{http://dx.doi.org/10.1145/634067.634320}
{doi: {{%
10\hspace{.1pt}\discretionary{.}{%
}{.}\hspace{.4pt}1145\discretionary{/}{%
}{/}634067\hspace{.1pt}\discretionary{.}{%
}{.}\hspace{.4pt}634320}}}


\bibitem[Wiebe and Vu, 2009]{wiebe09}
\href{http://dx.doi.org/10.1007/978-3-642-02556-3_81}{Wiebe, J. and Vu,
  K.-P.~L.} (2009).
\newblock \href{http://dx.doi.org/10.1007/978-3-642-02556-3_81}{Application of
  population stereotypes to computerized tasks}.
\newblock \href{http://dx.doi.org/10.1007/978-3-642-02556-3_81}{In {\em Proc.\
  Human Interface and the Management of Information}},
  \href{http://dx.doi.org/10.1007/978-3-642-02556-3_81}{pages 718--725}.
  \href{http://dx.doi.org/10.1007/978-3-642-02556-3_81}{Springer},
  \href{http://dx.doi.org/10.1007/978-3-642-02556-3_81}{Berlin\discretionary{/}{}{/}Heidelberg}.
  \href{http://dx.doi.org/10.1007/978-3-642-02556-3_81}
{doi: {{%
10\hspace{.1pt}\discretionary{.}{%
}{.}\hspace{.4pt}1007\discretionary{/}{%
}{/}978\discretionary{%
}{-}{-}3\discretionary{%
}{-}{-}642\discretionary{%
}{-}{-}02556\discretionary{%
}{-}{-}3\_81}}}


\bibitem[Yee, 2003]{yee03}
\href{http://dx.doi.org/10.1145/642611.642613}{Yee, K.-P.} (2003).
\newblock \href{http://dx.doi.org/10.1145/642611.642613}{Peephole displays: Pen
  interaction on spatially aware handheld computers}.
\newblock \href{http://dx.doi.org/10.1145/642611.642613}{In {\em Proc.\ CHI}},
  \href{http://dx.doi.org/10.1145/642611.642613}{pages 1--8}.
  \href{http://dx.doi.org/10.1145/642611.642613}{ACM},
  \href{http://dx.doi.org/10.1145/642611.642613}{New York}.
  \href{http://dx.doi.org/10.1145/642611.642613}
{doi: {{%
10\hspace{.1pt}\discretionary{.}{%
}{.}\hspace{.4pt}1145\discretionary{/}{%
}{/}642611\hspace{.1pt}\discretionary{.}{%
}{.}\hspace{.4pt}642613}}}


\bibitem[Zhai, 1995]{zhai95}
\href{http://citeseerx.ist.psu.edu/viewdoc/download?doi=10.1.1.71.2326&rep=rep1&type=pdf}{Zhai,
  S.} (1995).
\newblock
  \href{http://citeseerx.ist.psu.edu/viewdoc/download?doi=10.1.1.71.2326&rep=rep1&type=pdf}{{\em
  Human performance in six degree of freedom input control}}.
\newblock
  \href{http://citeseerx.ist.psu.edu/viewdoc/download?doi=10.1.1.71.2326&rep=rep1&type=pdf}{PhD
  thesis},
  \href{http://citeseerx.ist.psu.edu/viewdoc/download?doi=10.1.1.71.2326&rep=rep1&type=pdf}{University
  of Toronto},
  \href{http://citeseerx.ist.psu.edu/viewdoc/download?doi=10.1.1.71.2326&rep=rep1&type=pdf}{Canada}.

\bibitem[Zhai, 1998]{zhai98-2}
\href{http://dx.doi.org/10.1145/307710.307728}{Zhai, S.} (1998).
\newblock \href{http://dx.doi.org/10.1145/307710.307728}{User performance in
  relation to {3D} input device design}.
\newblock \href{http://dx.doi.org/10.1145/307710.307728}{{\em ACM SIGGRAPH
  Computer Graphics}},
  \href{http://dx.doi.org/10.1145/307710.307728}{32(4):50--54}.
  \href{http://dx.doi.org/10.1145/307710.307728}
{doi: {{%
10\hspace{.1pt}\discretionary{.}{%
}{.}\hspace{.4pt}1145\discretionary{/}{%
}{/}307710\hspace{.1pt}\discretionary{.}{%
}{.}\hspace{.4pt}307728}}}


\bibitem[Zhai and Milgram, 1998]{zhai98}
\href{http://dx.doi.org/10.1145/274644.274689}{Zhai, S. and Milgram, P.}
  (1998).
\newblock \href{http://dx.doi.org/10.1145/274644.274689}{Quantifying
  coordination in multiple {DOF} movement and its application to evaluating 6
  {DOF} input devices}.
\newblock \href{http://dx.doi.org/10.1145/274644.274689}{In {\em Proc.\ CHI}},
  \href{http://dx.doi.org/10.1145/274644.274689}{pages 320--327}.
  \href{http://dx.doi.org/10.1145/274644.274689}{ACM},
  \href{http://dx.doi.org/10.1145/274644.274689}{New York}.
  \href{http://dx.doi.org/10.1145/274644.274689}
{doi: {{%
10\hspace{.1pt}\discretionary{.}{%
}{.}\hspace{.4pt}1145\discretionary{/}{%
}{/}274644\hspace{.1pt}\discretionary{.}{%
}{.}\hspace{.4pt}274689}}}


\end{thebibliography}

\newpage
\section*{Appendix: Pseudocode}
\label{sec:app:pseudocode}

\algnewcommand{\LineComment}[1]{\State {\color{black!65}\(\triangleright\) \textit{#1}}}

\makeatletter
\renewcommand{\ALG@name}{Code}
\makeatother

\setlength{\intextsep}{.8mm}

\begin{algorithm}[H]
  \caption{Absolute position control mapping}
  \label{code:absolute-mapping}
  \begin{algorithmic}[1]
    \vspace{.5mm}
    \State \Call {Init}{}()
    \Loop
    \State $\vec{v} \gets \mathrm{tracker\_pos}()-\vec{pc_0}$
    \State $\quat{r} \gets \mathrm{tracker\_rot}()\times\mathrm{inv}(\quat{qc_0})$
    \State $(\vec{v}, \quat{r}) \gets$ \Call {Transform}{$\vec{v}, \quat{r}, 0, t$}
    \State object\_pos $\gets \vec{v} + \vec{pd_0}$
    \State object\_rot $\gets \quat{r} \times \quat{qd_0}$
    \vspace{.25mm}
    \State $t \gets t+1$
    \EndLoop
  \end{algorithmic}
\end{algorithm}

\begin{algorithm}[H]
  \caption{Relative position control mapping}
  \label{code:relative-mapping}
  \begin{algorithmic}[1]
    \vspace{.5mm}
    \State \Call {Init}{}()
    \Loop
    \State $\vec{v} \gets \vec{pc_t}-\vec{pc_{t-1}}$
    \State $\quat{r} \gets \quat{qc_t}\times\mathrm{inv}(\quat{qc_{t-1}})$
    \State $(\vec{v}, \quat{r}) \gets$ \Call {Transform}{$\vec{v}, \quat{r}, t, t{-}1$}
    \State object\_pos $\gets \vec{v} + {}$object\_pos
    \State object\_rot $\gets \quat{r} \times {}$object\_rot
    \vspace{.25mm}
    \State $\vec{pc_t} \gets \mathrm{tracker\_pos}()$
    \State $\quat{qc_t} \gets \mathrm{tracker\_rot}()$
    \State $t \gets t+1$
    \EndLoop
  \end{algorithmic}
\end{algorithm}

\begin{algorithm}[H]
  \caption{Rate control mapping}
  \label{code:rate-mapping}
  \begin{algorithmic}[1]
    \vspace{.5mm}
    \State \Call {Init}{}()
    \Loop
    \State $\vec{v} \gets \vec{pc_t}-\vec{pc_0}$
    \State $\quat{r} \gets \quat{qc_t}\times\mathrm{inv}(\quat{qc_0})$
    \State $(\vec{v}, \quat{r}) \gets$ \Call {Transform}{$\vec{v}, \quat{r}, t, 0$}
    \State object\_pos $\gets \vec{v} + {}$object\_pos
    \State object\_rot $\gets \quat{r} \times {}$object\_rot
    \vspace{.25mm}
    \State $\vec{pc_t} \gets \mathrm{tracker\_pos}()$
    \State $\quat{qc_t} \gets \mathrm{tracker\_rot}()$
    \State $t \gets t+1$
    \EndLoop
  \end{algorithmic}
\end{algorithm}

\begin{algorithm}[H]
  \caption{Common code\vspace{.6mm}}
  \label{code:common}
  \begin{algorithmic}[1]
    \vspace{.5mm}
    \Procedure {Init}{}()
    \State $t \gets 0$
    \State $\vec{pc_0} \gets \mathrm{tracker\_pos}()$
    \State $\quat{qc_0} \gets \mathrm{tracker\_rot}()$
    \State $\vec{pd_0} \gets$ object\_pos
    \State $\quat{qd_0} \gets$ object\_rot
    \State $t \gets t+1$
    \EndProcedure
    \vspace{.5mm}
    \Function {Transform} {$\vec{v}, \quat{r}, t, \mathit{from\_t}$}
    \LineComment{Conversion to screen space}
    \vspace{-.25mm}
    \State $\quat{q} \gets \quat{qc_\mathit{from\_t}}$
    \State $\vec{v} \gets (\mathrm{inv}(\quat{q}) \times \mathrm{quat}(v_x,v_y,v_z,0) \times \quat{q})_{xyz}$
    \State $\quat{r} \gets \mathrm{inv}(\quat{q}) \times r \times \quat{q}$
    \vspace{.5mm}
    \LineComment{Control-Display gain}
    \vspace{-.25mm}
    \State $k_t \gets \gain(t)$
    \State $\vec{v} \gets k_t \times \vec{v}$
    \State $\quat{r} \gets \slerp(\mathrm{identity}, \quat{r}, k_t)$
    \vspace{.5mm}
    \State \Return $(\vec{v}, \vec{r})$
    \EndFunction
  \end{algorithmic}
\end{algorithm}

\end{document}